\documentclass[aps,prl,showpacs,twocolumn,superscriptaddress,amsmath,amssymb]{revtex4}

\usepackage{epsfig,afterpage}
\usepackage{graphicx}
\usepackage{amsfonts,bbold}
\usepackage{amssymb}
\usepackage{indentfirst}
\usepackage{amsmath,amsthm}
\usepackage{dsfont}

\usepackage{epsfig}
\usepackage{subfigure}

\usepackage{multirow}
\usepackage{pstricks}
\usepackage{psfrag}
\usepackage{lscape}

\usepackage{wasysym} 

\def\beq{\begin{equation}}
\def\eeq{\end{equation}}
\def\bea{\begin{eqnarray}}
\def\eea{\end{eqnarray}}
\def\bq{\begin{quote}}
\def\eq{\end{quote}}
\def\ben{\begin{enumerate}}
\def\een{\end{enumerate}}
\def\bit{\begin{itemize}}
\def\eit{\end{itemize}}

\begin{document}

\title{Strong and weak thermalization of infinite non-integrable quantum systems}

 \author{M. C. Ba\~nuls}\email{banulsm@mpq.mpg.de} 
 \affiliation{Max-Planck-Institut f\"ur Quantenoptik,
   Hans-Kopfermann-Str. 1, 85748 Garching, Germany.}
 \author{J. I. Cirac}
 \affiliation{Max-Planck-Institut f\"ur Quantenoptik,
   Hans-Kopfermann-Str. 1, 85748 Garching, Germany.}
 \author{M. B. Hastings}
 \affiliation{Microsoft Research, Station Q, CNSI Building, 
 University of California, 
   Santa Barbara, CA, 93106.}

\begin{abstract}
When a non-integrable system evolves out of equilibrium for a long 
 time, local observables 
are expected to attain stationary expectation values, 
independent of the details of the initial state. 
However, intriguing experimental results with ultracold 
gases~\cite{kinoshita06,hofferberth07} have shown no thermalization in
non-integrable settings, triggering an intense
theoretical effort 
to decide the question~\cite{rigol07gge,kollath07quench,manmana07,cramer08exact,rigol08nat,rigol09break}.  Here we show that
the phenomenology of thermalization in a quantum system is much richer
than its classical counterpart.  Using a new numerical technique, we
identify two 
distinct thermalization regimes, strong and weak,
occurring 
for different initial states.
Strong thermalization, intrinsically quantum, 
happens when instantaneous local expectation
values 
converge to the thermal ones.
Weak thermalization, well-known in classical systems,
happens when local expectation values converge to the thermal ones
only after time averaging.
 Remarkably, 
we find a third group of states 
showing no
thermalization, neither strong nor weak, 
to the time scales one can reliably simulate.
\end{abstract}

\maketitle



In the 19th century, statistical mechanics was developed as a microscopic
model explaining the fundamental results of thermodynamics.  Starting
in the early 20th century,
quantum statistical mechanics has been developed to describe thermodynamics
of quantum systems.  However, current experiments ~\cite{kinoshita06,hofferberth07}
 with ultracold
atoms have attained a level of isolation and control of parameters that
forces us to address not just the equilibrium properties
of quantum systems but also the question of how and why these systems relax to
equilibrium starting from a nonthermal state.
For classical systems, Boltzmann's molecular chaos assumption provides
a quantitative tool to describe this relaxation, and
ergodic properties of the system give the explanation.  If the time dynamics
of a system explores all states with a given energy with uniform probability,
then the long-time {\it average} of any observed quantity will approach the
expectation value for that quantity in the microcanonical ensemble.  

Similarly, the emergence of a thermal bath~\cite{deutsch91} in a
quantum system has been justified under the assumption that the
long-time average of typical initial states produces a density matrix
that approaches the thermal
average~\cite{deutsch91,berges01T,rigol08nat}.  It has been proposed,
however, that thermalization may happen without any time average in
the quantum case~\cite{srednicki94}; indeed, it is possible that, due
to quantum entanglement, starting from fixed, non-thermal initial
conditions, the reduced density matrix at a given time $t$ on a given
region $A$, such that $A$ is small compared to the system size, will
converge to the thermal expectation value at long times $t$.  Such a
phenomenon, which we call ``strong thermalization", cannot occur for a
classical system as it relies on the quantum mechanical fact that even
if the global density matrix is a pure state, the reduced density
matrix may be a mixed state.


For integrable systems the existence of local conserved quantities
prevents relaxation to the thermal state, which is constrained only by
energy.  Instead, it has been suggested that these systems will relax
to a state described by a generalized thermal ensemble, compatible
with the set of conserved quantities~\cite{rigol07gge,rigol08nat,cramer08exact}.  In
an infinite non-integrable system, there exist an infinite number of
conserved quantities, such as the powers of the Hamiltonian. However, not
being local, they are not expected to prevent the
thermalization of
local observables~\cite{deutsch91,srednicki94,rigol08nat}.

Remarkably enough, a recent experiment~\cite{kinoshita06} observed no
signs of thermalization after long evolution in a nearly integrable
case.  The results could be in part understood with the hypothesis of
relaxation to a generalized thermal ensemble~\cite{rigol07gge} in an
integrable case, but from the theoretical point of view it is not
clear why no thermalization was present away from true integrability.

Various theoretical studies have tried to elucidate the question of
whether or under which conditions thermalization will occur in non-integrable
models~\cite{kollath07quench,manmana07,rigol08nat,flesch08probing,moeckel08,rigol09break,keilmann09}. 
 The study 
has to resort to numerical techniques, given
the lack of analytical solutions.  Moreover, the
intrinsic computational complexity of simulating large quantum systems
limits the affordable studies to finite systems or short times,
so that results showing non-thermalization cannot be extrapolated to infinite 
times or system sizes.

Most of the studies have 
 tried to decide
whether a given model thermalizes or not as a function of
the Hamiltonian parameters~\cite{manmana07,kollath07quench,flesch08probing,rigol09break}. 
Recent works~\cite{cassidy09threshold,santos10onset}.
have also analyzed 
the link between appearance of quantum chaos and thermalization in
these systems.


Here we consider the question of relaxation using a fixed,
non-integrable Hamiltonian, but a range of initial states.  We
discover a rich phenomenology of different relaxation
regimes. 
The analyzed system is an infinite translationally invariant spin
chain with a nearest-neighbor interaction.  
Thanks to a recently
developed algorithm~\cite{banuls09fold}, we are able to explore 
its dynamics 
at relatively long times.
Furthermore, the method gives us access to the whole reduced density
matrix for a block of few particles.  Instead of analyzing the
behavior of individual expectation values, as previous numerical
studies, we may then 
quantify the degree of thermalization as the 
distance between the reduced density matrix of the evolved state and
the thermal state, $\rho_{th}(\beta)=\frac{\mathrm{e}^{-\beta
    H}}{\mathrm{tr}(\mathrm{e}^{-\beta H})}$, corresponding to the
same energy.  
 We estimate this distance
both for the instantaneous reduced state, $\rho(t)$, and for the time
averaged one
$\bar\rho(t)\equiv\frac{1}{t/2}\int_{\frac{t}{2}}^t\rho(\tau)
d\tau$~\footnote{We average from half time in order to eliminate the
  shortest time dynamics and study the long time behavior.}.


We find two clearly distinct thermalization regimes. Some states
show \emph{strong thermalization}. Their instantaneous expectation
values converge (fast) to the thermal ones, without the need to
consider the time average.  For other states we encounter instead
\emph{weak thermalization}. We have strong numerical evidence that
such states do not relax even at very long times.  Nevertheless, the
time averaged observables do relax to the thermal values.  Finally, we
also obtain weaker numerical evidence that for some third group of
states no relaxation occurs, at least to the longest time scales we
can reliably simulate.  
Strikingly, this situation occurs even in regions of energy for which
the spectrum of the system is shown to be chaotic (see Supplementary material).
Since our analysis is numerical, it may be
argued that the conclusions are only valid for a limited range of
time.
However, the study of the properties of the
method~\cite{banuls09fold}, as well as the detailed analysis of errors
in the current study (described in the Supplementary material) ensure that the
reliability of our results extends to longer times than that of
previous studies, enough to give strong evidence about the existence of
these different quantum thermalization regimes.

\begin{figure}[floatfix]
%
%
\begin{psfrags}%
\psfragscanon%
%
\psfrag{s01}[t][t]{\color[rgb]{0,0,0}\setlength{\tabcolsep}{0pt}\begin{tabular}{c}t\end{tabular}}%
\psfrag{s02}[b][b]{\color[rgb]{0,0,0}\setlength{\tabcolsep}{0pt}\begin{tabular}{c}$d(\rho(t),\rho_{th})$\end{tabular}}%
\psfrag{s07}[B][b]{\color[rgb]{0,0,0}\setlength{\tabcolsep}{0pt}\begin{tabular}{c}$ d(\bar(\rho)(t),\rho_{th})$\end{tabular}}%
\psfrag{s11}[t][b]{\color[rgb]{0,0,0}\setlength{\tabcolsep}{0pt}\begin{tabular}{c}$ \langle O(t)\rangle-\langle O\rangle_{th}$\end{tabular}}%
%
\psfrag{x01}[t][t]{0}%
\psfrag{x02}[t][t]{0.1}%
\psfrag{x03}[t][t]{0.2}%
\psfrag{x04}[t][t]{0.3}%
\psfrag{x05}[t][t]{0.4}%
\psfrag{x06}[t][t]{0.5}%
\psfrag{x07}[t][t]{0.6}%
\psfrag{x08}[t][t]{0.7}%
\psfrag{x09}[t][t]{0.8}%
\psfrag{x10}[t][t]{0.9}%
\psfrag{x11}[t][t]{1}%
\psfrag{x12}[t][t]{0}%
\psfrag{x13}[t][t]{5}%
\psfrag{x14}[t][t]{10}%
\psfrag{x15}[t][t]{2}%
\psfrag{x16}[t][t]{4}%
\psfrag{x17}[t][t]{6}%
\psfrag{x18}[t][t]{8}%
\psfrag{x19}[t][t]{10}%
\psfrag{x20}[t][t]{12}%
\psfrag{x21}[t][t]{2}%
\psfrag{x22}[t][t]{3}%
\psfrag{x23}[t][t]{4}%
\psfrag{x24}[t][t]{5}%
\psfrag{x25}[t][t]{6}%
\psfrag{x26}[t][t]{7}%
\psfrag{x27}[t][t]{8}%
\psfrag{x28}[t][t]{9}%
\psfrag{x29}[t][t]{10}%
\psfrag{x30}[t][t]{11}%
\psfrag{x31}[t][t]{12}%
%
\psfrag{v01}[r][r]{0}%
\psfrag{v02}[r][r]{0.1}%
\psfrag{v03}[r][r]{0.2}%
\psfrag{v04}[r][r]{0.3}%
\psfrag{v05}[r][r]{0.4}%
\psfrag{v06}[r][r]{0.5}%
\psfrag{v07}[r][r]{0.6}%
\psfrag{v08}[r][r]{0.7}%
\psfrag{v09}[r][r]{0.8}%
\psfrag{v10}[r][r]{0.9}%
\psfrag{v11}[r][r]{1}%
\psfrag{v12}[r][r]{-0.5}%
\psfrag{v13}[r][r]{0}%
\psfrag{v14}[r][r]{0.5}%
\psfrag{v15}[r][r]{0}%
\psfrag{v16}[r][r]{0.1}%
\psfrag{v17}[r][r]{0.2}%
\psfrag{v18}[r][r]{0}%
\psfrag{v19}[r][r]{0.05}%
\psfrag{v20}[r][r]{0.1}%
\psfrag{v21}[r][r]{0.15}%
\psfrag{v22}[r][r]{0.2}%
%
\resizebox{\columnwidth}{!}{\includegraphics{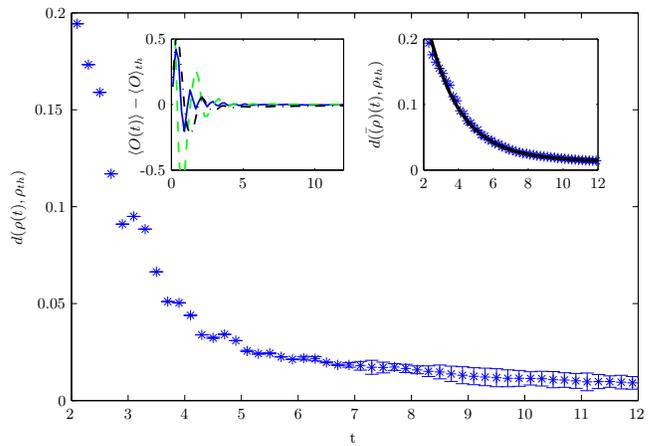}}%
\end{psfrags}%
%

\caption{Strong thermalization: initial state $|Y+\rangle$.  The main
  plot shows the distance between the three-body reduced density
  matrix at instant $t$ and the corresponding thermal state
  ($\beta=0$).  The error bars show the difference between the result
  with the largest bond dimension ($D=120$) and with a lower one
  ($D=60$), which gives us an estimate of the truncation error.  The
  right inset shows the distance between the time-averaged reduced
  density matrix and the thermal state. We superimpose a fit to a curve
  decaying like $b/\sqrt{t}$ for long times (solid black line).  The left inset
  shows the difference between the thermal expectation values and the
  time dependent single body observables $\langle \sigma_x\rangle$
  (blue solid line), $\langle \sigma_y\rangle$ (dashed green) and
  $\langle \sigma_z\rangle$ (dash-dotted black line). Convergence to
  the thermal state is observed in all three plots.  }
\label{fig:instant}
\label{fig:Yall}
\end{figure}

The first regime, or \emph{strong thermalization}, is illustrated by
the initial state $|Y+\rangle$  (Fig.~\ref{fig:Yall}), in which all spins are aligned along
the positive $\hat y$ direction, corresponding to zero initial energy, and thus
$\beta=0$. The density matrix at any time converges fast to the
thermal distribution at the same energy. 
As a consequence, the time averaged density matrix also converges to the 
thermal values.

\begin{figure}[floatfix]
%
%
\begin{psfrags}%
\psfragscanon%
%
\psfrag{s01}[t][t]{\color[rgb]{0,0,0}\setlength{\tabcolsep}{0pt}\begin{tabular}{c}$t$\end{tabular}}%
\psfrag{s02}[b][b]{\color[rgb]{0,0,0}\setlength{\tabcolsep}{0pt}\begin{tabular}{c}$d(\rho(t),\rho_{th})$\end{tabular}}%
\psfrag{s07}[b][b]{\color[rgb]{0,0,0}\setlength{\tabcolsep}{0pt}\begin{tabular}{c}$ d(\bar(\rho)(t),\rho_{th})$\end{tabular}}%
\psfrag{s11}[t][B]{\color[rgb]{0,0,0}\setlength{\tabcolsep}{0pt}\begin{tabular}{c}$ \langle O(t)\rangle-\langle O\rangle_{th}$\end{tabular}}%
%
\psfrag{x01}[t][t]{0}%
\psfrag{x02}[t][t]{0.1}%
\psfrag{x03}[t][t]{0.2}%
\psfrag{x04}[t][t]{0.3}%
\psfrag{x05}[t][t]{0.4}%
\psfrag{x06}[t][t]{0.5}%
\psfrag{x07}[t][t]{0.6}%
\psfrag{x08}[t][t]{0.7}%
\psfrag{x09}[t][t]{0.8}%
\psfrag{x10}[t][t]{0.9}%
\psfrag{x11}[t][t]{1}%
\psfrag{x12}[t][t]{0}%
\psfrag{x13}[t][t]{5}%
\psfrag{x14}[t][t]{10}%
\psfrag{x15}[t][t]{15}%
\psfrag{x16}[t][t]{0}%
\psfrag{x17}[t][t]{5}%
\psfrag{x18}[t][t]{10}%
\psfrag{x19}[t][t]{15}%
\psfrag{x20}[t][t]{0}%
\psfrag{x21}[t][t]{2}%
\psfrag{x22}[t][t]{4}%
\psfrag{x23}[t][t]{6}%
\psfrag{x24}[t][t]{8}%
\psfrag{x25}[t][t]{10}%
\psfrag{x26}[t][t]{12}%
\psfrag{x27}[t][t]{14}%
\psfrag{x28}[t][t]{16}%
\psfrag{x29}[t][t]{18}%
%
\psfrag{v01}[r][r]{0}%
\psfrag{v02}[r][r]{0.1}%
\psfrag{v03}[r][r]{0.2}%
\psfrag{v04}[r][r]{0.3}%
\psfrag{v05}[r][r]{0.4}%
\psfrag{v06}[r][r]{0.5}%
\psfrag{v07}[r][r]{0.6}%
\psfrag{v08}[r][r]{0.7}%
\psfrag{v09}[r][r]{0.8}%
\psfrag{v10}[r][r]{0.9}%
\psfrag{v11}[r][r]{1}%
\psfrag{v12}[r][r]{-0.2}%
\psfrag{v13}[r][r]{0}%
\psfrag{v14}[r][r]{0.2}%
\psfrag{v15}[r][r]{0}%
\psfrag{v16}[r][r]{0.05}%
\psfrag{v17}[r][r]{0.1}%
\psfrag{v18}[r][r]{0}%
\psfrag{v19}[r][r]{0.1}%
\psfrag{v20}[r][r]{0.2}%
\psfrag{v21}[r][r]{0.3}%
\psfrag{v22}[r][r]{0.4}%
\psfrag{v23}[r][r]{0.5}%
%
\resizebox{\columnwidth}{!}{\includegraphics{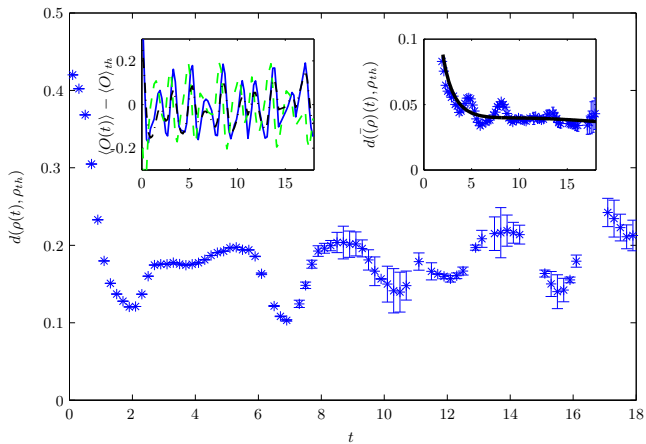}}%
\end{psfrags}%
%

\caption{ Weak thermalization: initial state $|Z+\rangle$.  The
  distance between the reduced density matrix for three sites and the
  thermal state of the same energy ($\beta=0.7275$) oscillates
  strongly with time. The right inset shows the distance between the
  time averaged reduced density matrix and the thermal ensemble. The
  superimposed solid line, fit to a curve which behaves as $b/\sqrt{t}$ for long times, shows that the
  behavior is compatible with a slow convergence only in average.  The
  left inset shows the oscillations of the individual single-body time
  dependent expectation values around the thermal ones. No sign of
  damping is observed for the longest times ($t\sim 18$) we have
  simulated.  These results were obtained with bond dimension $D=240$,
  and the error bars show the difference to $D=120$ results.}
\label{fig:average}
\label{fig:Zall}
\end{figure}

We identify an example of the second regime, or \emph{weak
  thermalization} in the initial state $|Z+\rangle$
(Fig.~\ref{fig:Zall}), in which all spins are initially aligned along
the positive $\hat z$ direction. For this state, whose energy matches
that of a thermal state with $\beta\approx0.7275$, no relaxation at
all is observed for long times. Instead, the distance between the
evolved and the thermal reduced density matrices seems to oscillate
strongly for arbitrarily long times.  All expectation values of the
form $\langle\sigma_a\otimes\sigma_b\otimes\sigma_c\rangle$ show
similar non-damped irregular oscillating behavior.  Since the entropy
increases linearly with time in the $|Z+\rangle$ state, the behavior
is {\it not} due to a lack of propagating excitations (see
Supplementary material).  The time averaged density matrix shows
instead an only slightly oscillating behavior, compatible with a very
slow convergence to the thermal state, at a rate $1/\sqrt{t}$.  This
rate is characteristic of diffusive relaxation in one dimension, and
has been seen elsewhere~\cite{flesch08probing}.

\begin{figure}[floatfix]
%
%
\begin{psfrags}%
\psfragscanon%
%
\psfrag{s17}[t][t]{\color[rgb]{0,0,0}\setlength{\tabcolsep}{0pt}\begin{tabular}{c}$t$\end{tabular}}%
\psfrag{s18}[b][b]{\color[rgb]{0,0,0}\setlength{\tabcolsep}{0pt}\begin{tabular}{c}$ d(\rho(t),\rho_{th})$\end{tabular}}%
\psfrag{s23}[b][b]{\color[rgb]{0,0,0}\setlength{\tabcolsep}{0pt}\begin{tabular}{c}$ d(\bar(\rho)(t),\rho_{th})$\end{tabular}}%
\psfrag{s27}[t][b]{\color[rgb]{0,0,0}\setlength{\tabcolsep}{0pt}\begin{tabular}{c}$ \langle O(t)\rangle-\langle O\rangle_{th}$\end{tabular}}%
%
\psfrag{x01}[t][t]{0}%
\psfrag{x02}[t][t]{0.1}%
\psfrag{x03}[t][t]{0.2}%
\psfrag{x04}[t][t]{0.3}%
\psfrag{x05}[t][t]{0.4}%
\psfrag{x06}[t][t]{0.5}%
\psfrag{x07}[t][t]{0.6}%
\psfrag{x08}[t][t]{0.7}%
\psfrag{x09}[t][t]{0.8}%
\psfrag{x10}[t][t]{0.9}%
\psfrag{x11}[t][t]{1}%
\psfrag{x12}[t][t]{0}%
\psfrag{x13}[t][t]{5}%
\psfrag{x14}[t][t]{10}%
\psfrag{x15}[t][t]{0}%
\psfrag{x16}[t][t]{5}%
\psfrag{x17}[t][t]{10}%
\psfrag{x18}[t][t]{0}%
\psfrag{x19}[t][t]{2}%
\psfrag{x20}[t][t]{4}%
\psfrag{x21}[t][t]{6}%
\psfrag{x22}[t][t]{8}%
\psfrag{x23}[t][t]{10}%
\psfrag{x24}[t][t]{12}%
\psfrag{x25}[t][t]{0}%
\psfrag{x26}[t][t]{0.1}%
\psfrag{x27}[t][t]{0.2}%
\psfrag{x28}[t][t]{0.3}%
\psfrag{x29}[t][t]{0.4}%
\psfrag{x30}[t][t]{0.5}%
\psfrag{x31}[t][t]{0.6}%
\psfrag{x32}[t][t]{0.7}%
\psfrag{x33}[t][t]{0.8}%
\psfrag{x34}[t][t]{0.9}%
\psfrag{x35}[t][t]{1}%
%
\psfrag{v01}[r][r]{0}%
\psfrag{v02}[r][r]{0.1}%
\psfrag{v03}[r][r]{0.2}%
\psfrag{v04}[r][r]{0.3}%
\psfrag{v05}[r][r]{0.4}%
\psfrag{v06}[r][r]{0.5}%
\psfrag{v07}[r][r]{0.6}%
\psfrag{v08}[r][r]{0.7}%
\psfrag{v09}[r][r]{0.8}%
\psfrag{v10}[r][r]{0.9}%
\psfrag{v11}[r][r]{1}%
\psfrag{v12}[r][r]{-0.2}%
\psfrag{v13}[r][r]{0}%
\psfrag{v14}[r][r]{0.2}%
\psfrag{v15}[r][r]{0}%
\psfrag{v16}[r][r]{0.02}%
\psfrag{v17}[r][r]{0.04}%
\psfrag{v18}[r][r]{0.06}%
\psfrag{v19}[r][r]{0.08}%
\psfrag{v20}[r][r]{0}%
\psfrag{v21}[r][r]{0.05}%
\psfrag{v22}[r][r]{0.1}%
\psfrag{v23}[r][r]{0.15}%
\psfrag{v24}[r][r]{0.2}%
\psfrag{v25}[r][r]{0.25}%
\psfrag{v26}[r][r]{0.3}%
\psfrag{v27}[r][r]{0}%
\psfrag{v28}[r][r]{0.1}%
\psfrag{v29}[r][r]{0.2}%
\psfrag{v30}[r][r]{0.3}%
\psfrag{v31}[r][r]{0.4}%
\psfrag{v32}[r][r]{0.5}%
\psfrag{v33}[r][r]{0.6}%
\psfrag{v34}[r][r]{0.7}%
\psfrag{v35}[r][r]{0.8}%
\psfrag{v36}[r][r]{0.9}%
\psfrag{v37}[r][r]{1}%
%
\resizebox{\columnwidth}{!}{\includegraphics{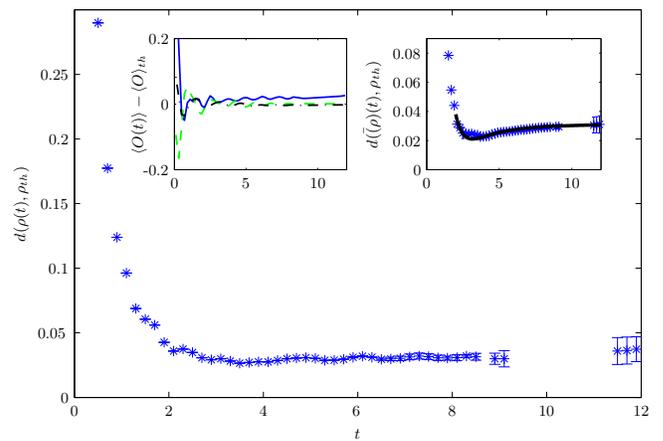}}%
\end{psfrags}%
%

\caption{No thermalization observed: initial state $|X+\rangle$.  The
  plot shows the time dependence of the distance between the evolved
  reduced density matrix for three sites and the thermal state
  ($\beta=-0.7180$).  No thermalization is observed in the
  instantaneous density matrix, nor in the time averaged one (right
  inset). On the latter, we superimpose a fit of the computed values
  to a time dependent function, which asymptotically tends to a
  constant ($0.03$). The plotted results correspond to a bond dimension $D=240$,
  while the distance to the results with $D=120$ is shown as error bars.
  In the left inset we plot, for the observables that determine the
  one site reduced density matrix, $\langle \sigma_x\rangle$ (blue
  solid line), $\langle \sigma_y\rangle$ (dashed green) and $\langle
  \sigma_z\rangle$ (dash-dotted black), the difference with respect to
  the thermal values. We observe that $\langle\sigma_x\rangle$ is
  the responsible for the lack of thermalization, while all 
  the other expectation values converge to the thermal averages.
  Studying the
  evolution of only a few local expectation values may not
  suffice then to detect the nonthermalization.
}
\label{fig:correls}
\label{fig:Xall}
\end{figure}

Finally, if the system is started in the $|X+\rangle$ state (Fig.~\ref{fig:Xall}), with all
spins aligned along the positive $\hat x$ direction, positive energy per particle, 
and corresponding to a Gibbs state with
$\beta\approx-0.7180$, the reduced density matrix shows initially a
fast relaxation. However at long times the distance between the evolved
state and the thermal one is different from zero, signaling that one
or more expectation values of local observables have not converged to
the thermal averages. We observe that the density matrix has not reached
a stationary value, either. Given the numerical character of the 
study,  this observation could indicate an absence of
thermalization, but also a much slower one. 
It is an intriguing case, worthy 
of further, ideally experimental, study.
Even in average, we detect no thermalization
of the $|X+\rangle$ state, up to the time
scales we are able to explore.

\begin{figure}[floatfix]
   \hspace{-.05\columnwidth}
   \begin{minipage}[c]{.3\columnwidth}
     \subfigure{
%
%
\begin{psfrags}%
\psfragscanon%
\Large
%
\psfrag{s13}[b][b]{\color[rgb]{0,0,0}\setlength{\tabcolsep}{0pt}\begin{tabular}{c}$|Z+\rangle$\end{tabular}}%
\psfrag{s22}[b][b]{\color[rgb]{0,0,0}\setlength{\tabcolsep}{0pt}\begin{tabular}{c}\end{tabular}}%
\psfrag{s23}[l][l]{\color[rgb]{0,0,0}\setlength{\tabcolsep}{0pt}\begin{tabular}{c}$|Y+\rangle$\end{tabular}}%
\psfrag{s24}[t][t]{\color[rgb]{0,0,0}\setlength{\tabcolsep}{0pt}\begin{tabular}{c}$|X+\rangle$\end{tabular}}%
%
\psfrag{x01}[t][t]{0}%
\psfrag{x02}[t][t]{0.1}%
\psfrag{x03}[t][t]{0.2}%
\psfrag{x04}[t][t]{0.3}%
\psfrag{x05}[t][t]{0.4}%
\psfrag{x06}[t][t]{0.5}%
\psfrag{x07}[t][t]{0.6}%
\psfrag{x08}[t][t]{0.7}%
\psfrag{x09}[t][t]{0.8}%
\psfrag{x10}[t][t]{0.9}%
\psfrag{x11}[t][t]{1}%
\psfrag{x12}[t][t]{-1}%
\psfrag{x13}[t][t]{-0.5}%
\psfrag{x14}[t][t]{0}%
\psfrag{x15}[t][t]{0.5}%
\psfrag{x16}[t][t]{1}%
%
\psfrag{v01}[r][r]{0}%
\psfrag{v02}[r][r]{0.1}%
\psfrag{v03}[r][r]{0.2}%
\psfrag{v04}[r][r]{0.3}%
\psfrag{v05}[r][r]{0.4}%
\psfrag{v06}[r][r]{0.5}%
\psfrag{v07}[r][r]{0.6}%
\psfrag{v08}[r][r]{0.7}%
\psfrag{v09}[r][r]{0.8}%
\psfrag{v10}[r][r]{0.9}%
\psfrag{v11}[r][r]{1}%
\psfrag{v12}[r][r]{-1}%
\psfrag{v13}[r][r]{-0.5}%
\psfrag{v14}[r][r]{0}%
\psfrag{v15}[r][r]{0.5}%
\psfrag{v16}[r][r]{1}%
%
\psfrag{z01}[r][r]{-1}%
\psfrag{z02}[r][r]{-0.5}%
\psfrag{z03}[r][r]{0}%
\psfrag{z04}[r][r]{0.5}%
\psfrag{z05}[r][r]{1}%
%
\resizebox{\columnwidth}{!}{\includegraphics{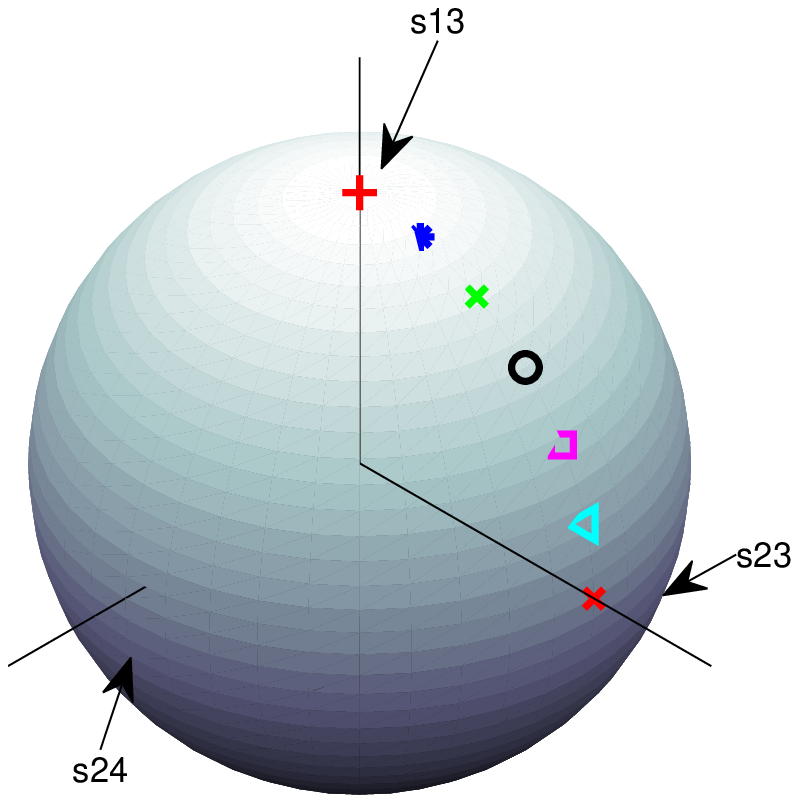}}%
\end{psfrags}%
%

     }
    
 \end{minipage}
 \hspace{.05\columnwidth}
 \begin{minipage}[c]{.6\columnwidth}
   \subfigure{
%
%
\begin{psfrags}%
\Large
\psfragscanon%
%
\psfrag{s05}[t][t]{\color[rgb]{0,0,0}\setlength{\tabcolsep}{0pt}\begin{tabular}{c}$t$\end{tabular}}%
\psfrag{s06}[b][t]{\color[rgb]{0,0,0}\setlength{\tabcolsep}{0pt}\begin{tabular}{c}$ d(\bar(\rho)(t),\rho_{th})$\end{tabular}}%
%
\psfrag{x01}[t][t]{0}%
\psfrag{x02}[t][t]{0.1}%
\psfrag{x03}[t][t]{0.2}%
\psfrag{x04}[t][t]{0.3}%
\psfrag{x05}[t][t]{0.4}%
\psfrag{x06}[t][t]{0.5}%
\psfrag{x07}[t][t]{0.6}%
\psfrag{x08}[t][t]{0.7}%
\psfrag{x09}[t][t]{0.8}%
\psfrag{x10}[t][t]{0.9}%
\psfrag{x11}[t][t]{1}%
\psfrag{x12}[t][t]{0}%
\psfrag{x13}[t][t]{1}%
\psfrag{x14}[t][t]{2}%
\psfrag{x15}[t][t]{3}%
\psfrag{x16}[t][t]{4}%
\psfrag{x17}[t][t]{5}%
\psfrag{x18}[t][t]{6}%
\psfrag{x19}[t][t]{7}%
\psfrag{x20}[t][t]{8}%
\psfrag{x21}[t][t]{9}%
\psfrag{x22}[t][t]{10}%
%
\psfrag{v01}[r][r]{0}%
\psfrag{v02}[r][r]{0.1}%
\psfrag{v03}[r][r]{0.2}%
\psfrag{v04}[r][r]{0.3}%
\psfrag{v05}[r][r]{0.4}%
\psfrag{v06}[r][r]{0.5}%
\psfrag{v07}[r][r]{0.6}%
\psfrag{v08}[r][r]{0.7}%
\psfrag{v09}[r][r]{0.8}%
\psfrag{v10}[r][r]{0.9}%
\psfrag{v11}[r][r]{1}%
\psfrag{v12}[r][r]{0}%
\psfrag{v13}[r][r]{0.02}%
\psfrag{v14}[r][r]{0.04}%
\psfrag{v15}[r][r]{0.06}%
\psfrag{v16}[r][r]{0.08}%
\psfrag{v17}[r][r]{0.1}%
\psfrag{v18}[r][r]{0.12}%
\psfrag{v19}[r][r]{0.14}%
\psfrag{v20}[r][r]{0.16}%
\psfrag{v21}[r][r]{0.18}%
\psfrag{v22}[r][r]{0.2}%
%
\resizebox{\columnwidth}{!}{\includegraphics{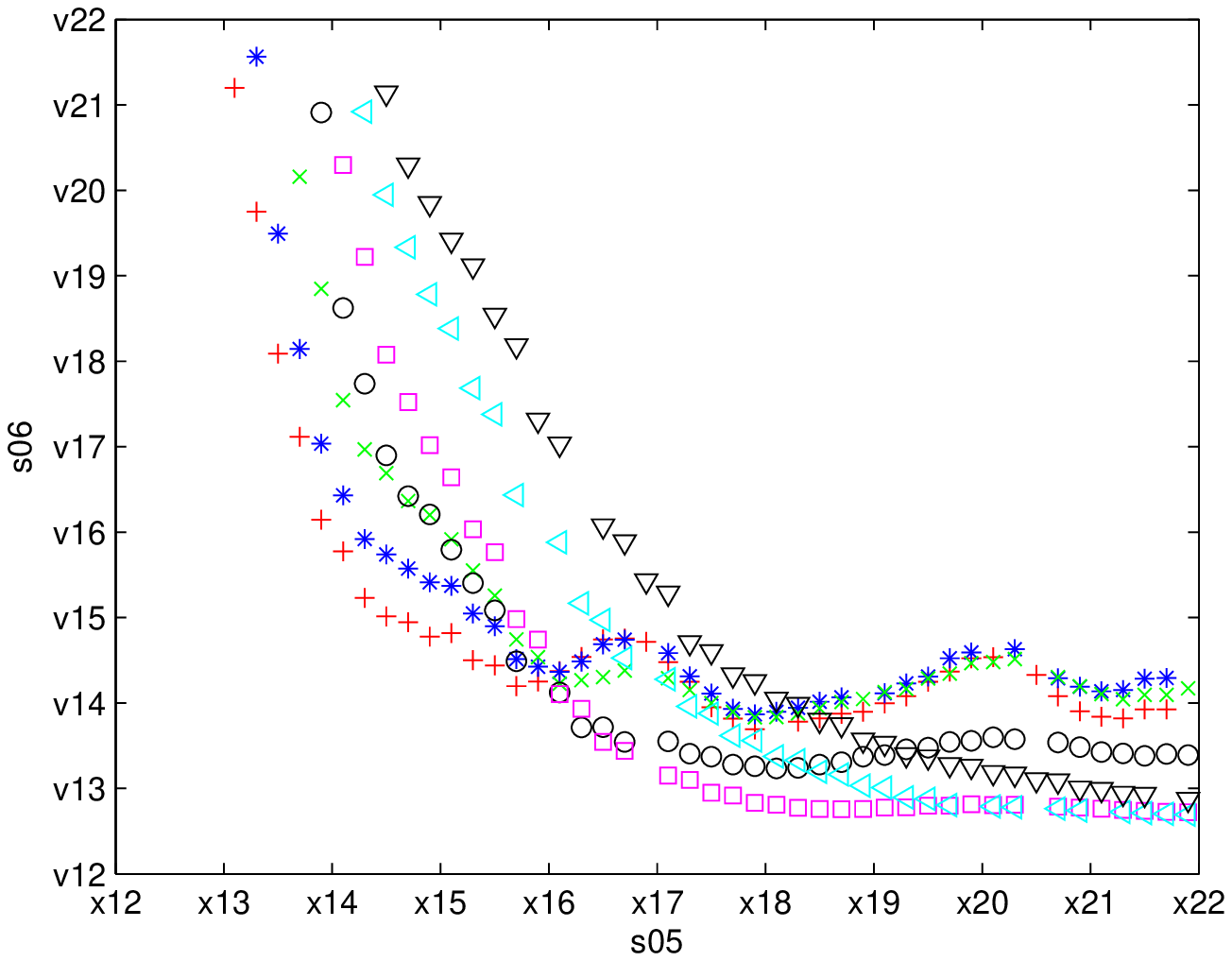}}%
\end{psfrags}%
%

   }
 \end{minipage}
 \caption{Distance between the averaged 3-body reduced density matrix
   and the thermal state as a function of time for various product
   initial states, from $|Z+\rangle$ ($\beta=0.7275$) to $|Y+\rangle$ ($\beta=0$),as indicated on the Bloch sphere on the left.}
 \label{fig:averageSetYZ}
\end{figure}

The appearance of the different thermalization regimes does not
require any fine tuning of the initial conditions or Hamiltonian parameters.
On the contrary, by 
analyzing the evolution of different sets of product initial states,
rotating the initial polarization of the spins from $\hat{y}$ to
$\hat{z}$, we see both strong and weak thermalization regimes over
a range of parameters, separated by a transition.
As shown in Fig.~\ref{fig:averageSetYZ}, we observe that weak thermalization
appears approximately half-way
(corresponding to $\beta\approx 0.3502$), with strong thermalization for those
states closer to $|Y+\rangle$. 
Similarly, we observe a transition
between the strong and the non-thermalizing cases (see Supplementary material).

The different regimes survive also under changes to the Hamiltonian 
parameters, 
 showing that they are also not an
isolated phenomenon for the particular values we chose. 
We have checked (see Supplementary material) that 
weak thermalization becomes prevalent if we decrease the magnitude of the 
transverse magnetic field $g$, while as we increase it, strong thermalization
shows up in a larger fraction of the initial states.
The same effect occurs when we decrease the magnitude of the parallel magnetic field, $h$.

Our data show that the thermalization
process of a quantum system is a much richer phenomenon than its
classical analogue. 
 The appearance of the various thermalization regimes 
cannot be linked exclusively to
the integrability of the Hamiltonian, since we observe these
different regimes for a
fixed model, not close to any integrable limit.  
In particular, some initial states thermalize
strongly, i.e. at the level of the instantaneous expectation values,
while others do it weakly, or only in average.  Some other states seem
to retain memory of the initial configuration for much longer and no 
type of relaxation could be proved. 
 The non-relaxing configurations would be good candidates for an
experimental study of thermalization.

\section*{Methods}

We consider an infinite translationally invariant spin chain with
nearest-neighbour interactions of the Ising type, plus a local
magnetic field with transverse ($g$) and parallel ($h$) components to
the two-body interaction.  
\beq 
H=-\sum_i
\sigma_z^{[i]}\otimes\sigma_z^{[i+1]}-g\sum_i
\sigma_x^{[i]}-h\sum_i\sigma_z^{[i]}.  
\label{eq:HIsingPar}
\eeq 
With a parallel ($g=0$) or transverse
($h=0$) magnetic field, the model is exactly solvable, but at a
different angle, we have a non-integrable model, i.e. the energy density
is the only conserved local quantity~\footnote{The non-integrability
  of the Hamiltonian can be assessed also from the point of view of its
  spectral statistics~\cite{izrailev90rep} (see Supplementary
  Material).}.  We simulate numerically the time evolution of various
initial configurations under fixed Hamiltonian parameters, $g=-1.05$
and $h=0.5$, far from any integrability limits. 
As initial configurations, we choose translationally invariant
product states. They are determined by the state of an individual spin,
$$
|\Psi\rangle=\cos\frac{\theta}{2}|0\rangle + \mathrm{e}^{i\phi} \sin\frac{\theta}{2}|1\rangle.
$$

In particular, the representative states discussed above correspond to parameters
$\theta=\frac{\pi}{2}$, $\phi=0$ ($|X+\rangle$),
$\theta=\phi=\frac{\pi}{2}$ ($|Y+\rangle$)
and $\theta=0$ ($|Z+\rangle$).

Using the recently developed folding method~\cite{banuls09fold}, we
compute all the time dependent expectation values of one-, two- and
three-body operators, for each initial state, what allows us to
reconstruct the whole reduced density matrix for up to three sites.
The thermal state $\rho_{th}(\beta)$ with the same energy as the
initial state is also calculated numerically with the same method (see
Supplementary Material).  The distance between the evolved and thermal
reduced density operators is then measured by the operator norm of
their difference, $d(\rho_1,\rho_2) \equiv \| \rho_1-\rho_2 \|_{op}$,
which in this case coincides with the maximum eigenvalue in absolute
value of the difference $\rho_1-\rho_2$.

\acknowledgments
We are grateful to Maciej Lewenstein and Jens Eisert 
for suggestions.
We acknowledge support by the 
DFG through Excellence Cluster MAP, the SFB
631, and the DFG Forschergruppe 635.

\bibliographystyle{naturemag}
\bibliography{thermalization}

\clearpage

\section{Supplementary material}

\section{The Hamiltonian} 
  
The model we consider is an infinite translationally invariant spin
chain with an Ising type nearest neighbour interaction, plus a
magnetic field 
\beq 
H=-\sum_i
\sigma_z^{[i]}\otimes\sigma_z^{[i+1]}-g\sum_i
\sigma_x^{[i]}-h\sum_i\sigma_z^{[i]}.  
\label{eq:HIsingPar}
\eeq 
When $g=0$, the Hamiltonian is trivially solvable, while for $h=0$ it
reduces to the integrable Ising model with a transverse field, for
which the exact solution can be found by fermionization.  In any other
case, the model is non-integrable.

\begin{figure}[floatfix]
\hspace{.05\columnwidth}
\begin{minipage}[c]{.5\columnwidth}
\subfigure[$g=-1.05$, $h=0.5$]{
 \label{fig:specNI}
%
%
\begin{psfrags}%
\psfragscanon%
\Large
%
\psfrag{s06}[t][t]{\color[rgb]{0,0,0}\setlength{\tabcolsep}{0pt}\begin{tabular}{c}s\end{tabular}}%
%
\psfrag{x01}[t][t]{0}%
\psfrag{x02}[t][t]{0.1}%
\psfrag{x03}[t][t]{0.2}%
\psfrag{x04}[t][t]{0.3}%
\psfrag{x05}[t][t]{0.4}%
\psfrag{x06}[t][t]{0.5}%
\psfrag{x07}[t][t]{0.6}%
\psfrag{x08}[t][t]{0.7}%
\psfrag{x09}[t][t]{0.8}%
\psfrag{x10}[t][t]{0.9}%
\psfrag{x11}[t][t]{1}%
\psfrag{x12}[t][t]{0}%
\psfrag{x13}[t][t]{1}%
\psfrag{x14}[t][t]{2}%
\psfrag{x15}[t][t]{3}%
\psfrag{x16}[t][t]{4}%
\psfrag{x17}[t][t]{5}%
%
\psfrag{v01}[r][r]{0}%
\psfrag{v02}[r][r]{0.1}%
\psfrag{v03}[r][r]{0.2}%
\psfrag{v04}[r][r]{0.3}%
\psfrag{v05}[r][r]{0.4}%
\psfrag{v06}[r][r]{0.5}%
\psfrag{v07}[r][r]{0.6}%
\psfrag{v08}[r][r]{0.7}%
\psfrag{v09}[r][r]{0.8}%
\psfrag{v10}[r][r]{0.9}%
\psfrag{v11}[r][r]{1}%
\psfrag{v12}[r][r]{0}%
\psfrag{v13}[r][r]{0.1}%
\psfrag{v14}[r][r]{0.2}%
\psfrag{v15}[r][r]{0.3}%
\psfrag{v16}[r][r]{0.4}%
\psfrag{v17}[r][r]{0.5}%
\psfrag{v18}[r][r]{0.6}%
\psfrag{v19}[r][r]{0.7}%
\psfrag{v20}[r][r]{0.8}%
\psfrag{v21}[r][r]{0.9}%
\psfrag{v22}[r][r]{1}%
%
\resizebox{\columnwidth}{!}{\includegraphics{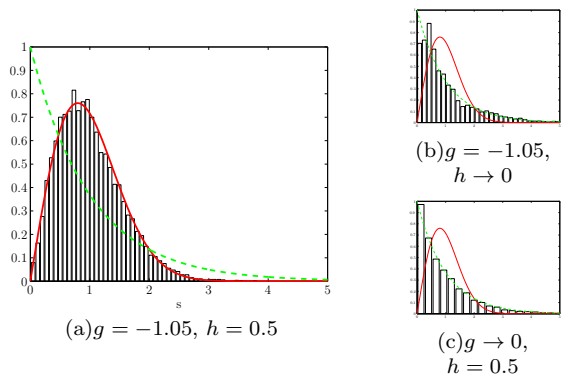}}%
\end{psfrags}%
%

} 
\end{minipage}
\hspace{.01\columnwidth}
\begin{minipage}[c]{.4\columnwidth}
\subfigure[$g=-1.05$, $h\rightarrow0$]{
 \label{fig:specInth}
  \input{figures/levelspacInth0.tex}
}
\subfigure[$g\rightarrow0$, $h=0.5$]{
 \label{fig:specIntg}
  \input{figures/levelspacIntg0.tex}
}
\end{minipage}
\caption{ Level spacing distribution of the unfolded
  spectrum~\cite{mehta} for the Hamiltonian~\ref{eq:HIsingPar} in a
  finite ($L=14$) system, for different Hamiltonian parameters. the
  superimposed curves show a Poissonian (dashed green) and Wigner
  (solid red) distribution, characterizing integrable and chaotic
  systems, respectively.}
\label{fig:levels}
\end{figure}

We fix the initial values of the Hamiltonian parameters for our study
to $g=-1.05$ and $h=0.5$, which is not close to either of the
integrable situations.  In this situation, the energy density is the only
conserved quantity.  We have additionally analyzed the spectral
properties of the Hamiltonian, for a finite system of length $L$, to
check the non-integrability in the sense of a spectrum with the
characteristics of a random matrix ensemble.  As shown in
Fig.~\ref{fig:levels}, already for $L=14$ the level spacing
distribution evidences the non-integrability of the chosen Hamiltonian
also from the point of view of its spectral properties. For
comparison, the level spacing distributions in both integrability
limits are also shown.

\begin{figure}[floatfix]
   \hspace{-.05\columnwidth}
   \begin{minipage}[c]{.45\columnwidth}
     \subfigure{
%
%
\begin{psfrags}%
\psfragscanon%
\LARGE
%
\psfrag{s05}[t][t]{\color[rgb]{0,0,0}\setlength{\tabcolsep}{0pt}\begin{tabular}{c}$s$\end{tabular}}%
\psfrag{s06}[b][b]{\color[rgb]{0,0,0}\setlength{\tabcolsep}{0pt}\begin{tabular}{c}$P(s)$\end{tabular}}%
%
\psfrag{x01}[t][t]{0}%
\psfrag{x02}[t][t]{0.1}%
\psfrag{x03}[t][t]{0.2}%
\psfrag{x04}[t][t]{0.3}%
\psfrag{x05}[t][t]{0.4}%
\psfrag{x06}[t][t]{0.5}%
\psfrag{x07}[t][t]{0.6}%
\psfrag{x08}[t][t]{0.7}%
\psfrag{x09}[t][t]{0.8}%
\psfrag{x10}[t][t]{0.9}%
\psfrag{x11}[t][t]{1}%
\psfrag{x12}[t][t]{0}%
\psfrag{x13}[t][t]{0.5}%
\psfrag{x14}[t][t]{1}%
\psfrag{x15}[t][t]{1.5}%
\psfrag{x16}[t][t]{2}%
\psfrag{x17}[t][t]{2.5}%
\psfrag{x18}[t][t]{3}%
\psfrag{x19}[t][t]{3.5}%
%
\psfrag{v01}[r][r]{0}%
\psfrag{v02}[r][r]{0.1}%
\psfrag{v03}[r][r]{0.2}%
\psfrag{v04}[r][r]{0.3}%
\psfrag{v05}[r][r]{0.4}%
\psfrag{v06}[r][r]{0.5}%
\psfrag{v07}[r][r]{0.6}%
\psfrag{v08}[r][r]{0.7}%
\psfrag{v09}[r][r]{0.8}%
\psfrag{v10}[r][r]{0.9}%
\psfrag{v11}[r][r]{1}%
\psfrag{v12}[r][r]{0}%
\psfrag{v13}[r][r]{0.1}%
\psfrag{v14}[r][r]{0.2}%
\psfrag{v15}[r][r]{0.3}%
\psfrag{v16}[r][r]{0.4}%
\psfrag{v17}[r][r]{0.5}%
\psfrag{v18}[r][r]{0.6}%
\psfrag{v19}[r][r]{0.7}%
\psfrag{v20}[r][r]{0.8}%
\psfrag{v21}[r][r]{0.9}%
\psfrag{v22}[r][r]{1}%
%
\resizebox{\columnwidth}{!}{\includegraphics{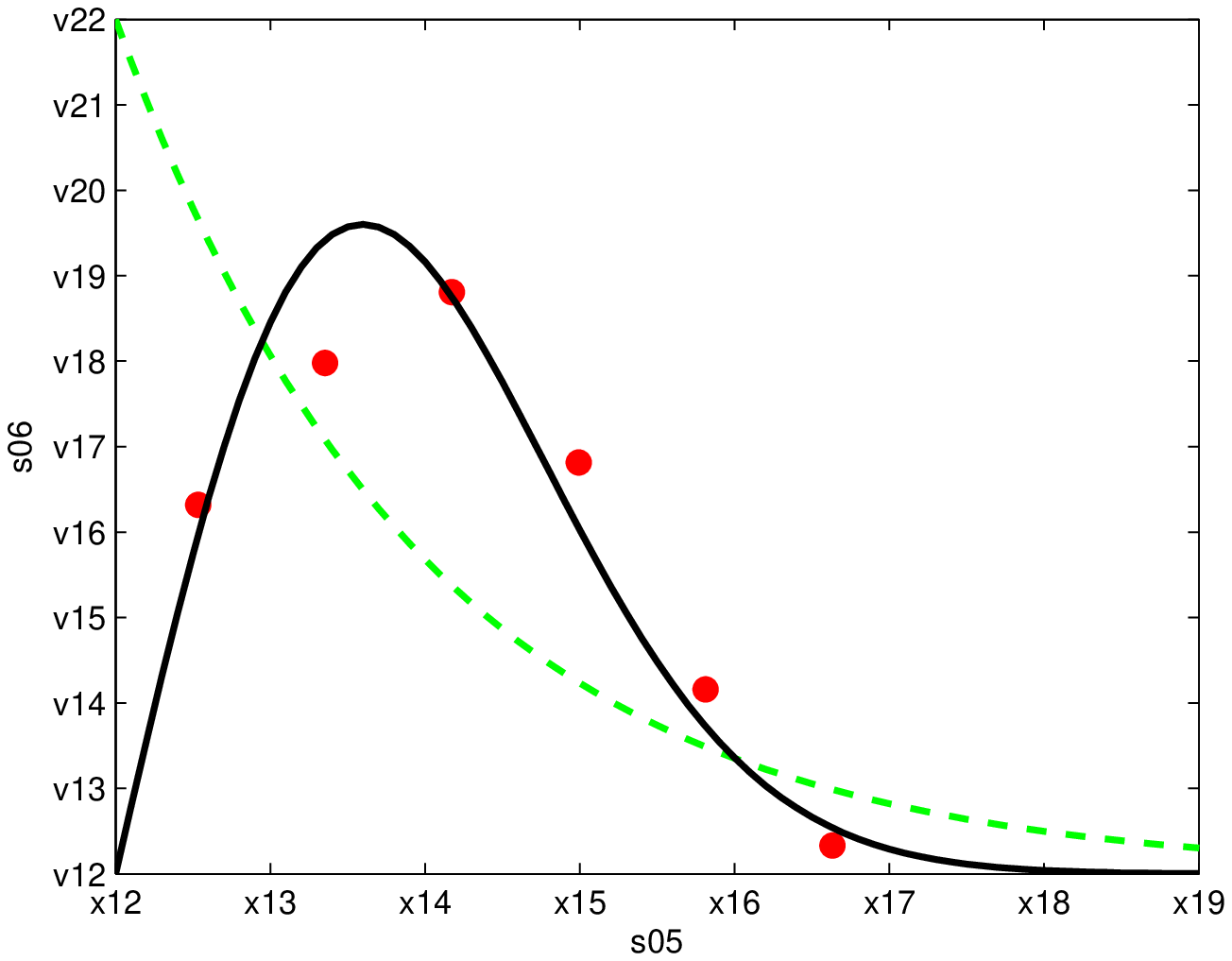}}%
\end{psfrags}%
%

     }
    
 \end{minipage}
 \hspace{.05\columnwidth}
 \begin{minipage}[c]{.45\columnwidth}
   \subfigure{
%
%
\begin{psfrags}%
\psfragscanon%
\LARGE
%
\psfrag{s01}[t][t]{\color[rgb]{0,0,0}\setlength{\tabcolsep}{0pt}\begin{tabular}{c}$t$\end{tabular}}%
\psfrag{s02}[b][B]{\color[rgb]{0,0,0}\setlength{\tabcolsep}{0pt}\begin{tabular}{c}$d(\bar(\rho)(t),\rho_{th})$\end{tabular}}%
%
\psfrag{x01}[t][t]{0}%
\psfrag{x02}[t][t]{0.1}%
\psfrag{x03}[t][t]{0.2}%
\psfrag{x04}[t][t]{0.3}%
\psfrag{x05}[t][t]{0.4}%
\psfrag{x06}[t][t]{0.5}%
\psfrag{x07}[t][t]{0.6}%
\psfrag{x08}[t][t]{0.7}%
\psfrag{x09}[t][t]{0.8}%
\psfrag{x10}[t][t]{0.9}%
\psfrag{x11}[t][t]{1}%
\psfrag{x12}[t][t]{2}%
\psfrag{x13}[t][t]{3}%
\psfrag{x14}[t][t]{4}%
\psfrag{x15}[t][t]{5}%
\psfrag{x16}[t][t]{6}%
\psfrag{x17}[t][t]{7}%
\psfrag{x18}[t][t]{8}%
\psfrag{x19}[t][t]{9}%
\psfrag{x20}[t][t]{10}%
\psfrag{x21}[t][t]{11}%
\psfrag{x22}[t][t]{12}%
%
\psfrag{v01}[r][r]{0}%
\psfrag{v02}[r][r]{0.1}%
\psfrag{v03}[r][r]{0.2}%
\psfrag{v04}[r][r]{0.3}%
\psfrag{v05}[r][r]{0.4}%
\psfrag{v06}[r][r]{0.5}%
\psfrag{v07}[r][r]{0.6}%
\psfrag{v08}[r][r]{0.7}%
\psfrag{v09}[r][r]{0.8}%
\psfrag{v10}[r][r]{0.9}%
\psfrag{v11}[r][r]{1}%
\psfrag{v12}[r][r]{0}%
\psfrag{v13}[r][r]{0.05}%
\psfrag{v14}[r][r]{0.1}%
\psfrag{v15}[r][r]{0.15}%
\psfrag{v16}[r][r]{0.2}%
\psfrag{v17}[r][r]{0.25}%
%
\resizebox{\columnwidth}{.8\columnwidth}{\includegraphics{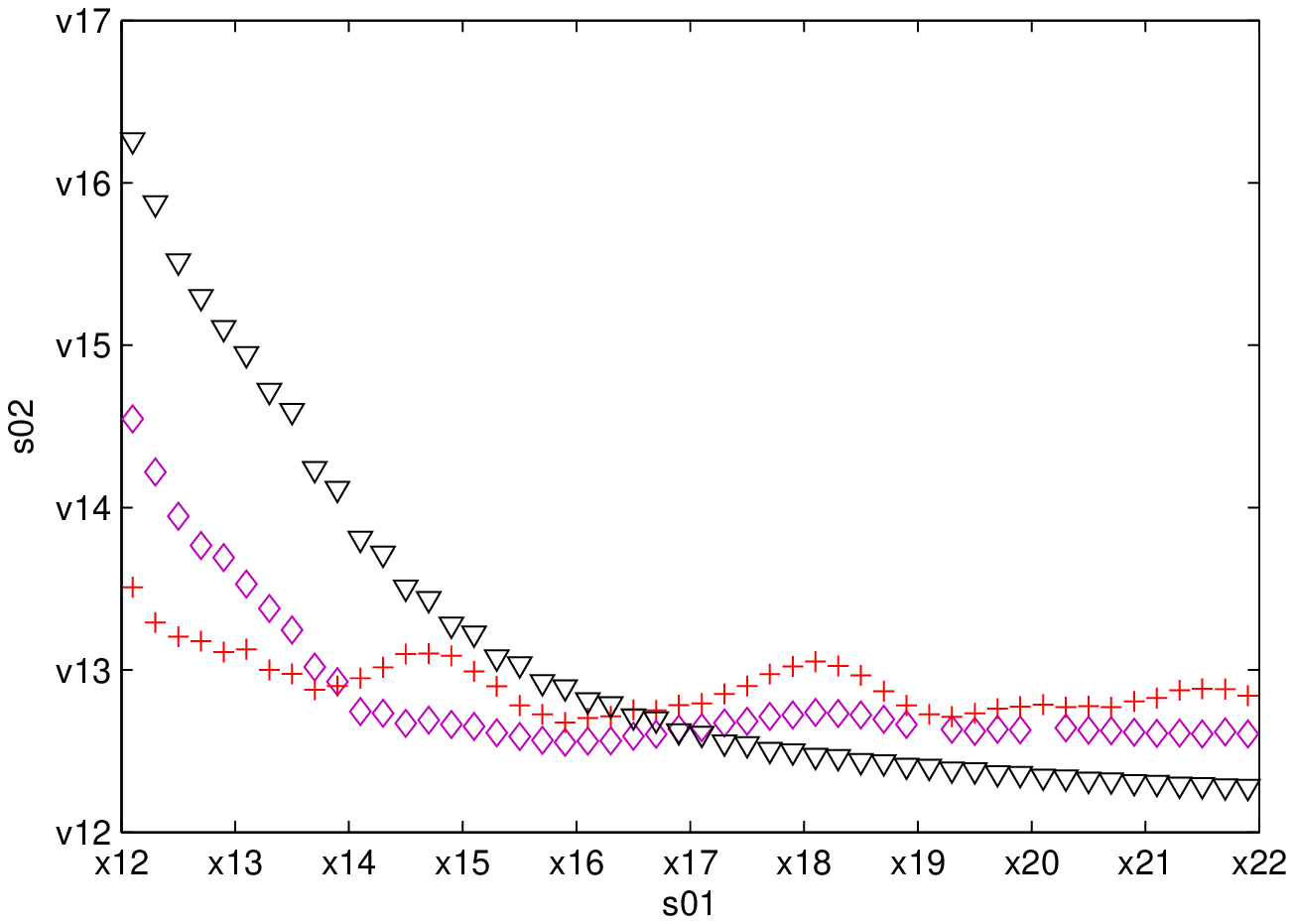}}%
\end{psfrags}%
%

   }
 \end{minipage}
 \caption{Non integrability for a weak thermalizing state. The left
   plot shows the level spacing distribution, in the finite system of
   length $L=14$, for a window of energy around $E/N=-0.93$, of width
   $\Delta E/N=0.21$.  We observe that the statistics is closer to
   Wigner-Dyson (black) than Poisson (green dashed). Nevertheless, the
   product state with the same energy per site in the infinite chain
   shows weak thermalization. The right plot illustrates this, by
   showing the distance between the thermal reduced density matrix and
   the time averaged one as a function of time. The magenta diamonds
   show the data for this particular state. The strongly thermalizing
   state $|Y+\rangle$ (black triangles) and the weakly thermalizing
   $|Z+\rangle$ (red crosses) are also shown for reference.}
 \label{fig:lowEn}
\end{figure}

\begin{figure}[floatfix]
   \hspace{-.05\columnwidth}
   \begin{minipage}[c]{.45\columnwidth}
     \subfigure{
%
%
\begin{psfrags}%
\psfragscanon%
\LARGE
%
\psfrag{s05}[t][t]{\color[rgb]{0,0,0}\setlength{\tabcolsep}{0pt}\begin{tabular}{c}$s$\end{tabular}}%
\psfrag{s06}[b][b]{\color[rgb]{0,0,0}\setlength{\tabcolsep}{0pt}\begin{tabular}{c}$P(s)$\end{tabular}}%
%
\psfrag{x01}[t][t]{0}%
\psfrag{x02}[t][t]{0.1}%
\psfrag{x03}[t][t]{0.2}%
\psfrag{x04}[t][t]{0.3}%
\psfrag{x05}[t][t]{0.4}%
\psfrag{x06}[t][t]{0.5}%
\psfrag{x07}[t][t]{0.6}%
\psfrag{x08}[t][t]{0.7}%
\psfrag{x09}[t][t]{0.8}%
\psfrag{x10}[t][t]{0.9}%
\psfrag{x11}[t][t]{1}%
\psfrag{x12}[t][t]{0}%
\psfrag{x13}[t][t]{0.5}%
\psfrag{x14}[t][t]{1}%
\psfrag{x15}[t][t]{1.5}%
\psfrag{x16}[t][t]{2}%
\psfrag{x17}[t][t]{2.5}%
\psfrag{x18}[t][t]{3}%
\psfrag{x19}[t][t]{3.5}%
%
\psfrag{v01}[r][r]{0}%
\psfrag{v02}[r][r]{0.1}%
\psfrag{v03}[r][r]{0.2}%
\psfrag{v04}[r][r]{0.3}%
\psfrag{v05}[r][r]{0.4}%
\psfrag{v06}[r][r]{0.5}%
\psfrag{v07}[r][r]{0.6}%
\psfrag{v08}[r][r]{0.7}%
\psfrag{v09}[r][r]{0.8}%
\psfrag{v10}[r][r]{0.9}%
\psfrag{v11}[r][r]{1}%
\psfrag{v12}[r][r]{0}%
\psfrag{v13}[r][r]{0.1}%
\psfrag{v14}[r][r]{0.2}%
\psfrag{v15}[r][r]{0.3}%
\psfrag{v16}[r][r]{0.4}%
\psfrag{v17}[r][r]{0.5}%
\psfrag{v18}[r][r]{0.6}%
\psfrag{v19}[r][r]{0.7}%
\psfrag{v20}[r][r]{0.8}%
\psfrag{v21}[r][r]{0.9}%
\psfrag{v22}[r][r]{1}%
%
\resizebox{\columnwidth}{!}{\includegraphics{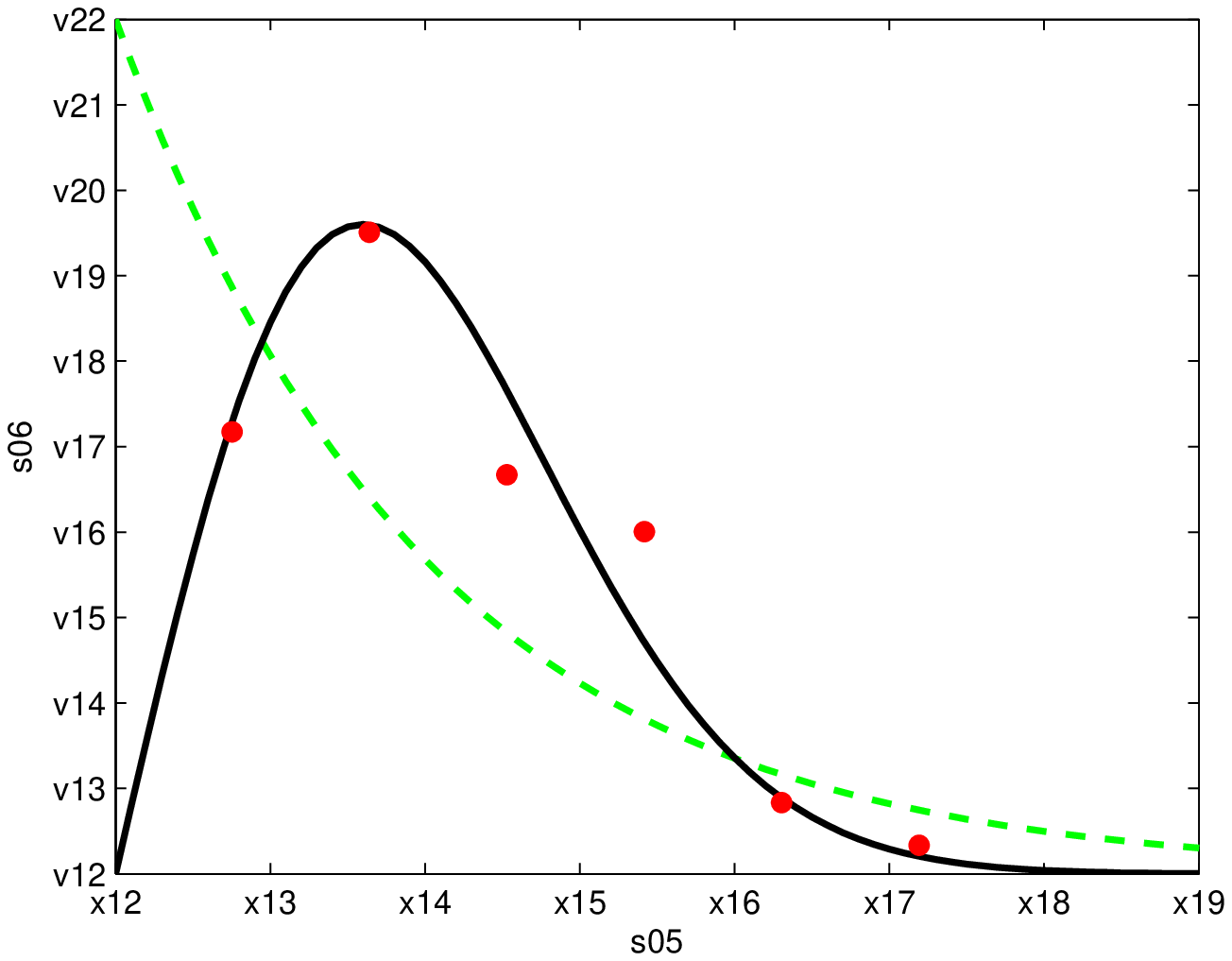}}%
\end{psfrags}%
%

     }
    
 \end{minipage}
 \hspace{.05\columnwidth}
 \begin{minipage}[c]{.45\columnwidth}
   \subfigure{
%
%
\begin{psfrags}%
\psfragscanon%
\LARGE
%
\psfrag{s01}[t][t]{\color[rgb]{0,0,0}\setlength{\tabcolsep}{0pt}\begin{tabular}{c}$t$\end{tabular}}%
\psfrag{s02}[b][B]{\color[rgb]{0,0,0}\setlength{\tabcolsep}{0pt}\begin{tabular}{c}$d(\bar(\rho)(t),\rho_{th})$\end{tabular}}%
%
\psfrag{x01}[t][t]{0}%
\psfrag{x02}[t][t]{0.1}%
\psfrag{x03}[t][t]{0.2}%
\psfrag{x04}[t][t]{0.3}%
\psfrag{x05}[t][t]{0.4}%
\psfrag{x06}[t][t]{0.5}%
\psfrag{x07}[t][t]{0.6}%
\psfrag{x08}[t][t]{0.7}%
\psfrag{x09}[t][t]{0.8}%
\psfrag{x10}[t][t]{0.9}%
\psfrag{x11}[t][t]{1}%
\psfrag{x12}[t][t]{2}%
\psfrag{x13}[t][t]{3}%
\psfrag{x14}[t][t]{4}%
\psfrag{x15}[t][t]{5}%
\psfrag{x16}[t][t]{6}%
\psfrag{x17}[t][t]{7}%
\psfrag{x18}[t][t]{8}%
\psfrag{x19}[t][t]{9}%
\psfrag{x20}[t][t]{10}%
%
\psfrag{v01}[r][r]{0}%
\psfrag{v02}[r][r]{0.1}%
\psfrag{v03}[r][r]{0.2}%
\psfrag{v04}[r][r]{0.3}%
\psfrag{v05}[r][r]{0.4}%
\psfrag{v06}[r][r]{0.5}%
\psfrag{v07}[r][r]{0.6}%
\psfrag{v08}[r][r]{0.7}%
\psfrag{v09}[r][r]{0.8}%
\psfrag{v10}[r][r]{0.9}%
\psfrag{v11}[r][r]{1}%
\psfrag{v12}[r][r]{0}%
\psfrag{v13}[r][r]{0.05}%
\psfrag{v14}[r][r]{0.1}%
\psfrag{v15}[r][r]{0.15}%
\psfrag{v16}[r][r]{0.2}%
%
\resizebox{\columnwidth}{.8\columnwidth}{\includegraphics{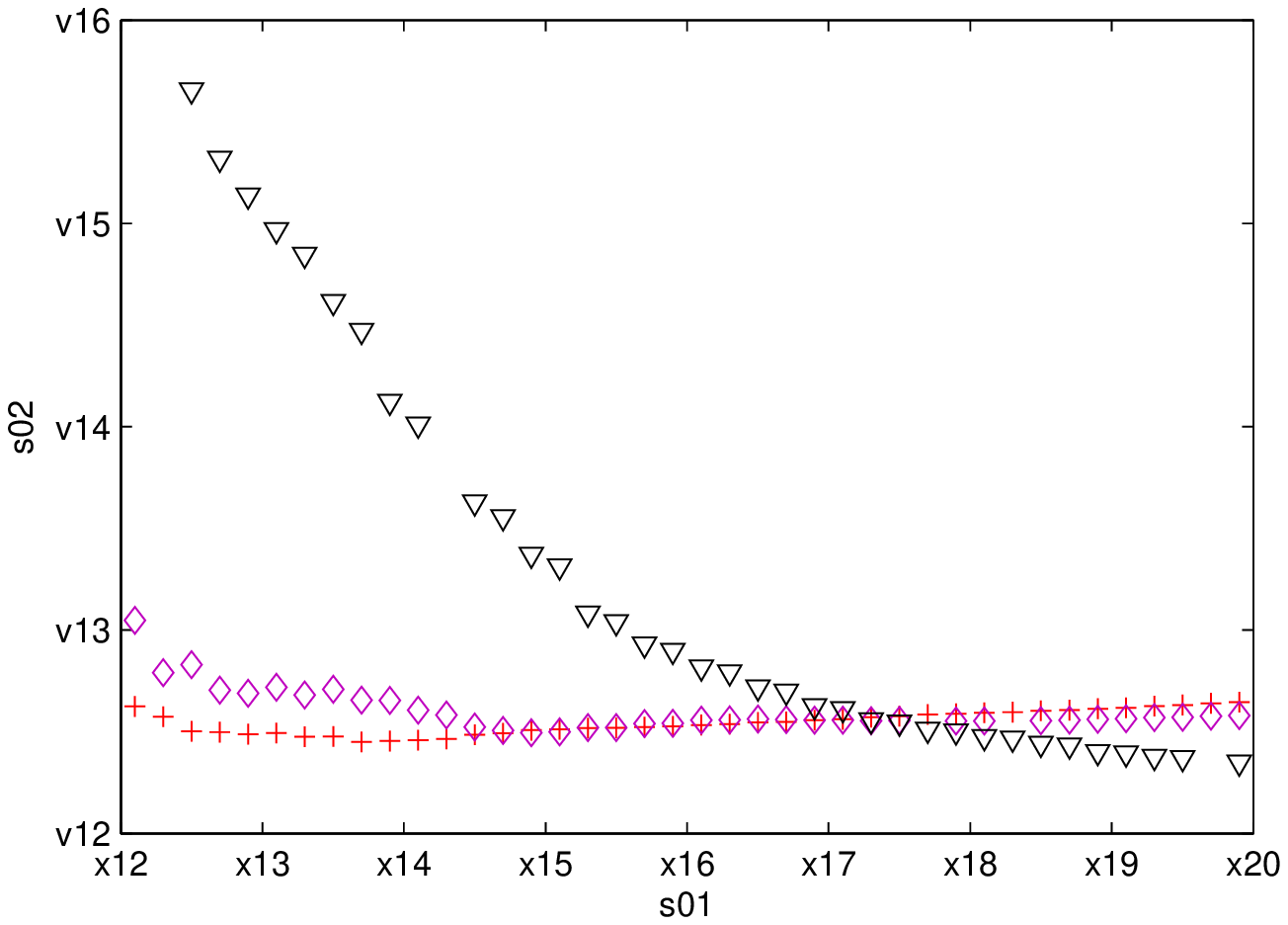}}%
\end{psfrags}%
%

   }
 \end{minipage}
 \caption{Non integrability for a non-thermalizing state. The left
   plot shows that the level spacing distribution, in the finite
   system of length $L=14$, for a window of energy centered on
   $E/N=0.91$ and of width $\Delta E/N=0.21$, is closer to
   Wigner-Dyson than Poisson. Nevertheless, the product state with the
   same energy per site in the infinite chain shows no
   thermalization, as the right plot illustrates.
   The magenta diamonds represent the
   distance between the thermal reduced density matrix and the time
   averaged $\rho$ for this particular state as a function of time. 
   The strongly thermalizing state
   $|Y+\rangle$ (black triangles) and the non thermalizing
   $|X+\rangle$ (red crosses) are also shown for comparison.}
 \label{fig:highEn}
\end{figure}

It could be argued that the non thermalization we observe occurs for
states which lie close to the edges of the spectrum, as $|Z+\rangle$
and $|X+\rangle$, and at these energies there could be an integrable
effective model~\cite{distasio95}, even for $g=-1.05$, $h=0.5$. 
The spectral properties discussed in Fig.~\ref{fig:levels} would
be dominated by the central part of the spectrum and not reflect
the properties at very low or high energies, while the spectrum 
in an energy interval in these regions should show a very different behavior.
 To discard the integrability of
the system in the interesting cases, we have checked the level
statistics 
of a small energy window
around the energy per site of a weak thermalizing state
(Fig.~\ref{fig:lowEn}) and a non-thermalizing one (Fig.~\ref{fig:highEn})
for the case of a finite system ($L=14$) which can be exactly solved.
In both cases we have found that the level spacing distribution is 
typical of a non-integrable system, also in these regions of energy.

\section{The numerical method}


The time evolution of an infinite 1D quantum system is simulated
numerically within the Matrix Product
States~\cite{aklt88,kluemper91,kluemper92,fannes92fcs,verstraete04dmrg,perez07mps}
(MPS) formalism, using the new technique introduced
in~\cite{banuls09fold}.  With this folding method, it is possible to study the
out-of-equilibrium dynamics of the system after longer times than
with other similar methods.

\begin{figure}[floatfix]
\hspace{-.05\columnwidth}
\begin{minipage}[c]{.4\columnwidth}
\subfigure[2D network representing a local expectation value $\langle O(t)\rangle$]{
 \label{fig:tn-b}
\psfrag{contractL}[c][c]{contraction}
\psfrag{contractR}[c][c]{}
\psfrag{Eo}[c][]{$E_O$}
\psfrag{E}[c][]{$E$}
\psfrag{langle}[c][]{$\langle L |$}
\psfrag{rangle}[c][]{$|R\rangle$}
\psfrag{time}[bc][bc]{time}
\psfrag{space}[tc][tc]{space}
  \includegraphics[height=.9\columnwidth]{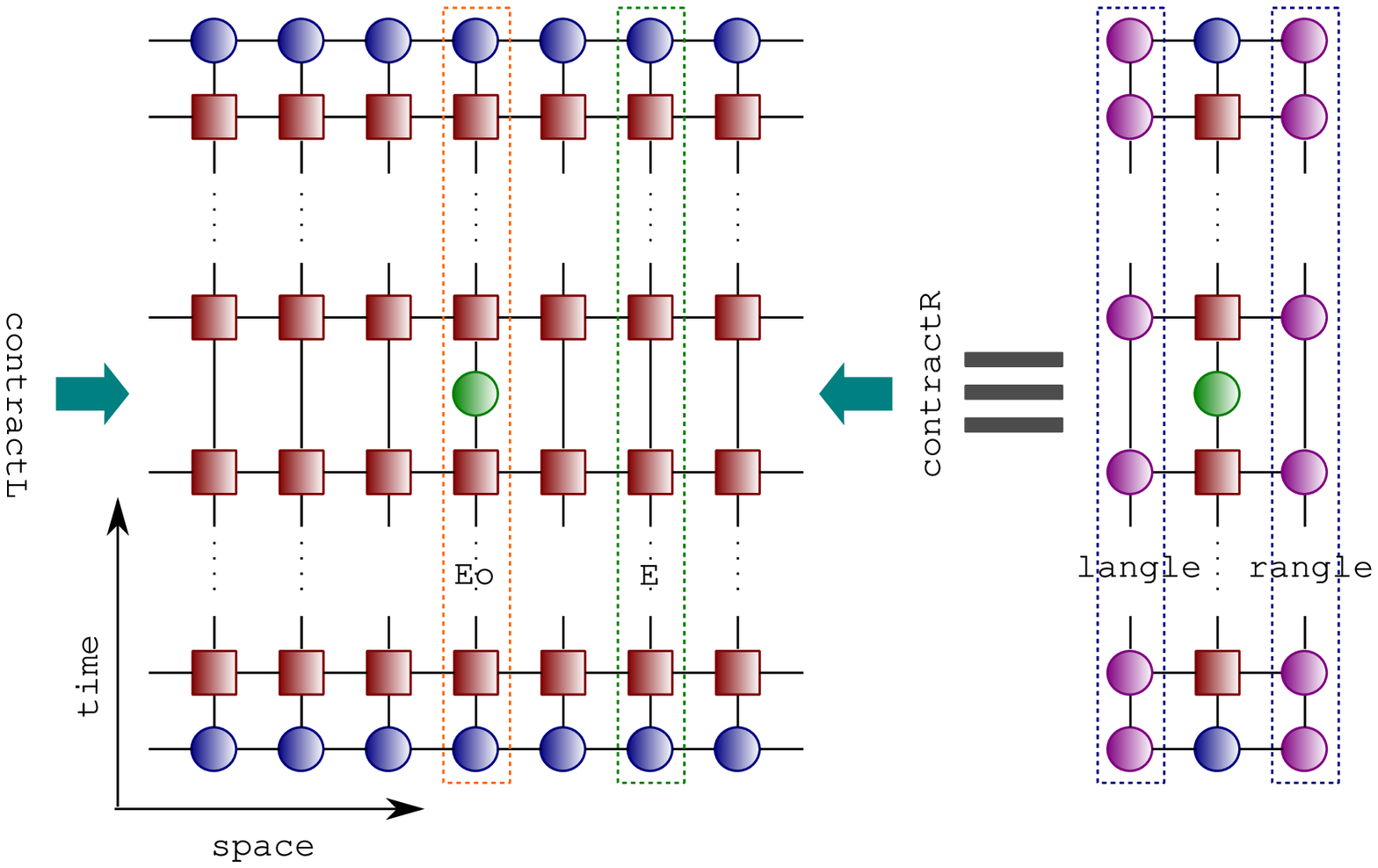}
} 
\end{minipage}
\hspace{.185\columnwidth}
\begin{minipage}[c]{.4\columnwidth}
\subfigure[Transverse  contraction of folded network.]{
 \label{fig:tn-fold}
\psfrag{Eo}[bc][bc]{$\tilde{E}_O$}
\psfrag{E}[bc][bc]{$\tilde{E}$}
\psfrag{langle}[c][c]{$\langle \tilde{L} |$}
\psfrag{rangle}[c][c]{$|\tilde{R}\rangle$}
\psfrag{fol}[cl][cl]{\parbox{.2\columnwidth}{folding axis}}
  \includegraphics[height=.82\columnwidth]{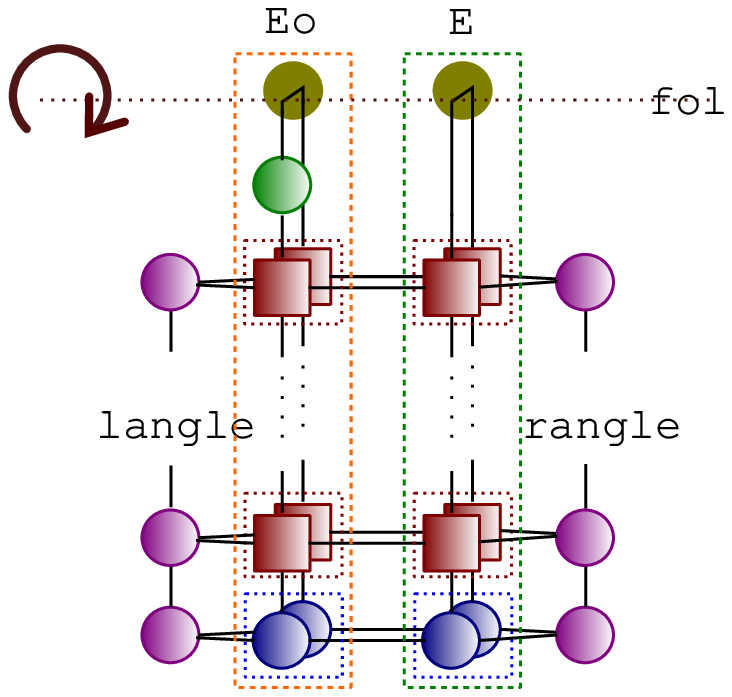}
}
\end{minipage}
\caption{
In the folding approach (b), operators for the same time 
step are grouped together in a double effective operator~\cite{banuls09fold}.
}
\end{figure}

In this method, any
time dependent expectation value $\langle \Psi(t)|O| \Psi(t) \rangle$
can then be expressed as a two dimensional tensor network
(Fig.~\ref{fig:tn-b}), constructed from the a Suzuki-Trotter
expansion of the evolution
operator~\cite{trotter59,suzuki90}.  Each discrete time step
corresponds then to a sequence of matrix product operators
(MPO)~\cite{murg08mpo}.  In contrast to the standard MPS algorithm, in which
the evolved state is approximated by an MPS after each time step by
means of successive truncations~\cite{vidal03eff,vidal07infinite}, 
the new algorithm performs the
contraction of the tensor network in the transverse direction,
i.e. along space.  The left and tight semi-infinite lattices can then
be effectively substituted by the left and right dominant eigenvectors
of the transfer matrix of the evolved state, $\langle L |$ and $| R
\rangle$.  Before contracting we apply a folding to the network along
the time direction (Fig.~\ref{fig:tn-fold}).  The folding operation
can be understood as performing the contraction $\langle
\Psi(t)|O|\Psi(t)\rangle =
\langle\Phi|(O|\Psi(t)\rangle\otimes|\bar{\Psi}(t)\rangle)$, where
$|\Phi\rangle$ is a product of unnormalized maximally entangled pairs
between each site of the chain and its conjugate
(see~\cite{banuls09fold} for a detailed discussion of the algorithm).
This is equivalent to grouping together tensors that correspond to the
same time step in $\Psi$ and its Hermitian conjugate, and achieves a
more efficient representation of the entanglement in the transverse
direction, which in turn gives access to the simulation of longer
evolution times.

\begin{figure}[floatfix]
%
%
\begin{psfrags}%
\psfragscanon%
%
\psfrag{s06}[b][b]{\color[rgb]{0,0,0}\setlength{\tabcolsep}{0pt}\begin{tabular}{c}\end{tabular}}%
\psfrag{s07}[b][b]{\color[rgb]{0,0,0}\setlength{\tabcolsep}{0pt}\begin{tabular}{c}\end{tabular}}%
%
\psfrag{x01}[t][t]{0}%
\psfrag{x02}[t][t]{0.1}%
\psfrag{x03}[t][t]{0.2}%
\psfrag{x04}[t][t]{0.3}%
\psfrag{x05}[t][t]{0.4}%
\psfrag{x06}[t][t]{0.5}%
\psfrag{x07}[t][t]{0.6}%
\psfrag{x08}[t][t]{0.7}%
\psfrag{x09}[t][t]{0.8}%
\psfrag{x10}[t][t]{0.9}%
\psfrag{x11}[t][t]{1}%
\psfrag{x12}[t][t]{-2}%
\psfrag{x13}[t][t]{-1.5}%
\psfrag{x14}[t][t]{-1}%
\psfrag{x15}[t][t]{-0.5}%
\psfrag{x16}[t][t]{0}%
\psfrag{x17}[t][t]{0.5}%
\psfrag{x18}[t][t]{1}%
\psfrag{x19}[t][t]{1.5}%
\psfrag{x20}[t][t]{2}%
%
\psfrag{v01}[r][r]{0}%
\psfrag{v02}[r][r]{0.1}%
\psfrag{v03}[r][r]{0.2}%
\psfrag{v04}[r][r]{0.3}%
\psfrag{v05}[r][r]{0.4}%
\psfrag{v06}[r][r]{0.5}%
\psfrag{v07}[r][r]{0.6}%
\psfrag{v08}[r][r]{0.7}%
\psfrag{v09}[r][r]{0.8}%
\psfrag{v10}[r][r]{0.9}%
\psfrag{v11}[r][r]{1}%
\psfrag{v12}[r][r]{-2}%
\psfrag{v13}[r][r]{-1.5}%
\psfrag{v14}[r][r]{-1}%
\psfrag{v15}[r][r]{-0.5}%
\psfrag{v16}[r][r]{0}%
\psfrag{v17}[r][r]{0.5}%
\psfrag{v18}[r][r]{1}%
\psfrag{v19}[r][r]{1.5}%
%
\resizebox{\columnwidth}{!}{\includegraphics{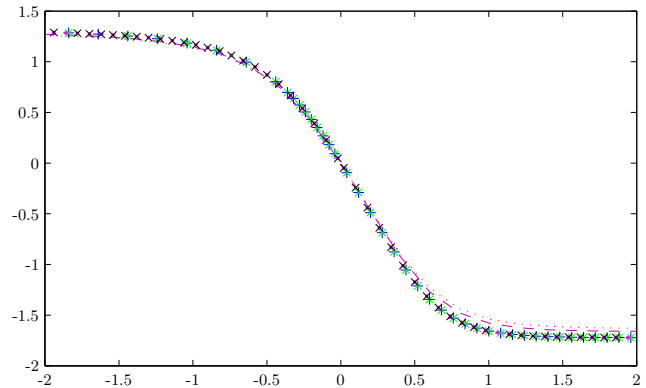}}%
\end{psfrags}%
%

\caption{Energy per particle as a function of the inverse temperature
  $\beta$ in the thermal state for the non-integrable model with
  $g=-1.05$, $h=0.5$.  The plot shows the results of the fourth order
  decomposition with $\delta=0.02$ and bond dimensions $D=10$ (blue
  crosses) and $D=20$ (green crosses), and with step $\delta=0.01$ and
  bond dimensions $D=20$ (pink dots) and $D=40$ (black x).  We also
  show the results of the exact numerical calculation for a finite
  system of size $L=8$ (dotted) and $L=12$ (dashed line).  }
\label{fig:thermIsingPar}
\end{figure}

The folding technique is appropriate for the simulation of time
evolution, but using imaginary time evolution, it is also possible to
efficiently compute any local expectation value in a thermal state.
We obtain in this way the dependency of the thermal state energy with
the inverse temperature $\beta$, $E_{th}(\beta)$
(Fig.~\ref{fig:thermIsingPar}).  From these data we may compute, for
each one of the initial states we have studied, the value of $\beta$
corresponding to the thermal state with the same energy.  This
determines the state to which the initial configuration would be
expected to relax, being energy density the only conserved quantity
constraining the relaxation.  It is then possible to compare the
$N$-particle reduced density matrix for the evolved state and for the
thermal state at $\beta$, to compute the desired distance between
density operators.

To determine the reduced density matrix for $N$ sites, both for the thermal and 
the evolved state, we need to compute the expectation
values of all $N$-term products of Pauli matrices and the identity.
Both in the cases of the thermal and the evolved state,
the computation of the dominant left and right eigenvectors
is common to all such expectation values.
We show here the results for $N=3$ sites.

\section{Accuracy of the results}

Due to the numerical nature of the study, all the results are subject to
errors. In order to assess the validity of our conclusions,
we discuss here the character and magnitude of these errors 
using different criteria.

The numerical method has two sources of error.  The first one is the
Trotter decomposition. The approximation of the evolution operator by
a product of exponentials introduces an error, which can be controlled
by either reducing the parameter $\delta$, which determines the size
of the time step, or by using a Suzuki-Trotter expansion which is
exact to a higher order in $\delta$.  Both decreasing the time step or
increasing the order of the expansion involve longer vectors in the
transverse direction, which will potentially worsen the truncation
error. It is then convenient to find a trade-off between both factors.
In our analysis we use a fourth order Suzuki-Trotter decomposition
with time step $\delta=0.1$. We have checked the convergence of our results
with this order and time step.

\begin{figure}[floatfix]
%
%
\begin{psfrags}%
\psfragscanon%
%
\psfrag{s06}[][]{\color[rgb]{0,0,0}\setlength{\tabcolsep}{0pt}\begin{tabular}{c}\end{tabular}}%
\psfrag{s07}[][]{\color[rgb]{0,0,0}\setlength{\tabcolsep}{0pt}\begin{tabular}{c} \end{tabular}}%
\psfrag{s16}[t][t]{\color[rgb]{0,0,0}\setlength{\tabcolsep}{0pt}\begin{tabular}{c}$\epsilon=10^{-6}$\end{tabular}}%
\psfrag{s17}[t][t]{\color[rgb]{0,0,0}\setlength{\tabcolsep}{0pt}\begin{tabular}{c}$\epsilon=10^{-4}$\end{tabular}}%
\psfrag{s18}[t][t]{\color[rgb]{0,0,0}\setlength{\tabcolsep}{0pt}\begin{tabular}{c}$\epsilon=10^{-2}$\end{tabular}}%
\psfrag{s19}[t][t]{\color[rgb]{0,0,0}\setlength{\tabcolsep}{0pt}\begin{tabular}{c}$\epsilon=10^{-8}$\end{tabular}}%
%
\psfrag{x01}[t][t]{0}%
\psfrag{x02}[t][t]{0.1}%
\psfrag{x03}[t][t]{0.2}%
\psfrag{x04}[t][t]{0.3}%
\psfrag{x05}[t][t]{0.4}%
\psfrag{x06}[t][t]{0.5}%
\psfrag{x07}[t][t]{0.6}%
\psfrag{x08}[t][t]{0.7}%
\psfrag{x09}[t][t]{0.8}%
\psfrag{x10}[t][t]{0.9}%
\psfrag{x11}[t][t]{1}%
\psfrag{x12}[t][t]{2}%
\psfrag{x13}[t][t]{4}%
\psfrag{x14}[t][t]{6}%
\psfrag{x15}[t][t]{8}%
\psfrag{x16}[t][t]{10}%
\psfrag{x17}[t][t]{0}%
\psfrag{x18}[t][t]{1}%
\psfrag{x19}[t][t]{2}%
\psfrag{x20}[t][t]{3}%
\psfrag{x21}[t][t]{4}%
\psfrag{x22}[t][t]{5}%
\psfrag{x23}[t][t]{6}%
\psfrag{x24}[t][t]{7}%
\psfrag{x25}[t][t]{8}%
\psfrag{x26}[t][t]{9}%
\psfrag{x27}[t][t]{10}%
%
\psfrag{v01}[r][r]{0}%
\psfrag{v02}[r][r]{0.1}%
\psfrag{v03}[r][r]{0.2}%
\psfrag{v04}[r][r]{0.3}%
\psfrag{v05}[r][r]{0.4}%
\psfrag{v06}[r][r]{0.5}%
\psfrag{v07}[r][r]{0.6}%
\psfrag{v08}[r][r]{0.7}%
\psfrag{v09}[r][r]{0.8}%
\psfrag{v10}[r][r]{0.9}%
\psfrag{v11}[r][r]{1}%
\psfrag{v12}[r][r]{20}%
\psfrag{v13}[r][r]{40}%
\psfrag{v14}[r][r]{60}%
\psfrag{v15}[r][r]{80}%
\psfrag{v16}[r][r]{100}%
\psfrag{v17}[r][r]{0.5}%
\psfrag{v18}[r][r]{0.6}%
\psfrag{v19}[r][r]{0.7}%
\psfrag{v20}[r][r]{0.8}%
\psfrag{v21}[r][r]{0.9}%
\psfrag{v22}[r][r]{1}%
%
\resizebox{\columnwidth}{!}{\includegraphics{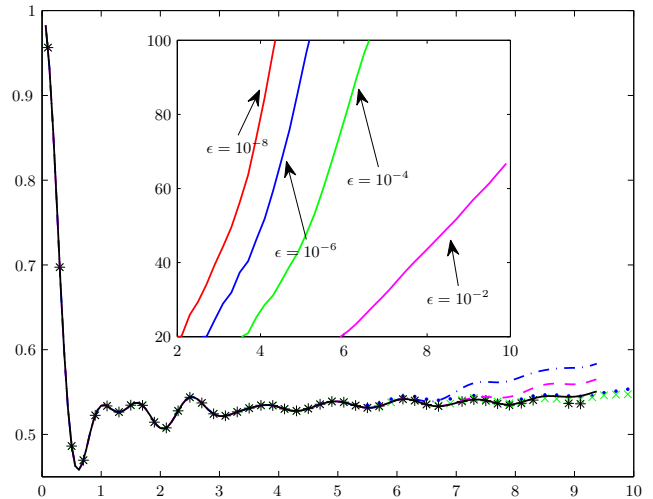}}%
\end{psfrags}%
%

\caption{As studied in detail in~\cite{banuls09fold},
  for the non-integrable model, and initial state $|X+\rangle$,
  magnetization ($\langle \sigma_x(t)\rangle$) results with the folded
  approach for D=60 (blue dots), 120 (green crosses), 240 (black
  stars) are compared to those of iTEBD (dash-dotted blue line for
  $D=256$, dashed magenta for $D=512$ and black solid line for
  $D=1024$). In the inset, we show the required value of $D$ as a
  function of time, for different levels of accuracy. Notice that a
  moderate bond dimension ($D<100$) suffices for a qualitative
  description ($\epsilon\sim1\%$).  }
\label{fig:IsingP}
\end{figure}

A second, generally less benign source of error is truncation, i.e.
the fact of approximating a given vector by a MPS of fixed bond
dimension, which constitutes the main source of numerical errors in
any MPS algorithm.  In our scheme the evolved state is not explicitly
truncated, but truncation takes place in the transverse direction,
when we approximate the dominant left and right eigenvectors of the
transfer matrix by MPS. In a standard MPS algorithm for time
evolution, truncation errors are dramatic, in the sense that, when
they appear, the results deviate abruptly from the exact ones, and it
becomes imposible to extract information from the computed quantities
as soon as truncation error sets in~\cite{schollwoeck06tDMRG}.
However, in the folding technique, we may extract information from
longer time simulations, because errors come in smoothly and, even
when some truncation error occurs, the method is still able to provide
a qualitative description of the physics, since we expect that our
predictions deviate smoothly from the exact values (see
Fig.~\ref{fig:IsingP} and discussion in~\cite{banuls09fold}).

To bound the numerical errors in the non-integrable model we may
compare our results to those from other approaches, check the
convergence of the results as we increase the bond dimension, or make
use of some external physical criterion to assess the consistency of
the computed numbers.  We have used all three kinds of tests.  First,
we have cross-checked the results of our simulations with some large
bond dimension simulations using the iTEBD
algorithm~\cite{vidal07infinite}, in which contraction is done in the
time direction. As shown in Fig.~\ref{fig:IsingP}, the folding results
with $D=120$ are accurate to the longest times we can simulate with
iTEBD, $t\simeq 9-10$.  Most remarkably, this bond dimension is enough
to get a qualitative description (precision $1\%$) of the dynamics to
even longer times.  Second, to witness the appearance of truncation 
errors, we have run the simulations with
increasing bond dimension.The comparison of
our results with highest bond dimension $D=240$ 
with those for $D=120$ gives us a bound on the error, which we
represent on the plots as an error bar.

\begin{figure}[floatfix]
\input{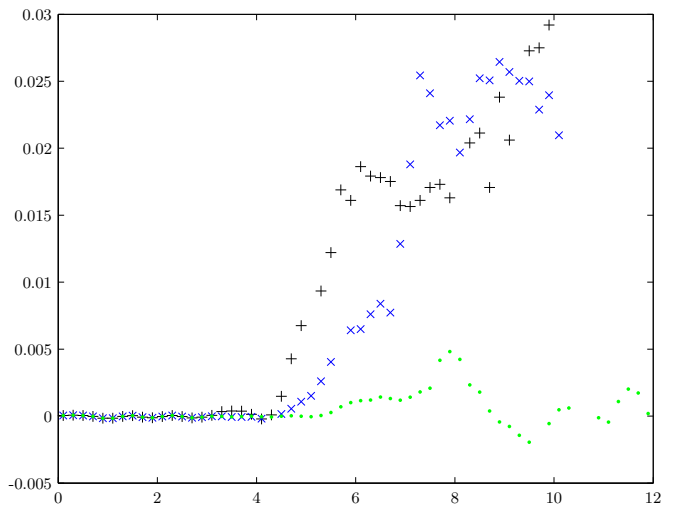}
\caption{Energy per particle as a function of time, for the initial
  state $|Y+\rangle$ (corresponding to $\beta=0$), for bond dimension
  $D=40$ (black crosses), $D=60$ (blue x) and $D=120$ (green dots).  }
\label{fig:errorEY}
\end{figure}

Finally, as a physical check of the consistency of the results, we
test the conservation of energy along the evolution. We study the
unitary evolution of a closed system, and the energy per particle must
be constant. However, the numerical implementation does not enforce
this condition. On the contrary, from the Suzuki-Trotter expansion to
the truncation of the dominant eigenvectors, the numerical errors will
in general violate this condition.  A very large deviation between the
time-evolved energy and the initial one would warn us about the
validity of the results.  As shown in Fig.~\ref{fig:errorEZ}
and~\ref{fig:errorEX} for the initial states $|Z+\rangle$ and
$|X+\rangle$, with bond dimension $D=240$, the relative error in the
energy for the range of times we are analysing (respectively $t\simeq
18$ and $t\simeq 12$) is kept to only a few percent, consistent with
the estimated truncation error.  For the initial state $|Y+\rangle$,
with zero initial energy, we plot instead the expectation value of
energy as a function of time (see Fig.~\ref{fig:errorEY}).

One may think that this deviation of the energy could also introduce
an error in the distance we are computing, as the thermal states
corresponding to the computed energy density and to the initial one
will be different. To bound this error, we have computed the distance
between such pair of thermal states, corresponding to the initial
state and to the largest value of the energy found in the evolution,
to the range of times we are showing.  We find that this distance is
significantly smaller than the one we observe during the dynamical
evolution. In particular, for the $|Z+\rangle$ initial state, the
deviation in energy at times $t\approx 10$ reaches a $1\%$, which
corresponds to a distance
$d(\rho(\beta_0),\rho(\beta'))\approx9\times10^{-3}$, while the
observed distance between the thermal state and the evolved one
oscillates around $0.15$.  For the longest times we show $t\approx
18$, the largest distance grows to a maximum value of $0.06$.  For the
$|X+\rangle$ initial state, the maximum deviation in energy, at
$t\approx 12$, corresponds to a distance
$d(\rho(\beta_0),\rho(\beta'))\approx11\times10^{-3}$, while the one
we find at this same long time is $0.04$

\begin{figure}[floatfix]
\input{figures/errorEZ+.tex}
\caption{Relative error in the energy per particle as a function of
  time, for the initial state $|Z+\rangle$ (corresponding to
  $\beta=0.7275$), for bond dimension $D=40$ (black crosses), $D=60$
  (blue x), $D=120$ (green dots) and $D=240$ (red stars).  }
\label{fig:errorEZ}
\end{figure}

\begin{figure}[floatfix]
\input{figures/errorEX+.tex}
\caption{Relative error in the energy per particle as a function of
  time, for the initial state $|X+\rangle$ (corresponding to
  $\beta=-0.7180$), for bond dimension $D=40$ (black crosses), $D=60$
  (blue x), $D=120$ (green dots) and $D=240$ (red stars).  }
\label{fig:errorEX}
\end{figure}

\section{Detailed results}

Here we compile our results using various initial states and
Hamiltonian parameters, to show the survival of the different
thermalization regimes over a range of parameters.

 \begin{figure}[floatfix]
   \hspace{-.05\columnwidth}
   \begin{minipage}[c]{.45\columnwidth}
     \subfigure{
%
%
\begin{psfrags}%
\psfragscanon%
\LARGE
%
\psfrag{s06}[][]{\color[rgb]{0,0,0}\setlength{\tabcolsep}{0pt}\begin{tabular}{c} \end{tabular}}%
\psfrag{s07}[][]{\color[rgb]{0,0,0}\setlength{\tabcolsep}{0pt}\begin{tabular}{c} \end{tabular}}%
\psfrag{s12}[t][t]{\color[rgb]{0,0,0}\setlength{\tabcolsep}{0pt}\begin{tabular}{c}D=64\end{tabular}}%
\psfrag{s13}[t][t]{\color[rgb]{0,0,0}\setlength{\tabcolsep}{0pt}\begin{tabular}{c}D=128\end{tabular}}%
\psfrag{s14}[t][t]{\color[rgb]{0,0,0}\setlength{\tabcolsep}{0pt}\begin{tabular}{c}D=256\end{tabular}}%
\psfrag{s15}[t][t]{\color[rgb]{0,0,0}\setlength{\tabcolsep}{0pt}\begin{tabular}{c}D=512\end{tabular}}%
%
\psfrag{x01}[t][t]{0}%
\psfrag{x02}[t][t]{0.1}%
\psfrag{x03}[t][t]{0.2}%
\psfrag{x04}[t][t]{0.3}%
\psfrag{x05}[t][t]{0.4}%
\psfrag{x06}[t][t]{0.5}%
\psfrag{x07}[t][t]{0.6}%
\psfrag{x08}[t][t]{0.7}%
\psfrag{x09}[t][t]{0.8}%
\psfrag{x10}[t][t]{0.9}%
\psfrag{x11}[t][t]{1}%
\psfrag{x12}[t][t]{0}%
\psfrag{x13}[t][t]{1}%
\psfrag{x14}[t][t]{2}%
\psfrag{x15}[t][t]{3}%
\psfrag{x16}[t][t]{4}%
\psfrag{x17}[t][t]{5}%
\psfrag{x18}[t][t]{6}%
\psfrag{x19}[t][t]{7}%
\psfrag{x20}[t][t]{8}%
\psfrag{x21}[t][t]{9}%
\psfrag{x22}[t][t]{10}%
%
\psfrag{v01}[r][r]{0}%
\psfrag{v02}[r][r]{0.1}%
\psfrag{v03}[r][r]{0.2}%
\psfrag{v04}[r][r]{0.3}%
\psfrag{v05}[r][r]{0.4}%
\psfrag{v06}[r][r]{0.5}%
\psfrag{v07}[r][r]{0.6}%
\psfrag{v08}[r][r]{0.7}%
\psfrag{v09}[r][r]{0.8}%
\psfrag{v10}[r][r]{0.9}%
\psfrag{v11}[r][r]{1}%
\psfrag{v12}[r][r]{0}%
\psfrag{v13}[r][r]{0.1}%
\psfrag{v14}[r][r]{0.2}%
\psfrag{v15}[r][r]{0.3}%
\psfrag{v16}[r][r]{0.4}%
\psfrag{v17}[r][r]{0.5}%
\psfrag{v18}[r][r]{0.6}%
\psfrag{v19}[r][r]{0.7}%
\psfrag{v20}[r][r]{0.8}%
\psfrag{v21}[r][r]{0.9}%
\psfrag{v22}[r][r]{1}%
%
\resizebox{\columnwidth}{!}{\includegraphics{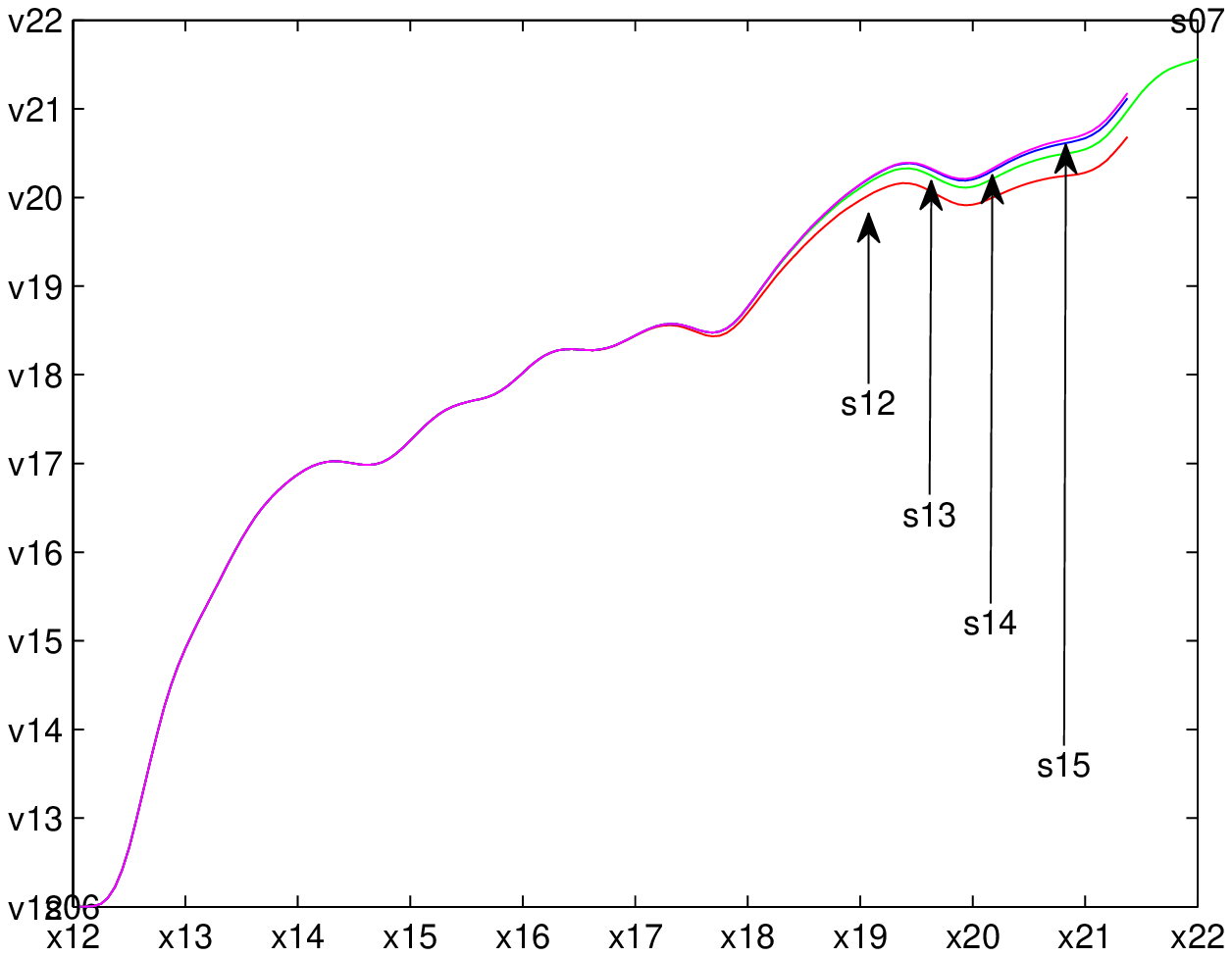}}%
\end{psfrags}%
%

     }
    
 \end{minipage}
 \hspace{.05\columnwidth}
 \begin{minipage}[c]{.45\columnwidth}
   \subfigure{
%
%
\begin{psfrags}%
\psfragscanon%
\LARGE
%
\psfrag{s09}[t][t]{\color[rgb]{0,0,0}\setlength{\tabcolsep}{0pt}\begin{tabular}{c}D=256\end{tabular}}%
\psfrag{s10}[t][t]{\color[rgb]{0,0,0}\setlength{\tabcolsep}{0pt}\begin{tabular}{c}D=512\end{tabular}}%
%
\psfrag{x01}[t][t]{0}%
\psfrag{x02}[t][t]{0.1}%
\psfrag{x03}[t][t]{0.2}%
\psfrag{x04}[t][t]{0.3}%
\psfrag{x05}[t][t]{0.4}%
\psfrag{x06}[t][t]{0.5}%
\psfrag{x07}[t][t]{0.6}%
\psfrag{x08}[t][t]{0.7}%
\psfrag{x09}[t][t]{0.8}%
\psfrag{x10}[t][t]{0.9}%
\psfrag{x11}[t][t]{1}%
\psfrag{x12}[t][t]{0}%
\psfrag{x13}[t][t]{1}%
\psfrag{x14}[t][t]{2}%
\psfrag{x15}[t][t]{3}%
\psfrag{x16}[t][t]{4}%
\psfrag{x17}[t][t]{5}%
\psfrag{x18}[t][t]{6}%
\psfrag{x19}[t][t]{7}%
\psfrag{x20}[t][t]{8}%
\psfrag{x21}[t][t]{9}%
\psfrag{x22}[t][t]{10}%
%
\psfrag{v01}[r][r]{0}%
\psfrag{v02}[r][r]{0.1}%
\psfrag{v03}[r][r]{0.2}%
\psfrag{v04}[r][r]{0.3}%
\psfrag{v05}[r][r]{0.4}%
\psfrag{v06}[r][r]{0.5}%
\psfrag{v07}[r][r]{0.6}%
\psfrag{v08}[r][r]{0.7}%
\psfrag{v09}[r][r]{0.8}%
\psfrag{v10}[r][r]{0.9}%
\psfrag{v11}[r][r]{1}%
\psfrag{v12}[r][r]{0}%
\psfrag{v13}[r][r]{0.5}%
\psfrag{v14}[r][r]{1}%
\psfrag{v15}[r][r]{1.5}%
\psfrag{v16}[r][r]{2}%
\psfrag{v17}[r][r]{2.5}%
\psfrag{v18}[r][r]{3}%
\psfrag{v19}[r][r]{3.5}%
\psfrag{v20}[r][r]{4}%
\psfrag{v21}[r][r]{4.5}%
%
\resizebox{\columnwidth}{!}{\includegraphics{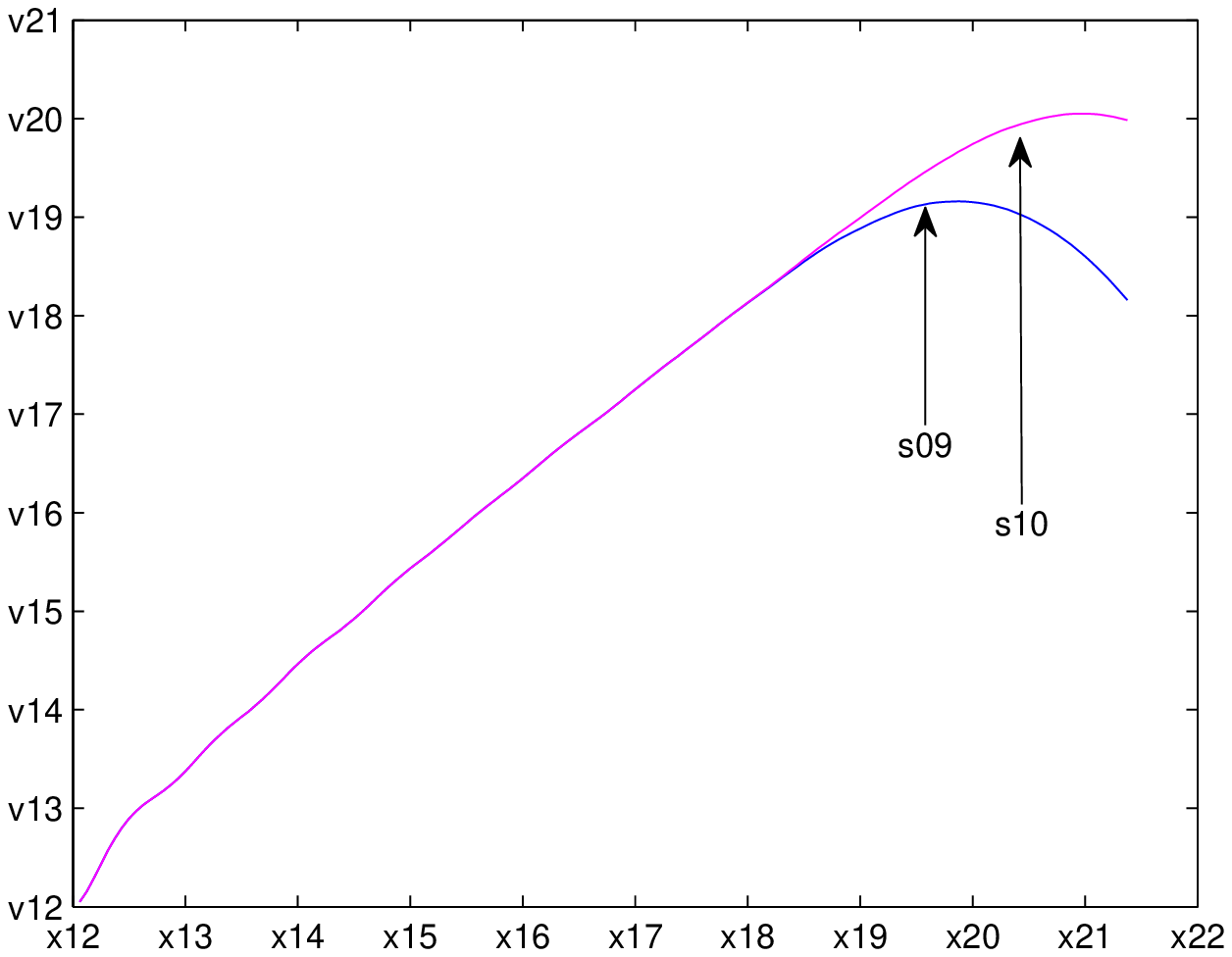}}%
\end{psfrags}%
%

   }
 \end{minipage}
 \caption{Entropy of the half-chain as a function of time for initial 
state $|Z+\rangle$ (left) and $|X+\rangle$ (right).
The entropy is computed from the iTEBD simulation with different
bond dimensions, indicated on the figures.
}
 \label{fig:entropy}
\end{figure}

 \begin{figure}[floatfix]
   \hspace{-.05\columnwidth}
   \begin{minipage}[c]{.3\columnwidth}
     \subfigure{
%
%
\begin{psfrags}%
\psfragscanon%
\Large
%
\psfrag{s13}[b][b]{\color[rgb]{0,0,0}\setlength{\tabcolsep}{0pt}\begin{tabular}{c}$|Z+\rangle$\end{tabular}}%
\psfrag{s22}[b][b]{\color[rgb]{0,0,0}\setlength{\tabcolsep}{0pt}\begin{tabular}{c}\end{tabular}}%
\psfrag{s23}[l][l]{\color[rgb]{0,0,0}\setlength{\tabcolsep}{0pt}\begin{tabular}{c}$|Y+\rangle$\end{tabular}}%
\psfrag{s24}[t][t]{\color[rgb]{0,0,0}\setlength{\tabcolsep}{0pt}\begin{tabular}{c}$|X+\rangle$\end{tabular}}%
%
\psfrag{x01}[t][t]{0}%
\psfrag{x02}[t][t]{0.1}%
\psfrag{x03}[t][t]{0.2}%
\psfrag{x04}[t][t]{0.3}%
\psfrag{x05}[t][t]{0.4}%
\psfrag{x06}[t][t]{0.5}%
\psfrag{x07}[t][t]{0.6}%
\psfrag{x08}[t][t]{0.7}%
\psfrag{x09}[t][t]{0.8}%
\psfrag{x10}[t][t]{0.9}%
\psfrag{x11}[t][t]{1}%
\psfrag{x12}[t][t]{-1}%
\psfrag{x13}[t][t]{-0.5}%
\psfrag{x14}[t][t]{0}%
\psfrag{x15}[t][t]{0.5}%
\psfrag{x16}[t][t]{1}%
%
\psfrag{v01}[r][r]{0}%
\psfrag{v02}[r][r]{0.1}%
\psfrag{v03}[r][r]{0.2}%
\psfrag{v04}[r][r]{0.3}%
\psfrag{v05}[r][r]{0.4}%
\psfrag{v06}[r][r]{0.5}%
\psfrag{v07}[r][r]{0.6}%
\psfrag{v08}[r][r]{0.7}%
\psfrag{v09}[r][r]{0.8}%
\psfrag{v10}[r][r]{0.9}%
\psfrag{v11}[r][r]{1}%
\psfrag{v12}[r][r]{-1}%
\psfrag{v13}[r][r]{-0.5}%
\psfrag{v14}[r][r]{0}%
\psfrag{v15}[r][r]{0.5}%
\psfrag{v16}[r][r]{1}%
%
\psfrag{z01}[r][r]{-1}%
\psfrag{z02}[r][r]{-0.5}%
\psfrag{z03}[r][r]{0}%
\psfrag{z04}[r][r]{0.5}%
\psfrag{z05}[r][r]{1}%
%
\resizebox{\columnwidth}{!}{\includegraphics{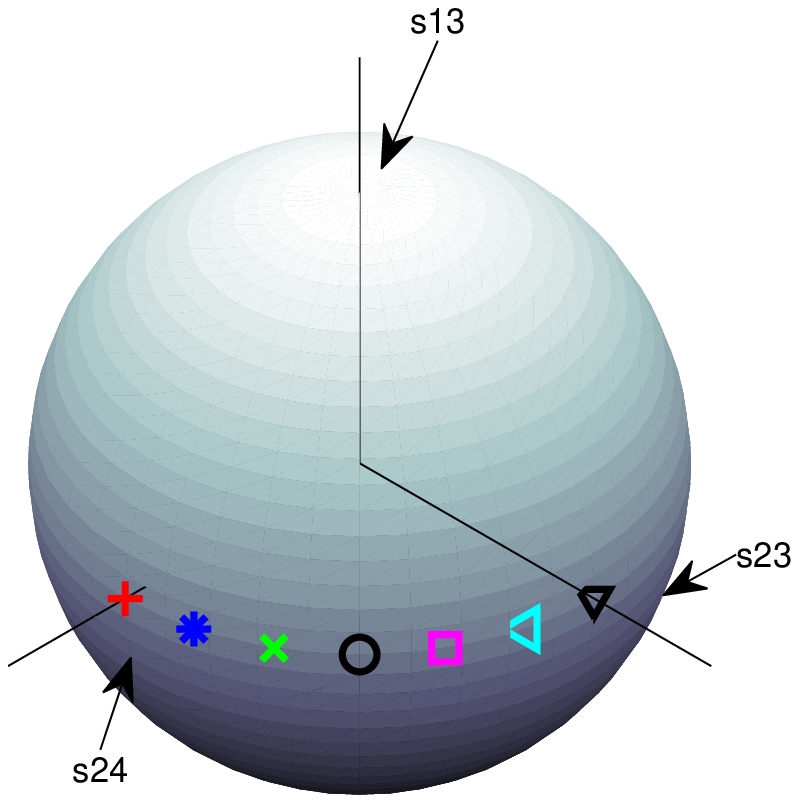}}%
\end{psfrags}%
%

     }
    
 \end{minipage}
 \hspace{.05\columnwidth}
 \begin{minipage}[c]{.6\columnwidth}
   \subfigure{
%
%
\begin{psfrags}%
\psfragscanon%
%
\psfrag{s05}[t][t]{\color[rgb]{0,0,0}\setlength{\tabcolsep}{0pt}\begin{tabular}{c}$t$\end{tabular}}%
\psfrag{s06}[b][b]{\color[rgb]{0,0,0}\setlength{\tabcolsep}{0pt}\begin{tabular}{c}$d(\bar{\rho}(t),\rho_{th})$\end{tabular}}%
%
\psfrag{x01}[t][t]{0}%
\psfrag{x02}[t][t]{0.1}%
\psfrag{x03}[t][t]{0.2}%
\psfrag{x04}[t][t]{0.3}%
\psfrag{x05}[t][t]{0.4}%
\psfrag{x06}[t][t]{0.5}%
\psfrag{x07}[t][t]{0.6}%
\psfrag{x08}[t][t]{0.7}%
\psfrag{x09}[t][t]{0.8}%
\psfrag{x10}[t][t]{0.9}%
\psfrag{x11}[t][t]{1}%
\psfrag{x12}[t][t]{0}%
\psfrag{x13}[t][t]{1}%
\psfrag{x14}[t][t]{2}%
\psfrag{x15}[t][t]{3}%
\psfrag{x16}[t][t]{4}%
\psfrag{x17}[t][t]{5}%
\psfrag{x18}[t][t]{6}%
\psfrag{x19}[t][t]{7}%
\psfrag{x20}[t][t]{8}%
\psfrag{x21}[t][t]{9}%
\psfrag{x22}[t][t]{10}%
%
\psfrag{v01}[r][r]{0}%
\psfrag{v02}[r][r]{0.1}%
\psfrag{v03}[r][r]{0.2}%
\psfrag{v04}[r][r]{0.3}%
\psfrag{v05}[r][r]{0.4}%
\psfrag{v06}[r][r]{0.5}%
\psfrag{v07}[r][r]{0.6}%
\psfrag{v08}[r][r]{0.7}%
\psfrag{v09}[r][r]{0.8}%
\psfrag{v10}[r][r]{0.9}%
\psfrag{v11}[r][r]{1}%
\psfrag{v12}[r][r]{0}%
\psfrag{v13}[r][r]{0.02}%
\psfrag{v14}[r][r]{0.04}%
\psfrag{v15}[r][r]{0.06}%
\psfrag{v16}[r][r]{0.08}%
\psfrag{v17}[r][r]{0.1}%
\psfrag{v18}[r][r]{0.12}%
\psfrag{v19}[r][r]{0.14}%
\psfrag{v20}[r][r]{0.16}%
\psfrag{v21}[r][r]{0.18}%
\psfrag{v22}[r][r]{0.2}%
%
\resizebox{\columnwidth}{!}{\includegraphics{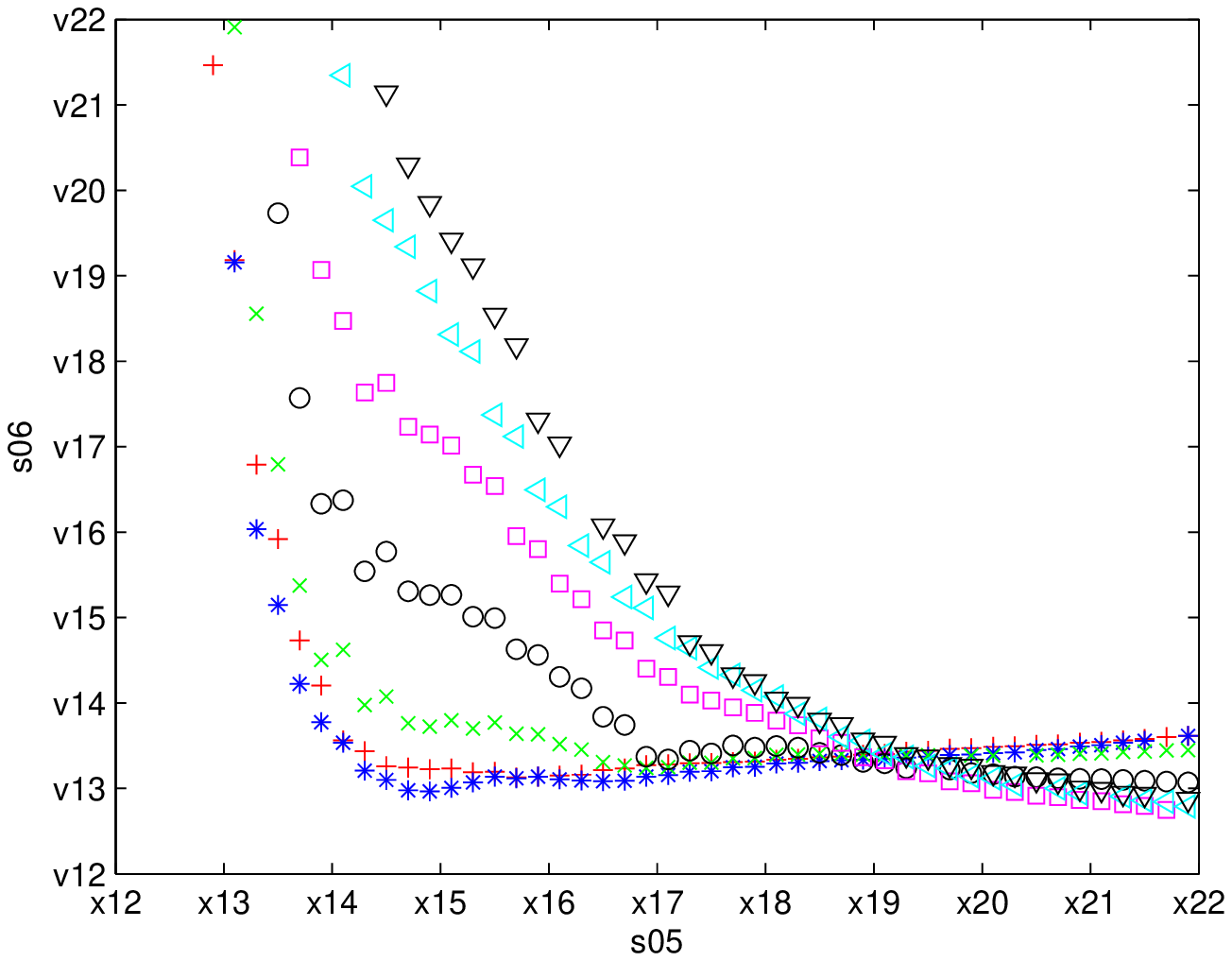}}%
\end{psfrags}%
%

   }
 \end{minipage}
 \caption{Distance between the averaged 3-body reduced density matrix
   and the thermal state as a function of time for various product
   initial states between $|X+\rangle$ ($\beta=-0.7180$) and 
   $|Y+\rangle$ ($\beta=0$).The single spin states are shown on
   the Bloch sphere on the left.}
  \label{fig:averageSetXY}
\end{figure}

 \begin{figure}[floatfix]
   \hspace{-.05\columnwidth}
   \begin{minipage}[c]{.3\columnwidth}
     \subfigure{
%
%
\begin{psfrags}%
\psfragscanon%
\Large
%
\psfrag{s13}[b][b]{\color[rgb]{0,0,0}\setlength{\tabcolsep}{0pt}\begin{tabular}{c}$|Z+\rangle$\end{tabular}}%
\psfrag{s22}[b][b]{\color[rgb]{0,0,0}\setlength{\tabcolsep}{0pt}\begin{tabular}{c}\end{tabular}}%
\psfrag{s23}[l][l]{\color[rgb]{0,0,0}\setlength{\tabcolsep}{0pt}\begin{tabular}{c}$|Y+\rangle$\end{tabular}}%
\psfrag{s24}[t][t]{\color[rgb]{0,0,0}\setlength{\tabcolsep}{0pt}\begin{tabular}{c}$|X+\rangle$\end{tabular}}%
%
\psfrag{x01}[t][t]{0}%
\psfrag{x02}[t][t]{0.1}%
\psfrag{x03}[t][t]{0.2}%
\psfrag{x04}[t][t]{0.3}%
\psfrag{x05}[t][t]{0.4}%
\psfrag{x06}[t][t]{0.5}%
\psfrag{x07}[t][t]{0.6}%
\psfrag{x08}[t][t]{0.7}%
\psfrag{x09}[t][t]{0.8}%
\psfrag{x10}[t][t]{0.9}%
\psfrag{x11}[t][t]{1}%
\psfrag{x12}[t][t]{-1}%
\psfrag{x13}[t][t]{-0.5}%
\psfrag{x14}[t][t]{0}%
\psfrag{x15}[t][t]{0.5}%
\psfrag{x16}[t][t]{1}%
%
\psfrag{v01}[r][r]{0}%
\psfrag{v02}[r][r]{0.1}%
\psfrag{v03}[r][r]{0.2}%
\psfrag{v04}[r][r]{0.3}%
\psfrag{v05}[r][r]{0.4}%
\psfrag{v06}[r][r]{0.5}%
\psfrag{v07}[r][r]{0.6}%
\psfrag{v08}[r][r]{0.7}%
\psfrag{v09}[r][r]{0.8}%
\psfrag{v10}[r][r]{0.9}%
\psfrag{v11}[r][r]{1}%
\psfrag{v12}[r][r]{-1}%
\psfrag{v13}[r][r]{-0.5}%
\psfrag{v14}[r][r]{0}%
\psfrag{v15}[r][r]{0.5}%
\psfrag{v16}[r][r]{1}%
%
\psfrag{z01}[r][r]{-1}%
\psfrag{z02}[r][r]{-0.5}%
\psfrag{z03}[r][r]{0}%
\psfrag{z04}[r][r]{0.5}%
\psfrag{z05}[r][r]{1}%
%
\resizebox{\columnwidth}{!}{\includegraphics{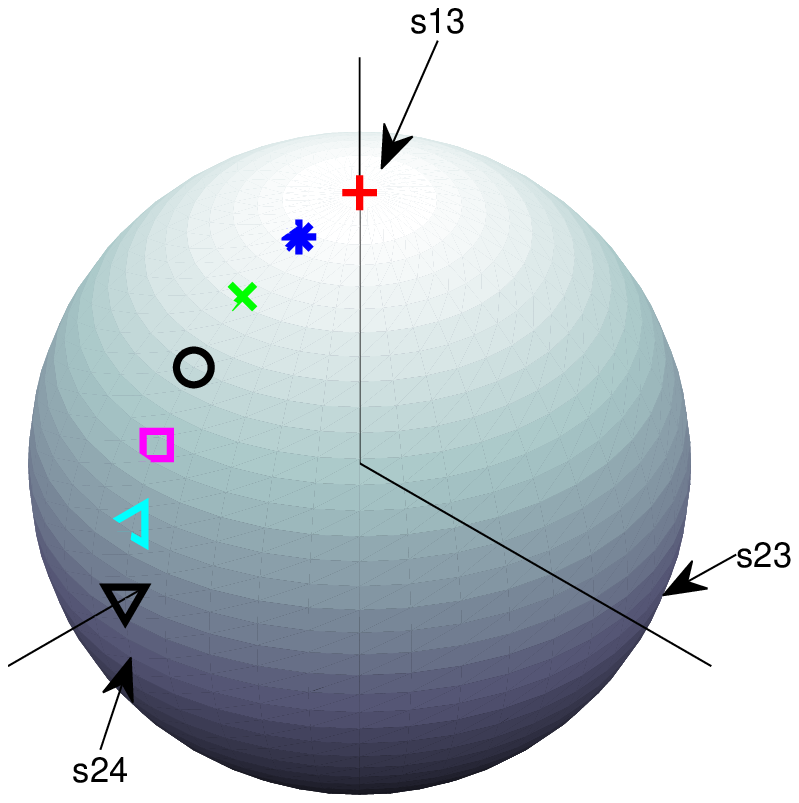}}%
\end{psfrags}%
%

     }
    
 \end{minipage}
 \hspace{.05\columnwidth}
 \begin{minipage}[c]{.6\columnwidth}
   \subfigure{
%
%
\begin{psfrags}%
\psfragscanon%
%
\psfrag{s05}[t][t]{\color[rgb]{0,0,0}\setlength{\tabcolsep}{0pt}\begin{tabular}{c}$t$\end{tabular}}%
\psfrag{s06}[b][b]{\color[rgb]{0,0,0}\setlength{\tabcolsep}{0pt}\begin{tabular}{c}$d(\bar{\rho}(t),\rho_{th})$\end{tabular}}%
%
\psfrag{x01}[t][t]{0}%
\psfrag{x02}[t][t]{0.1}%
\psfrag{x03}[t][t]{0.2}%
\psfrag{x04}[t][t]{0.3}%
\psfrag{x05}[t][t]{0.4}%
\psfrag{x06}[t][t]{0.5}%
\psfrag{x07}[t][t]{0.6}%
\psfrag{x08}[t][t]{0.7}%
\psfrag{x09}[t][t]{0.8}%
\psfrag{x10}[t][t]{0.9}%
\psfrag{x11}[t][t]{1}%
\psfrag{x12}[t][t]{0}%
\psfrag{x13}[t][t]{1}%
\psfrag{x14}[t][t]{2}%
\psfrag{x15}[t][t]{3}%
\psfrag{x16}[t][t]{4}%
\psfrag{x17}[t][t]{5}%
\psfrag{x18}[t][t]{6}%
\psfrag{x19}[t][t]{7}%
\psfrag{x20}[t][t]{8}%
\psfrag{x21}[t][t]{9}%
\psfrag{x22}[t][t]{10}%
%
\psfrag{v01}[r][r]{0}%
\psfrag{v02}[r][r]{0.1}%
\psfrag{v03}[r][r]{0.2}%
\psfrag{v04}[r][r]{0.3}%
\psfrag{v05}[r][r]{0.4}%
\psfrag{v06}[r][r]{0.5}%
\psfrag{v07}[r][r]{0.6}%
\psfrag{v08}[r][r]{0.7}%
\psfrag{v09}[r][r]{0.8}%
\psfrag{v10}[r][r]{0.9}%
\psfrag{v11}[r][r]{1}%
\psfrag{v12}[r][r]{0}%
\psfrag{v13}[r][r]{0.02}%
\psfrag{v14}[r][r]{0.04}%
\psfrag{v15}[r][r]{0.06}%
\psfrag{v16}[r][r]{0.08}%
\psfrag{v17}[r][r]{0.1}%
\psfrag{v18}[r][r]{0.12}%
\psfrag{v19}[r][r]{0.14}%
\psfrag{v20}[r][r]{0.16}%
\psfrag{v21}[r][r]{0.18}%
\psfrag{v22}[r][r]{0.2}%
%
\resizebox{\columnwidth}{!}{\includegraphics{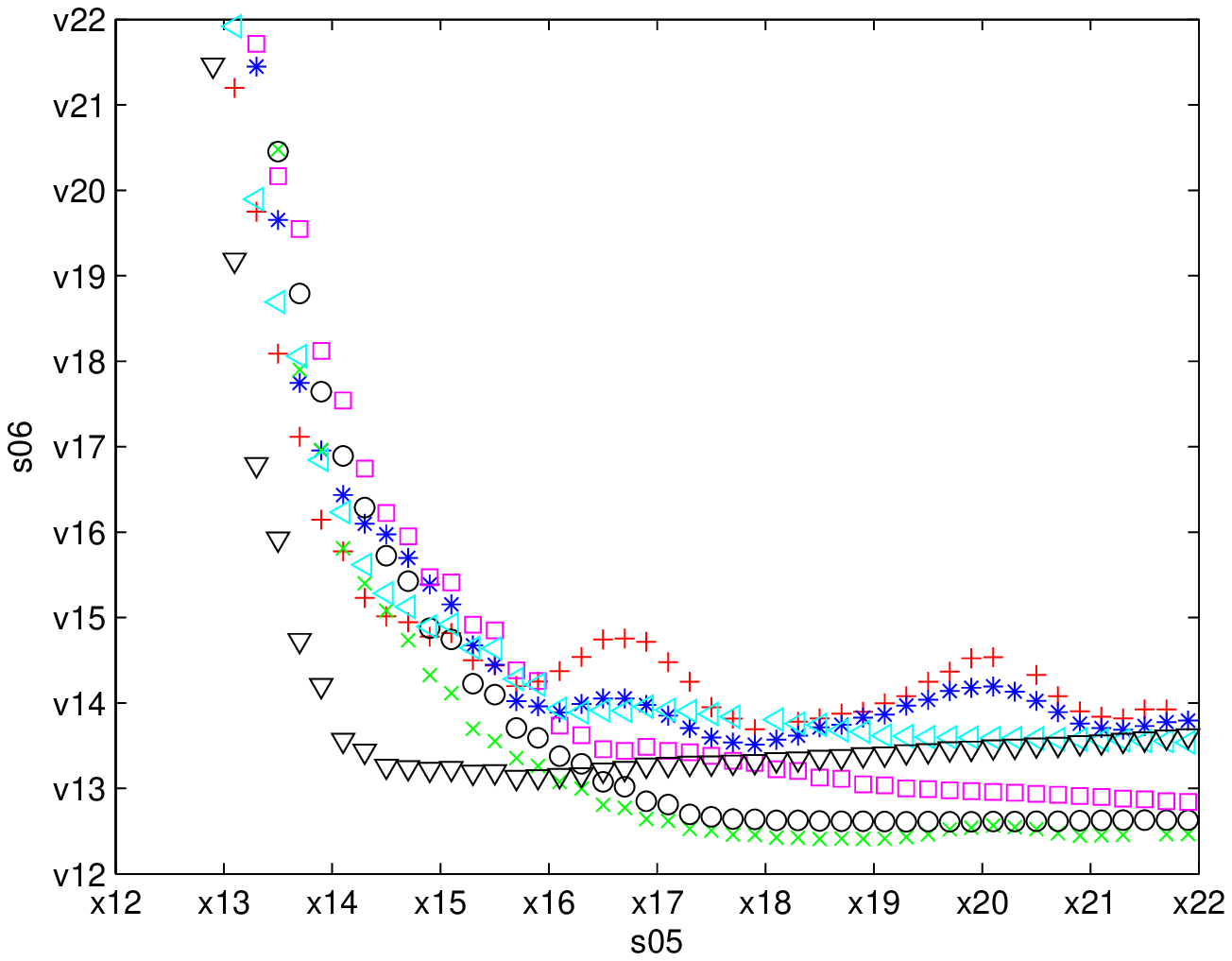}}%
\end{psfrags}%
%

   }
 \end{minipage}
 \caption{Distance between the averaged 3-body reduced density matrix
   and the thermal state as a function of time for various product
   initial states, indicated on the Bloch sphere on the left,
   and corresponding to inverse temperatures between
   $\beta=-0.7180$ and $\beta=0.7275$. }
 \label{fig:averageSetXZ}
\end{figure}

\subsection{Different initial states}


Our initial configurations, translationally invariant product states,
are specified by the state of a single spin,
$$
|\Psi\rangle=\cos\frac{\theta}{2}|0\rangle + \mathrm{e}^{i\phi} \sin\frac{\theta}{2}|1\rangle,
$$ which can be represented on the Bloch sphere by the point with
coordinates $(\theta,\phi)$.  This state has energy per spin
$E=-\left(\cos^2\theta +g \sin \theta \cos \phi + h \cos \theta
\right)$.  States with spins polarised in the three orthogonal
directions, $|X+\rangle$, $|Y+\rangle$ and $|Z+\rangle$, behave very
differently with respect to thermalization regarding the distance
between the reduced evolved density matrix and the thermal one.  We
may additionally check that their dynamics are essentially different
by looking at the individual expectation values
(see also Fig.~\ref{fig:correlsOther}).  For the initial state $|Y+\rangle$,
showing strong thermalization, all of the individual expectation
values converge fast to the thermal ones. Instead, for the initial
state $|Z+\rangle$, we observe irregular oscillations, showing no sign
of damping, to the longest times we are able to simulate.  For the
initial $|X+\rangle$ state, we check that only few expectation values
deviate from the thermal average, and are those preventing
thermalization of the whole reduced density matrix. In particular, for
$N=1$, only $\langle \sigma_x \rangle$ is responsible for the lack of
thermalization.  The behavior for larger reduced density matrices
$N=2,$ $3$ is qualitatively similar, although the time it takes for
$|Y+\rangle$ to thermalize becomes longer.

From the iTEBD simulations we may analyze how the entropy of the
half-chain increases in time for the non-thermalizing states (see
Fig.~\ref{fig:entropy}).  We notice that for the weak thermalizing
initial state $|Z+\rangle$ the entropy of the half-chain grows
linearly with time, so that the oscillatory behavior is not due to the
absence of propagating excitations.  For the initial state
$|X+\rangle$ the entropy also grows linearly with time, but it does so
at a faster pace. As shown in the figure, the iTEBD simulation can
reproduce this growth until the time when the truncation error becomes
dominant.  A closer look at the data of this simulation shows that the
distribution of Schmidt coefficients is very different for both
cases, even at times when they attain a comparable entropy. 
If we analyze the Schmidt decompositions of both states at the
time at which their entropy is $S\approx0.8$
($t=1.1875$ for $|X+\rangle$ and $t=8$ for $|Z+\rangle$), we observe
that the coefficient distribution for $|Z+\rangle$ has a much longer
tail.
This difference in the distribution can also be 
quantified by the 2-R\'enyi entropy 
$S_2=-\mathrm{log}(\mathrm{tr} \rho^2)$, which attains a lower value 
for the weak thermalizing state, $S_2(Z+)=0.348$, while $S_2(X+)=0.611$.

By rotating the inital state on the Bloch sphere, we observe a
transition from the strong thermalizing $|Y+\rangle$ to the weak one
$|Z+\rangle$, and also to the apparently non-thermalizing $|X+\rangle$
(Fig.~\ref{fig:averageSetXY}), the latter occurring around
$\phi\in[\frac{\pi}{6},\frac{\pi}{4}]$ ($\beta\in[-0.5382,-0.3915]$ ).
We have also analysed the transition from weak thermalization
$|Z+\rangle$ to non-thermalization $|X+\rangle$
(Fig.~\ref{fig:averageSetXZ}).  In this case we find some intermediate
states for which strong thermalization occurs.  By looking at the
energy, we may infer the corresponding $\beta$ of every initial
product state.
We discover that all the strong thermalizing states we have found have
energies, and thus $\beta$, close to zero.

 \begin{figure}[floatfix]
%
%
\begin{psfrags}%
\psfragscanon%
%
\psfrag{s05}[t][t]{\color[rgb]{0,0,0}\setlength{\tabcolsep}{0pt}\begin{tabular}{c}$t$\end{tabular}}%
\psfrag{s06}[b][b]{\color[rgb]{0,0,0}\setlength{\tabcolsep}{0pt}\begin{tabular}{c}$d(\bar{\rho}(t),\rho_{th})$\end{tabular}}%
%
\psfrag{x01}[t][t]{0}%
\psfrag{x02}[t][t]{0.1}%
\psfrag{x03}[t][t]{0.2}%
\psfrag{x04}[t][t]{0.3}%
\psfrag{x05}[t][t]{0.4}%
\psfrag{x06}[t][t]{0.5}%
\psfrag{x07}[t][t]{0.6}%
\psfrag{x08}[t][t]{0.7}%
\psfrag{x09}[t][t]{0.8}%
\psfrag{x10}[t][t]{0.9}%
\psfrag{x11}[t][t]{1}%
\psfrag{x12}[t][t]{0}%
\psfrag{x13}[t][t]{1}%
\psfrag{x14}[t][t]{2}%
\psfrag{x15}[t][t]{3}%
\psfrag{x16}[t][t]{4}%
\psfrag{x17}[t][t]{5}%
\psfrag{x18}[t][t]{6}%
\psfrag{x19}[t][t]{7}%
\psfrag{x20}[t][t]{8}%
\psfrag{x21}[t][t]{9}%
\psfrag{x22}[t][t]{10}%
%
\psfrag{v01}[r][r]{0}%
\psfrag{v02}[r][r]{0.1}%
\psfrag{v03}[r][r]{0.2}%
\psfrag{v04}[r][r]{0.3}%
\psfrag{v05}[r][r]{0.4}%
\psfrag{v06}[r][r]{0.5}%
\psfrag{v07}[r][r]{0.6}%
\psfrag{v08}[r][r]{0.7}%
\psfrag{v09}[r][r]{0.8}%
\psfrag{v10}[r][r]{0.9}%
\psfrag{v11}[r][r]{1}%
\psfrag{v12}[r][r]{0}%
\psfrag{v13}[r][r]{0.02}%
\psfrag{v14}[r][r]{0.04}%
\psfrag{v15}[r][r]{0.06}%
\psfrag{v16}[r][r]{0.08}%
\psfrag{v17}[r][r]{0.1}%
\psfrag{v18}[r][r]{0.12}%
\psfrag{v19}[r][r]{0.14}%
\psfrag{v20}[r][r]{0.16}%
\psfrag{v21}[r][r]{0.18}%
\psfrag{v22}[r][r]{0.2}%
%
\resizebox{\columnwidth}{!}{\includegraphics{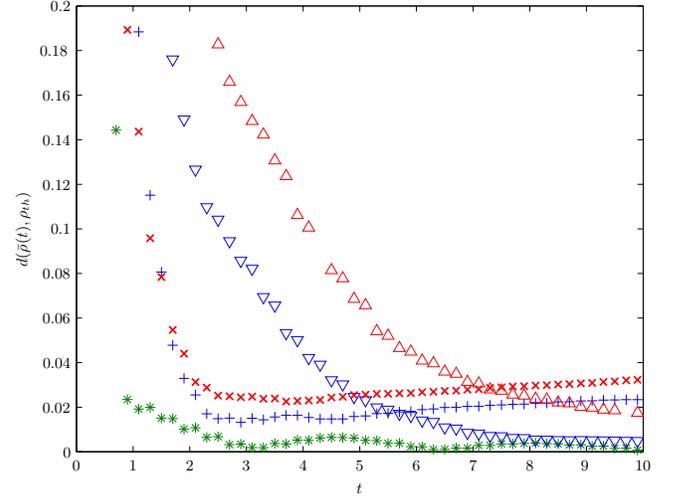}}%
\end{psfrags}%
%

   \caption{Distance between the averaged $N=3$ reduced density matrix
     and the thermal one for states with similar energy. For
     $\beta=0$, we compare $|Y+\rangle$ (red upwards-pointing
     triangles) and ($\theta=0.4398$,$\phi=\pi$) (blue
     downwards-pointing triangles).  For $\beta=-0.7180$, we compare
     $|X+\rangle$ (red x) to ($\theta=1.90$,$\phi=0.031$) (blue
     crosses).  We show also the behaviour of the state with maximal
     $\beta=1.5859$ (green stars), whose expectation values oscillate
     like for $|Z+\rangle$, but in average converges very fast to
     thermal.  All data correspond to $D=120$ simulation.  }
   \label{fig:extraStates}
\end{figure}

Instead, for larger $\beta>0$, we observe oscillations and weak
thermalization as in the $|Z+\rangle$ initial state.  An interesting
case is the state of maximum $\beta$ (the product state with minimal
energy), which we can identify, by studying the energy landscape over 
the Bloch sphere, at $\theta\approx 0.43$,
$\phi=\pi$.  The dynamics of this initial state shows also weak
thermalization, with strong oscillations of all the expectation
values, but fast thermalization in average
(Fig.~\ref{fig:extraStates}).  We may perform a similar analysis on
some extra states, with initial energy densities close to $|X+\rangle$
and $|Y+\rangle$, respectively, to check whether they relax to similar
thermal states.  We find that they seem to show the same regime of
thermalization (weak or strong) as the original states
(Fig.~\ref{fig:extraStates}), suggesting this has to do with the
initial energy. The relaxation curves however differ,
indicating that, even starting with the same $\beta$, 
the state of the system does not relax to the same state, at least
during the long range of times we simulate.


\subsection{Varying the Hamiltonian parameters}

\begin{figure}[floatfix]
  \hspace{-.05\columnwidth}
  \begin{minipage}[c]{.45\columnwidth}
    \subfigure{
%
%
\begin{psfrags}%
\psfragscanon%
\Large
%
\psfrag{s05}[t][t]{\color[rgb]{0,0,0}\setlength{\tabcolsep}{0pt}\begin{tabular}{c}\end{tabular}}%
%
\psfrag{x01}[t][t]{0}%
\psfrag{x02}[t][t]{0.1}%
\psfrag{x03}[t][t]{0.2}%
\psfrag{x04}[t][t]{0.3}%
\psfrag{x05}[t][t]{0.4}%
\psfrag{x06}[t][t]{0.5}%
\psfrag{x07}[t][t]{0.6}%
\psfrag{x08}[t][t]{0.7}%
\psfrag{x09}[t][t]{0.8}%
\psfrag{x10}[t][t]{0.9}%
\psfrag{x11}[t][t]{1}%
\psfrag{x12}[t][t]{0}%
\psfrag{x13}[t][t]{1}%
\psfrag{x14}[t][t]{2}%
\psfrag{x15}[t][t]{3}%
\psfrag{x16}[t][t]{4}%
\psfrag{x17}[t][t]{5}%
\psfrag{x18}[t][t]{6}%
\psfrag{x19}[t][t]{7}%
\psfrag{x20}[t][t]{8}%
\psfrag{x21}[t][t]{9}%
\psfrag{x22}[t][t]{10}%
%
\psfrag{v01}[r][r]{0}%
\psfrag{v02}[r][r]{0.1}%
\psfrag{v03}[r][r]{0.2}%
\psfrag{v04}[r][r]{0.3}%
\psfrag{v05}[r][r]{0.4}%
\psfrag{v06}[r][r]{0.5}%
\psfrag{v07}[r][r]{0.6}%
\psfrag{v08}[r][r]{0.7}%
\psfrag{v09}[r][r]{0.8}%
\psfrag{v10}[r][r]{0.9}%
\psfrag{v11}[r][r]{1}%
\psfrag{v12}[r][r]{0}%
\psfrag{v13}[r][r]{0.05}%
\psfrag{v14}[r][r]{0.1}%
\psfrag{v15}[r][r]{0.15}%
\psfrag{v16}[r][r]{0.2}%
\psfrag{v17}[r][r]{0.25}%
\psfrag{v18}[r][r]{0.3}%
%
\resizebox{\columnwidth}{!}{\includegraphics{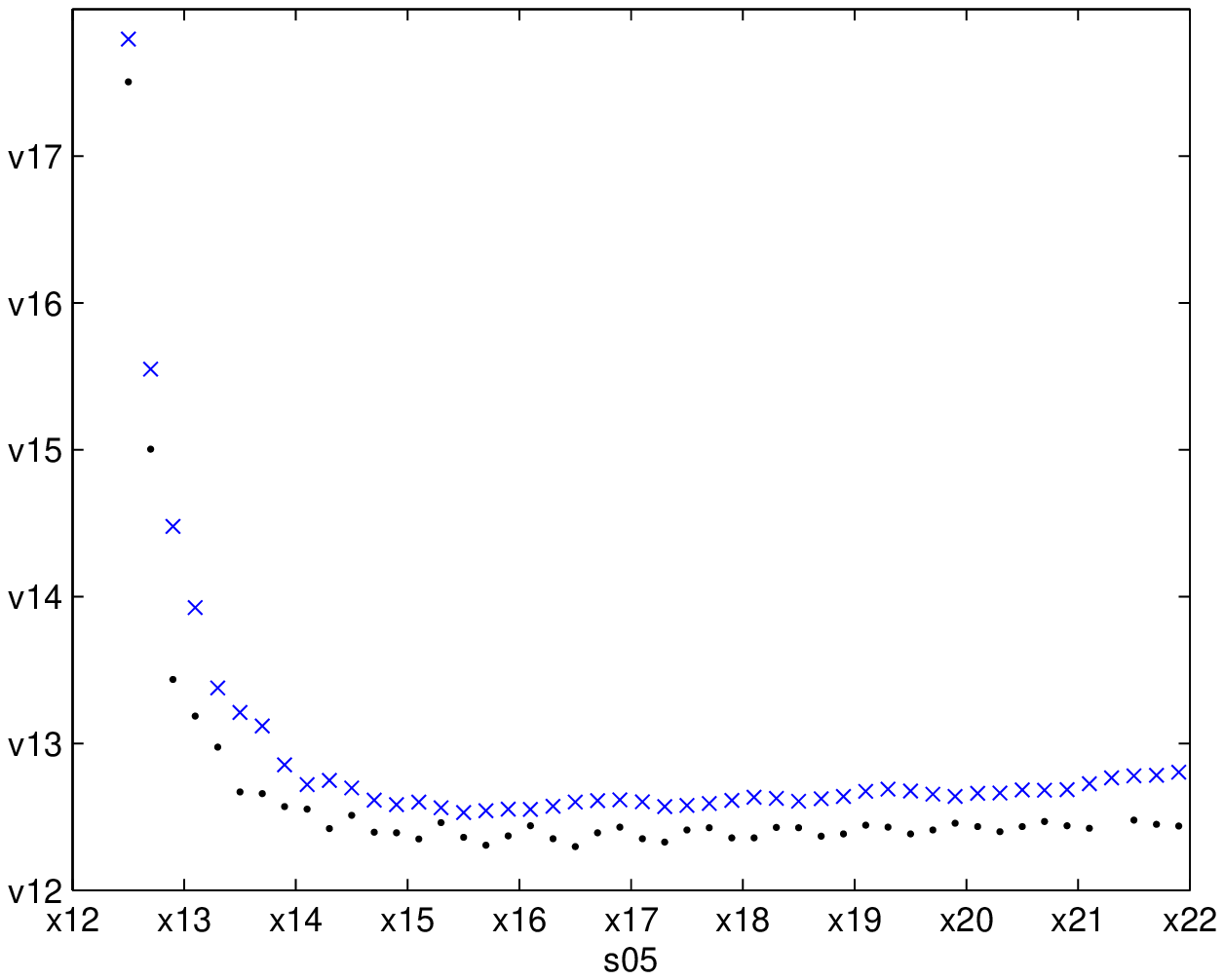}}%
\end{psfrags}%
%

    }
  \end{minipage}
  \hspace{.05\columnwidth}
  \begin{minipage}[c]{.45\columnwidth}
    \subfigure{
%
%
\begin{psfrags}%
\psfragscanon%
\Large
%
\psfrag{s09}[lt][lt]{\color[rgb]{0,0,0}\setlength{\tabcolsep}{0pt}\begin{tabular}{l}\end{tabular}}%
%
\psfrag{x01}[t][t]{0}%
\psfrag{x02}[t][t]{0.1}%
\psfrag{x03}[t][t]{0.2}%
\psfrag{x04}[t][t]{0.3}%
\psfrag{x05}[t][t]{0.4}%
\psfrag{x06}[t][t]{0.5}%
\psfrag{x07}[t][t]{0.6}%
\psfrag{x08}[t][t]{0.7}%
\psfrag{x09}[t][t]{0.8}%
\psfrag{x10}[t][t]{0.9}%
\psfrag{x11}[t][t]{1}%
\psfrag{x12}[t][t]{0}%
\psfrag{x13}[t][t]{1}%
\psfrag{x14}[t][t]{2}%
\psfrag{x15}[t][t]{3}%
\psfrag{x16}[t][t]{4}%
\psfrag{x17}[t][t]{5}%
\psfrag{x18}[t][t]{6}%
\psfrag{x19}[t][t]{7}%
\psfrag{x20}[t][t]{8}%
\psfrag{x21}[t][t]{9}%
\psfrag{x22}[t][t]{10}%
%
\psfrag{v01}[r][r]{0}%
\psfrag{v02}[r][r]{0.1}%
\psfrag{v03}[r][r]{0.2}%
\psfrag{v04}[r][r]{0.3}%
\psfrag{v05}[r][r]{0.4}%
\psfrag{v06}[r][r]{0.5}%
\psfrag{v07}[r][r]{0.6}%
\psfrag{v08}[r][r]{0.7}%
\psfrag{v09}[r][r]{0.8}%
\psfrag{v10}[r][r]{0.9}%
\psfrag{v11}[r][r]{1}%
\psfrag{v12}[r][r]{0}%
\psfrag{v13}[r][r]{0.1}%
\psfrag{v14}[r][r]{0.2}%
\psfrag{v15}[r][r]{0.3}%
\psfrag{v16}[r][r]{0.4}%
\psfrag{v17}[r][r]{0.5}%
\psfrag{v18}[r][r]{0.6}%
\psfrag{v19}[r][r]{0.7}%
\psfrag{v20}[r][r]{0.8}%
\psfrag{v21}[r][r]{0.9}%
%
\resizebox{\columnwidth}{!}{\includegraphics{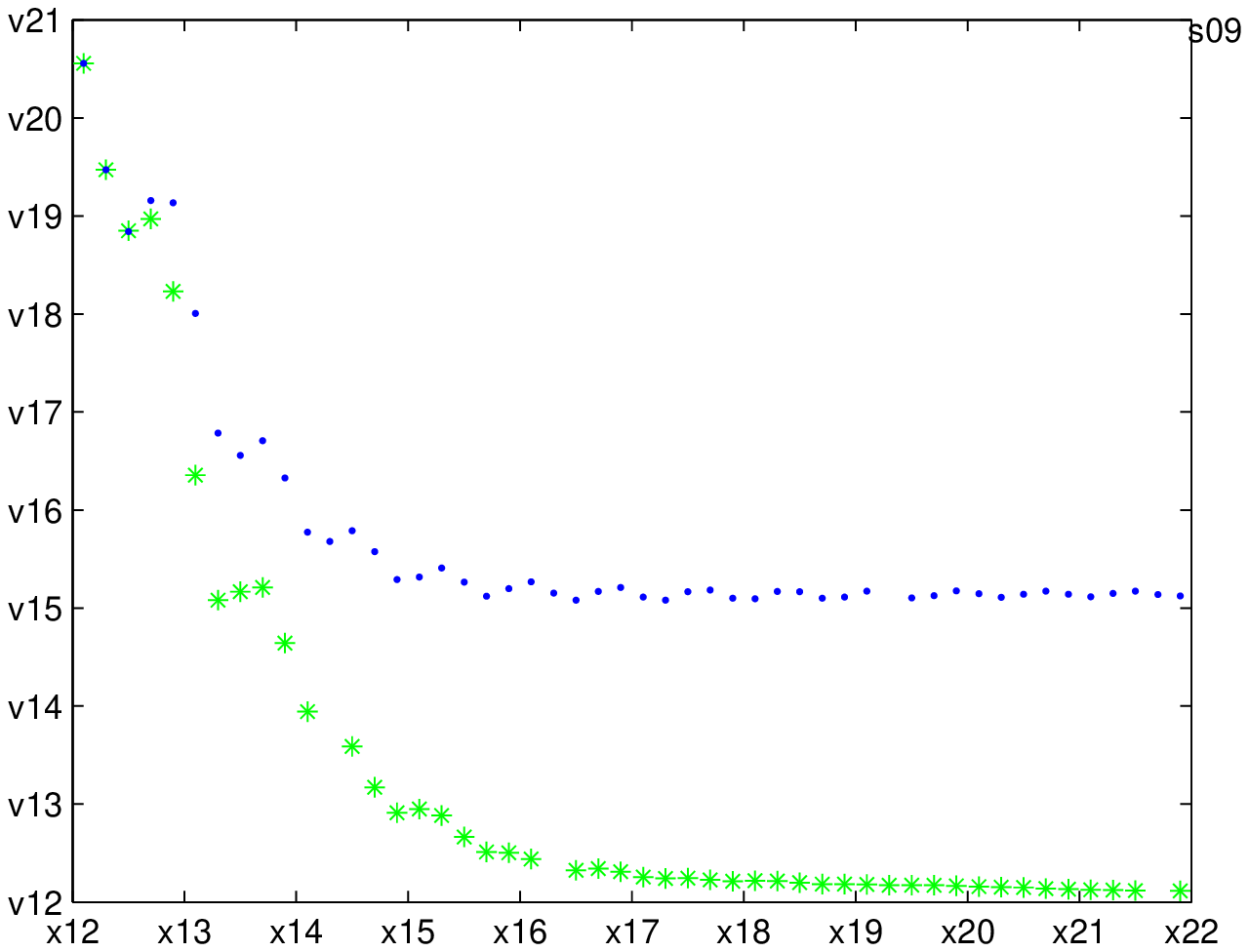}}%
\end{psfrags}%
%

    }
  \end{minipage}
  \\
  \begin{minipage}[c]{.45\columnwidth}
    \subfigure{
%
%
\begin{psfrags}%
\psfragscanon%
\Large
%
\psfrag{s05}[t][t]{\color[rgb]{0,0,0}\setlength{\tabcolsep}{0pt}\begin{tabular}{c}\end{tabular}}%
%
\psfrag{x01}[t][t]{0}%
\psfrag{x02}[t][t]{0.1}%
\psfrag{x03}[t][t]{0.2}%
\psfrag{x04}[t][t]{0.3}%
\psfrag{x05}[t][t]{0.4}%
\psfrag{x06}[t][t]{0.5}%
\psfrag{x07}[t][t]{0.6}%
\psfrag{x08}[t][t]{0.7}%
\psfrag{x09}[t][t]{0.8}%
\psfrag{x10}[t][t]{0.9}%
\psfrag{x11}[t][t]{1}%
\psfrag{x12}[t][t]{0}%
\psfrag{x13}[t][t]{1}%
\psfrag{x14}[t][t]{2}%
\psfrag{x15}[t][t]{3}%
\psfrag{x16}[t][t]{4}%
\psfrag{x17}[t][t]{5}%
\psfrag{x18}[t][t]{6}%
\psfrag{x19}[t][t]{7}%
\psfrag{x20}[t][t]{8}%
\psfrag{x21}[t][t]{9}%
\psfrag{x22}[t][t]{10}%
%
\psfrag{v01}[r][r]{0}%
\psfrag{v02}[r][r]{0.1}%
\psfrag{v03}[r][r]{0.2}%
\psfrag{v04}[r][r]{0.3}%
\psfrag{v05}[r][r]{0.4}%
\psfrag{v06}[r][r]{0.5}%
\psfrag{v07}[r][r]{0.6}%
\psfrag{v08}[r][r]{0.7}%
\psfrag{v09}[r][r]{0.8}%
\psfrag{v10}[r][r]{0.9}%
\psfrag{v11}[r][r]{1}%
\psfrag{v12}[r][r]{0}%
\psfrag{v13}[r][r]{0.1}%
\psfrag{v14}[r][r]{0.2}%
\psfrag{v15}[r][r]{0.3}%
\psfrag{v16}[r][r]{0.4}%
\psfrag{v17}[r][r]{0.5}%
\psfrag{v18}[r][r]{0.6}%
%
\resizebox{\columnwidth}{!}{\includegraphics{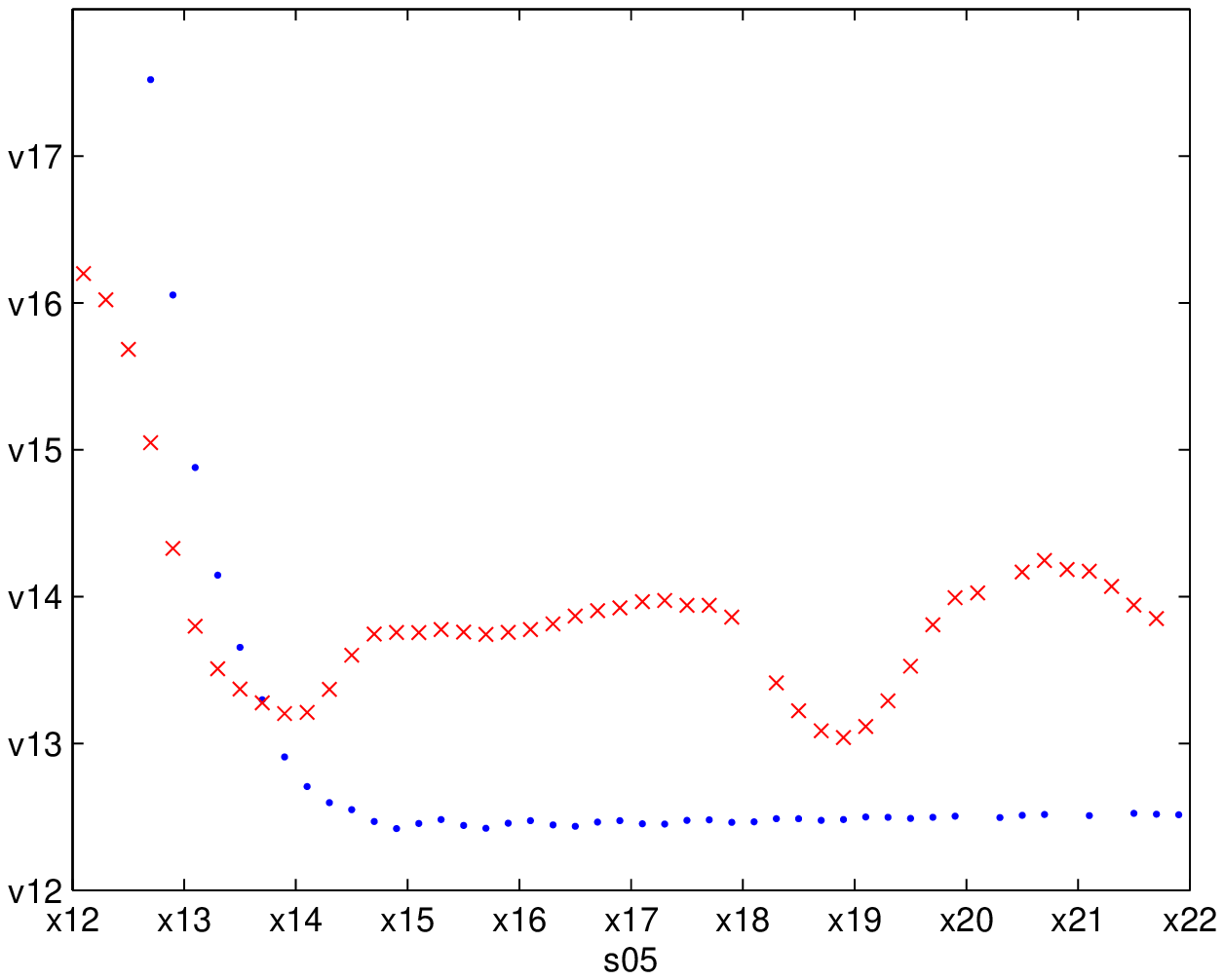}}%
\end{psfrags}%
%

    }
  \end{minipage}
  \caption{Comparison between the integrable ($g=1.05$, $h=0$) and
    non-integrable ($g=-1.05,$ $h=0.5$) models. The plots show the
    distance between the evolved $N=3$ reduced density matrix and the
    corresponding thermal one for the initial states $|X+\rangle$
    (upper left), $|Y+\rangle$ (upper right) and $|Z+\rangle$ (lower
    plot).  The integrable case is represented always by black
    dots. In the three cases, for the integrable limit we see the
    relaxation of the reduced density matrix to a state that is
    different from thermal. We notice also that the initial state
    $|X+\rangle$ shows in the non-integrable case a similar behaviour
    to the $h=0$ limit.}
\label{fig:integrable}
\end{figure}

 \begin{figure}[floatfix]
   \hspace{-.05\columnwidth}
   \begin{minipage}[c]{.45\columnwidth}
     \begin{minipage}[c]{.8\columnwidth}
       \subfigure{       
         
       }
   \end{minipage}
 \end{minipage}
 \hspace{.05\columnwidth}
 \begin{minipage}[c]{.45\columnwidth}
   \subfigure{
     \Large
%
%
\begin{psfrags}%
\psfragscanon%
%
\psfrag{s05}[t][t]{\color[rgb]{0,0,0}\setlength{\tabcolsep}{0pt}\begin{tabular}{c}\end{tabular}}%
%
\psfrag{x01}[t][t]{0}%
\psfrag{x02}[t][t]{0.1}%
\psfrag{x03}[t][t]{0.2}%
\psfrag{x04}[t][t]{0.3}%
\psfrag{x05}[t][t]{0.4}%
\psfrag{x06}[t][t]{0.5}%
\psfrag{x07}[t][t]{0.6}%
\psfrag{x08}[t][t]{0.7}%
\psfrag{x09}[t][t]{0.8}%
\psfrag{x10}[t][t]{0.9}%
\psfrag{x11}[t][t]{1}%
\psfrag{x12}[t][t]{3}%
\psfrag{x13}[t][t]{4}%
\psfrag{x14}[t][t]{5}%
\psfrag{x15}[t][t]{6}%
\psfrag{x16}[t][t]{7}%
\psfrag{x17}[t][t]{8}%
\psfrag{x18}[t][t]{9}%
\psfrag{x19}[t][t]{10}%
%
\psfrag{v01}[r][r]{0}%
\psfrag{v02}[r][r]{0.1}%
\psfrag{v03}[r][r]{0.2}%
\psfrag{v04}[r][r]{0.3}%
\psfrag{v05}[r][r]{0.4}%
\psfrag{v06}[r][r]{0.5}%
\psfrag{v07}[r][r]{0.6}%
\psfrag{v08}[r][r]{0.7}%
\psfrag{v09}[r][r]{0.8}%
\psfrag{v10}[r][r]{0.9}%
\psfrag{v11}[r][r]{1}%
\psfrag{v12}[r][r]{0}%
\psfrag{v13}[r][r]{0.05}%
\psfrag{v14}[r][r]{0.1}%
\psfrag{v15}[r][r]{0.15}%
\psfrag{v16}[r][r]{0.2}%
\psfrag{v17}[r][r]{0.25}%
\psfrag{v18}[r][r]{0.3}%
%
\resizebox{\columnwidth}{!}{\includegraphics{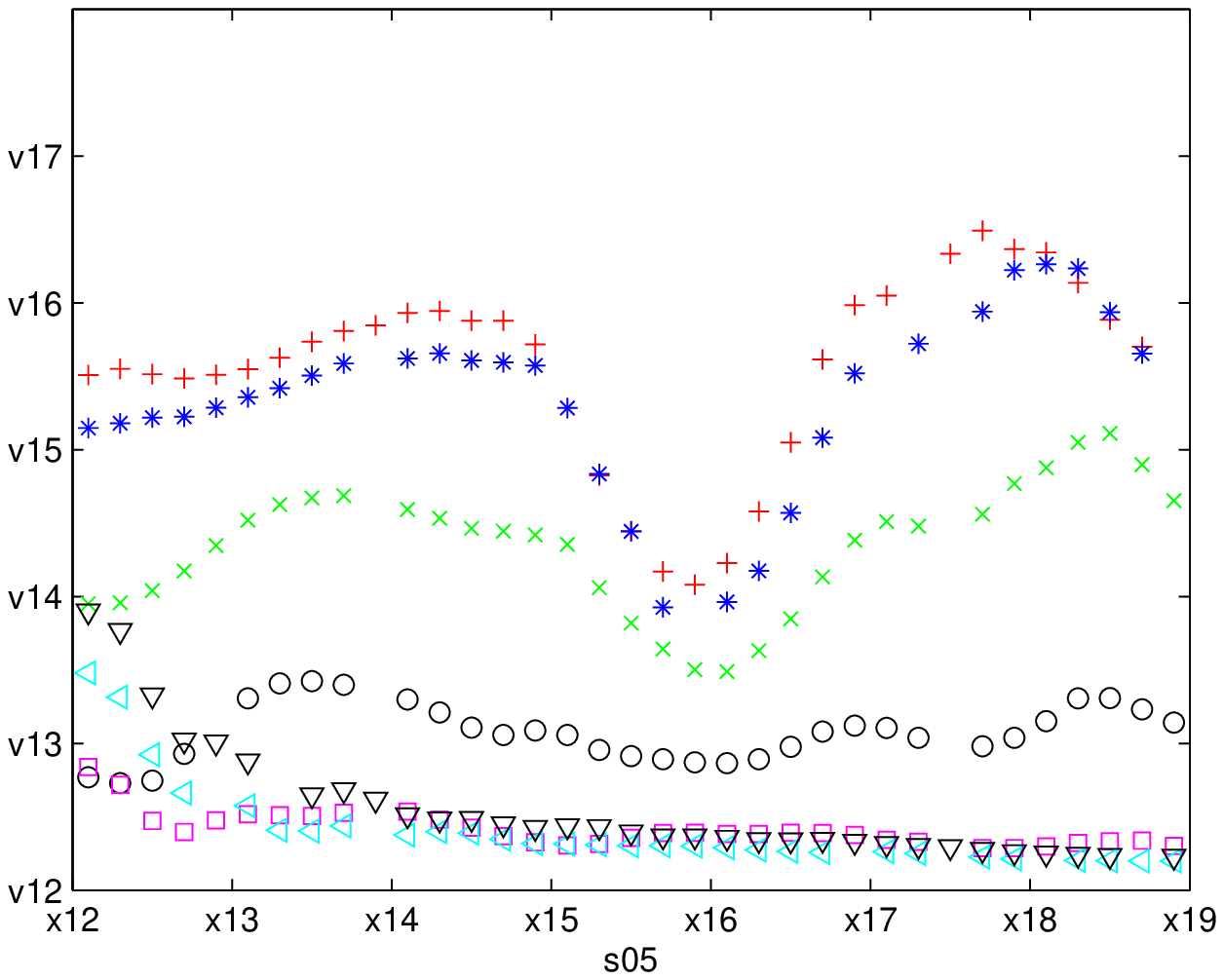}}%
\end{psfrags}%
%

}
\end{minipage}
\\
   \hspace{-.05\columnwidth}
   \begin{minipage}[c]{.45\columnwidth}
\subfigure{
  \Large
%
%
\begin{psfrags}%
\psfragscanon%
%
\psfrag{s05}[t][t]{\color[rgb]{0,0,0}\setlength{\tabcolsep}{0pt}\begin{tabular}{c}\end{tabular}}%
%
\psfrag{x01}[t][t]{0}%
\psfrag{x02}[t][t]{0.1}%
\psfrag{x03}[t][t]{0.2}%
\psfrag{x04}[t][t]{0.3}%
\psfrag{x05}[t][t]{0.4}%
\psfrag{x06}[t][t]{0.5}%
\psfrag{x07}[t][t]{0.6}%
\psfrag{x08}[t][t]{0.7}%
\psfrag{x09}[t][t]{0.8}%
\psfrag{x10}[t][t]{0.9}%
\psfrag{x11}[t][t]{1}%
\psfrag{x12}[t][t]{3}%
\psfrag{x13}[t][t]{4}%
\psfrag{x14}[t][t]{5}%
\psfrag{x15}[t][t]{6}%
\psfrag{x16}[t][t]{7}%
\psfrag{x17}[t][t]{8}%
\psfrag{x18}[t][t]{9}%
\psfrag{x19}[t][t]{10}%
%
\psfrag{v01}[r][r]{0}%
\psfrag{v02}[r][r]{0.1}%
\psfrag{v03}[r][r]{0.2}%
\psfrag{v04}[r][r]{0.3}%
\psfrag{v05}[r][r]{0.4}%
\psfrag{v06}[r][r]{0.5}%
\psfrag{v07}[r][r]{0.6}%
\psfrag{v08}[r][r]{0.7}%
\psfrag{v09}[r][r]{0.8}%
\psfrag{v10}[r][r]{0.9}%
\psfrag{v11}[r][r]{1}%
\psfrag{v12}[r][r]{0}%
\psfrag{v13}[r][r]{0.05}%
\psfrag{v14}[r][r]{0.1}%
\psfrag{v15}[r][r]{0.15}%
\psfrag{v16}[r][r]{0.2}%
\psfrag{v17}[r][r]{0.25}%
\psfrag{v18}[r][r]{0.3}%
\psfrag{v19}[r][r]{0.35}%
\psfrag{v20}[r][r]{0.4}%
%
\resizebox{\columnwidth}{!}{\includegraphics{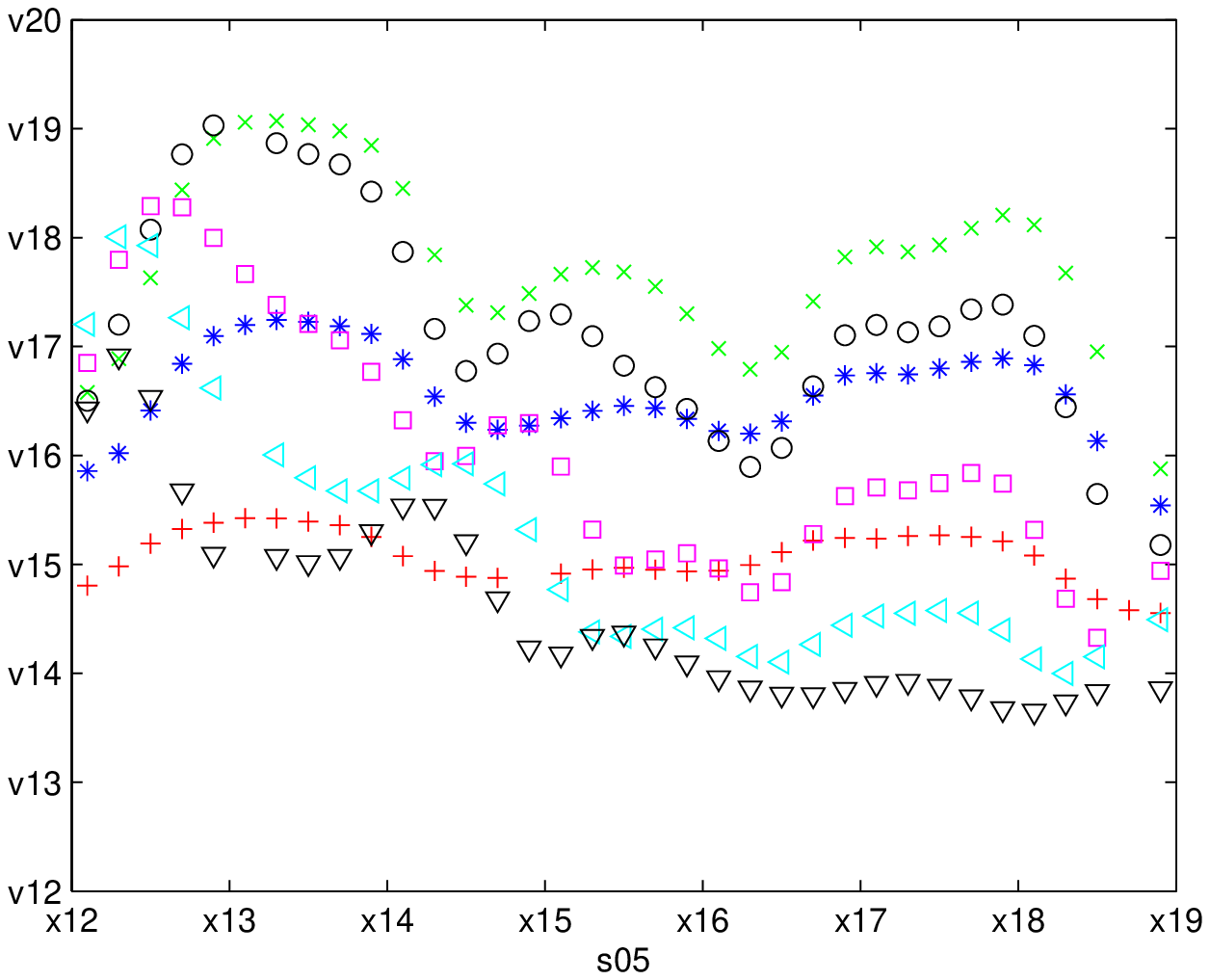}}%
\end{psfrags}%
%

}
\end{minipage}
   \hspace{.05\columnwidth}
   \begin{minipage}[c]{.45\columnwidth}
\subfigure{
  \Large
%
%
\begin{psfrags}%
\psfragscanon%
%
\psfrag{s05}[t][t]{\color[rgb]{0,0,0}\setlength{\tabcolsep}{0pt}\begin{tabular}{c}\end{tabular}}%
%
\psfrag{x01}[t][t]{0}%
\psfrag{x02}[t][t]{0.1}%
\psfrag{x03}[t][t]{0.2}%
\psfrag{x04}[t][t]{0.3}%
\psfrag{x05}[t][t]{0.4}%
\psfrag{x06}[t][t]{0.5}%
\psfrag{x07}[t][t]{0.6}%
\psfrag{x08}[t][t]{0.7}%
\psfrag{x09}[t][t]{0.8}%
\psfrag{x10}[t][t]{0.9}%
\psfrag{x11}[t][t]{1}%
\psfrag{x12}[t][t]{3}%
\psfrag{x13}[t][t]{4}%
\psfrag{x14}[t][t]{5}%
\psfrag{x15}[t][t]{6}%
\psfrag{x16}[t][t]{7}%
\psfrag{x17}[t][t]{8}%
\psfrag{x18}[t][t]{9}%
\psfrag{x19}[t][t]{10}%
%
\psfrag{v01}[r][r]{0}%
\psfrag{v02}[r][r]{0.1}%
\psfrag{v03}[r][r]{0.2}%
\psfrag{v04}[r][r]{0.3}%
\psfrag{v05}[r][r]{0.4}%
\psfrag{v06}[r][r]{0.5}%
\psfrag{v07}[r][r]{0.6}%
\psfrag{v08}[r][r]{0.7}%
\psfrag{v09}[r][r]{0.8}%
\psfrag{v10}[r][r]{0.9}%
\psfrag{v11}[r][r]{1}%
\psfrag{v12}[r][r]{0}%
\psfrag{v13}[r][r]{0.05}%
\psfrag{v14}[r][r]{0.1}%
\psfrag{v15}[r][r]{0.15}%
\psfrag{v16}[r][r]{0.2}%
\psfrag{v17}[r][r]{0.25}%
\psfrag{v18}[r][r]{0.3}%
%
\resizebox{\columnwidth}{!}{\includegraphics{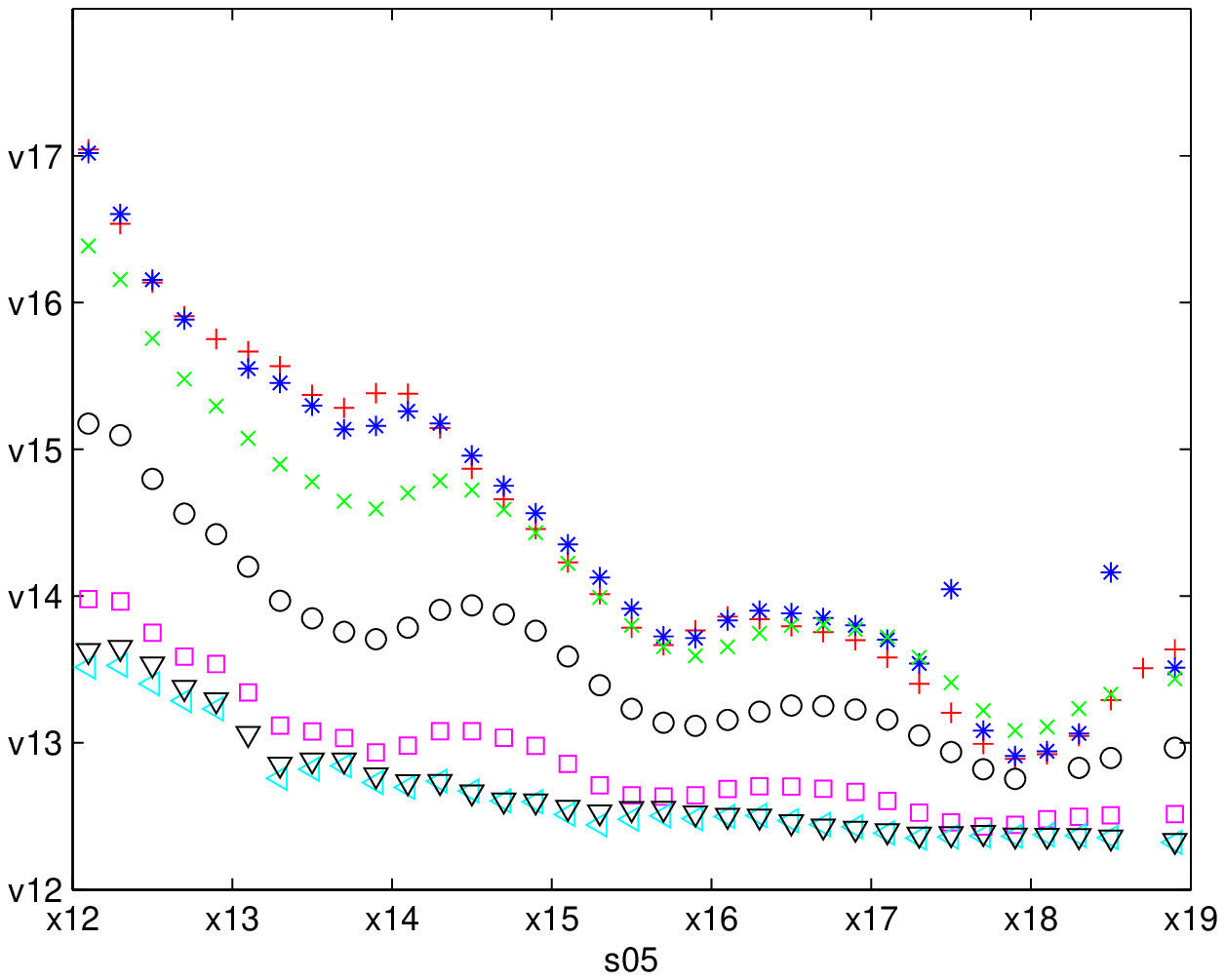}}%
\end{psfrags}%
%

}
\end{minipage}

\caption{Distance between instantaneous reduced density matrix and
  thermal state for $N=3$, and different strengths of the Hamiltonian
  parameter $g$. The initial states vary from $|Y+\rangle$ to
  $|Z+\rangle$, as depicted on the Bloch sphere.  Keeping $h=0.5$
  constant, we compare the case $g=-1.05$ (upper right pane) to
  $g=-0.5$ (lower left) and $g=-1.5$ (lower right).  }
\label{fig:differentG}
\end{figure}


 \begin{figure}[floatfix]
   \hspace{-.05\columnwidth}
   \begin{minipage}[c]{.45\columnwidth}
     \begin{minipage}[c]{.8\columnwidth}
     \subfigure{
       
     }
    
   \end{minipage}
 \end{minipage}
 \hspace{.05\columnwidth}
 \begin{minipage}[c]{.45\columnwidth}
   \subfigure{
     \Large
%
%
\begin{psfrags}%
\psfragscanon%
%
\psfrag{s05}[t][t]{\color[rgb]{0,0,0}\setlength{\tabcolsep}{0pt}\begin{tabular}{c}\end{tabular}}%
%
\psfrag{x01}[t][t]{0}%
\psfrag{x02}[t][t]{0.1}%
\psfrag{x03}[t][t]{0.2}%
\psfrag{x04}[t][t]{0.3}%
\psfrag{x05}[t][t]{0.4}%
\psfrag{x06}[t][t]{0.5}%
\psfrag{x07}[t][t]{0.6}%
\psfrag{x08}[t][t]{0.7}%
\psfrag{x09}[t][t]{0.8}%
\psfrag{x10}[t][t]{0.9}%
\psfrag{x11}[t][t]{1}%
\psfrag{x12}[t][t]{0}%
\psfrag{x13}[t][t]{1}%
\psfrag{x14}[t][t]{2}%
\psfrag{x15}[t][t]{3}%
\psfrag{x16}[t][t]{4}%
\psfrag{x17}[t][t]{5}%
\psfrag{x18}[t][t]{6}%
\psfrag{x19}[t][t]{7}%
\psfrag{x20}[t][t]{8}%
\psfrag{x21}[t][t]{9}%
\psfrag{x22}[t][t]{10}%
%
\psfrag{v01}[r][r]{0}%
\psfrag{v02}[r][r]{0.1}%
\psfrag{v03}[r][r]{0.2}%
\psfrag{v04}[r][r]{0.3}%
\psfrag{v05}[r][r]{0.4}%
\psfrag{v06}[r][r]{0.5}%
\psfrag{v07}[r][r]{0.6}%
\psfrag{v08}[r][r]{0.7}%
\psfrag{v09}[r][r]{0.8}%
\psfrag{v10}[r][r]{0.9}%
\psfrag{v11}[r][r]{1}%
\psfrag{v12}[r][r]{0}%
\psfrag{v13}[r][r]{0.05}%
\psfrag{v14}[r][r]{0.1}%
\psfrag{v15}[r][r]{0.15}%
\psfrag{v16}[r][r]{0.2}%
\psfrag{v17}[r][r]{0.25}%
\psfrag{v18}[r][r]{0.3}%
\psfrag{v19}[r][r]{0.35}%
\psfrag{v20}[r][r]{0.4}%
%
\resizebox{\columnwidth}{!}{\includegraphics{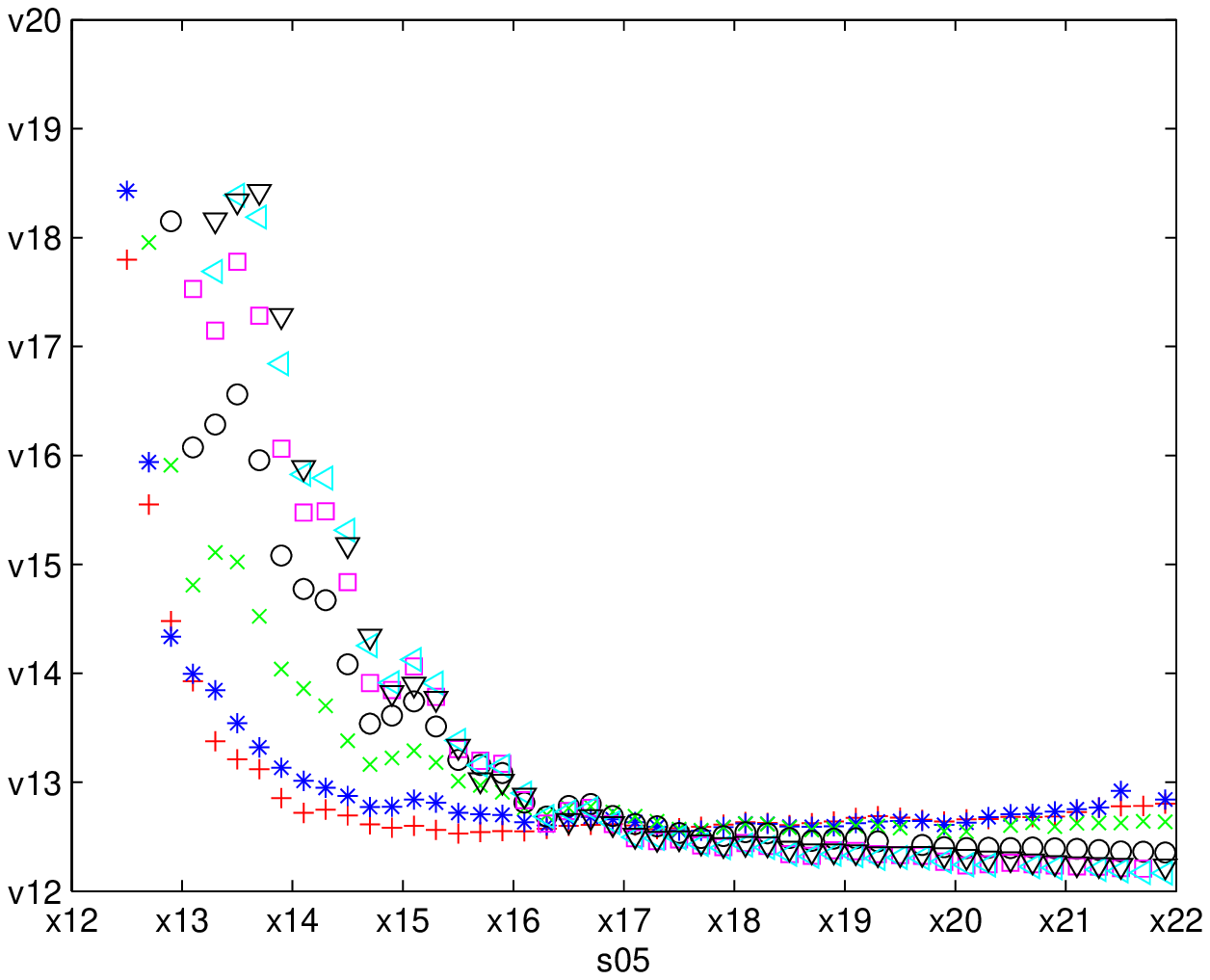}}%
\end{psfrags}%
%

}
\end{minipage}
\\
   \hspace{-.05\columnwidth}
   \begin{minipage}[c]{.45\columnwidth}
\subfigure{
  \Large
%
%
\begin{psfrags}%
\psfragscanon%
%
\psfrag{s05}[t][t]{\color[rgb]{0,0,0}\setlength{\tabcolsep}{0pt}\begin{tabular}{c}\end{tabular}}%
%
\psfrag{x01}[t][t]{0}%
\psfrag{x02}[t][t]{0.1}%
\psfrag{x03}[t][t]{0.2}%
\psfrag{x04}[t][t]{0.3}%
\psfrag{x05}[t][t]{0.4}%
\psfrag{x06}[t][t]{0.5}%
\psfrag{x07}[t][t]{0.6}%
\psfrag{x08}[t][t]{0.7}%
\psfrag{x09}[t][t]{0.8}%
\psfrag{x10}[t][t]{0.9}%
\psfrag{x11}[t][t]{1}%
\psfrag{x12}[t][t]{0}%
\psfrag{x13}[t][t]{1}%
\psfrag{x14}[t][t]{2}%
\psfrag{x15}[t][t]{3}%
\psfrag{x16}[t][t]{4}%
\psfrag{x17}[t][t]{5}%
\psfrag{x18}[t][t]{6}%
\psfrag{x19}[t][t]{7}%
\psfrag{x20}[t][t]{8}%
\psfrag{x21}[t][t]{9}%
\psfrag{x22}[t][t]{10}%
%
\psfrag{v01}[r][r]{0}%
\psfrag{v02}[r][r]{0.1}%
\psfrag{v03}[r][r]{0.2}%
\psfrag{v04}[r][r]{0.3}%
\psfrag{v05}[r][r]{0.4}%
\psfrag{v06}[r][r]{0.5}%
\psfrag{v07}[r][r]{0.6}%
\psfrag{v08}[r][r]{0.7}%
\psfrag{v09}[r][r]{0.8}%
\psfrag{v10}[r][r]{0.9}%
\psfrag{v11}[r][r]{1}%
\psfrag{v12}[r][r]{0}%
\psfrag{v13}[r][r]{0.05}%
\psfrag{v14}[r][r]{0.1}%
\psfrag{v15}[r][r]{0.15}%
\psfrag{v16}[r][r]{0.2}%
\psfrag{v17}[r][r]{0.25}%
\psfrag{v18}[r][r]{0.3}%
\psfrag{v19}[r][r]{0.35}%
\psfrag{v20}[r][r]{0.4}%
%
\resizebox{\columnwidth}{!}{\includegraphics{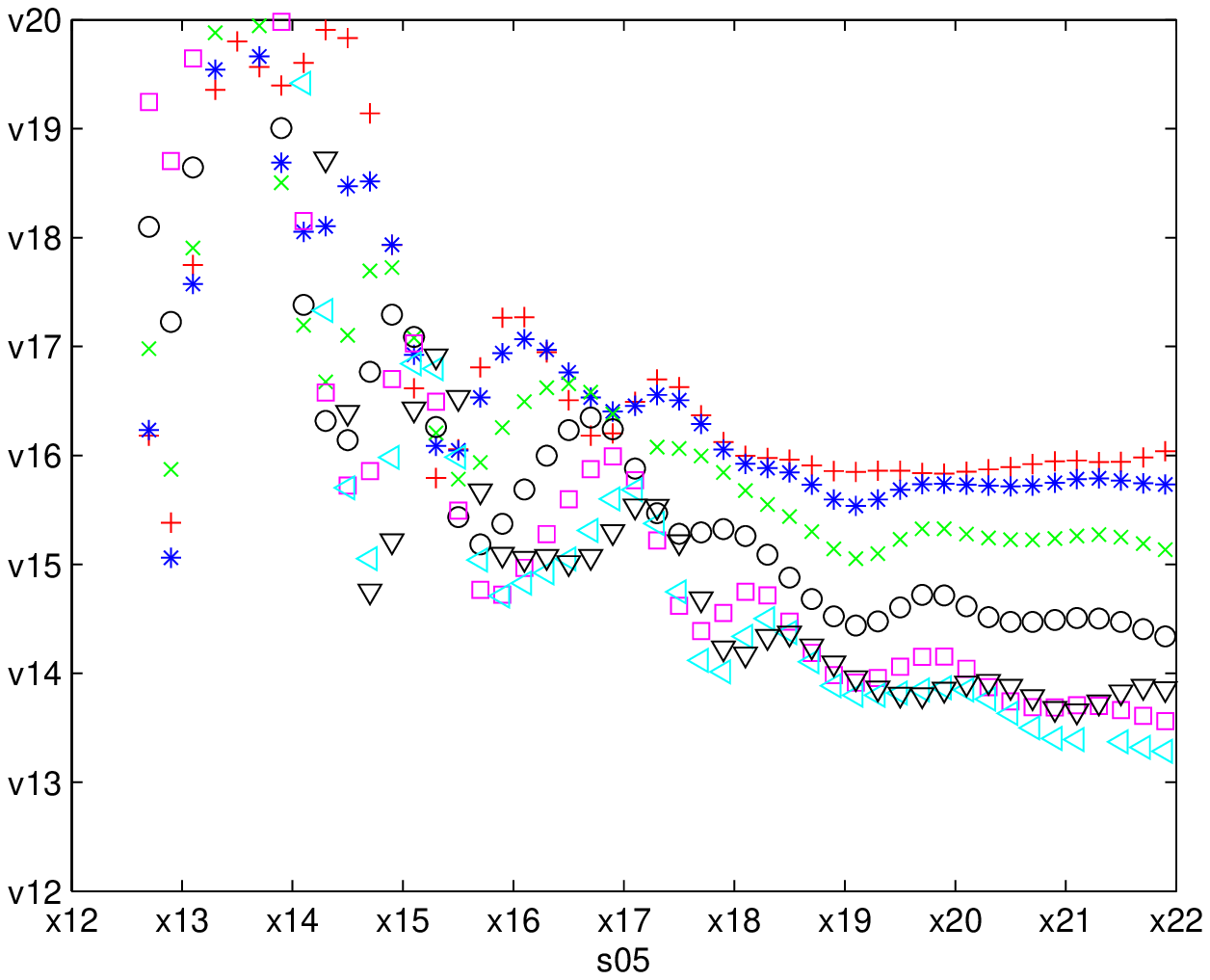}}%
\end{psfrags}%
%

}
\end{minipage}
   \hspace{.05\columnwidth}
   \begin{minipage}[c]{.45\columnwidth}
\subfigure{
  \Large
%
%
\begin{psfrags}%
\psfragscanon%
%
\psfrag{s05}[t][t]{\color[rgb]{0,0,0}\setlength{\tabcolsep}{0pt}\begin{tabular}{c}\end{tabular}}%
%
\psfrag{x01}[t][t]{0}%
\psfrag{x02}[t][t]{0.1}%
\psfrag{x03}[t][t]{0.2}%
\psfrag{x04}[t][t]{0.3}%
\psfrag{x05}[t][t]{0.4}%
\psfrag{x06}[t][t]{0.5}%
\psfrag{x07}[t][t]{0.6}%
\psfrag{x08}[t][t]{0.7}%
\psfrag{x09}[t][t]{0.8}%
\psfrag{x10}[t][t]{0.9}%
\psfrag{x11}[t][t]{1}%
\psfrag{x12}[t][t]{0}%
\psfrag{x13}[t][t]{2}%
\psfrag{x14}[t][t]{4}%
\psfrag{x15}[t][t]{6}%
\psfrag{x16}[t][t]{8}%
\psfrag{x17}[t][t]{10}%
%
\psfrag{v01}[r][r]{0}%
\psfrag{v02}[r][r]{0.1}%
\psfrag{v03}[r][r]{0.2}%
\psfrag{v04}[r][r]{0.3}%
\psfrag{v05}[r][r]{0.4}%
\psfrag{v06}[r][r]{0.5}%
\psfrag{v07}[r][r]{0.6}%
\psfrag{v08}[r][r]{0.7}%
\psfrag{v09}[r][r]{0.8}%
\psfrag{v10}[r][r]{0.9}%
\psfrag{v11}[r][r]{1}%
\psfrag{v12}[r][r]{0}%
\psfrag{v13}[r][r]{0.05}%
\psfrag{v14}[r][r]{0.1}%
\psfrag{v15}[r][r]{0.15}%
\psfrag{v16}[r][r]{0.2}%
\psfrag{v17}[r][r]{0.25}%
\psfrag{v18}[r][r]{0.3}%
\psfrag{v19}[r][r]{0.35}%
\psfrag{v20}[r][r]{0.4}%
%
\resizebox{\columnwidth}{!}{\includegraphics{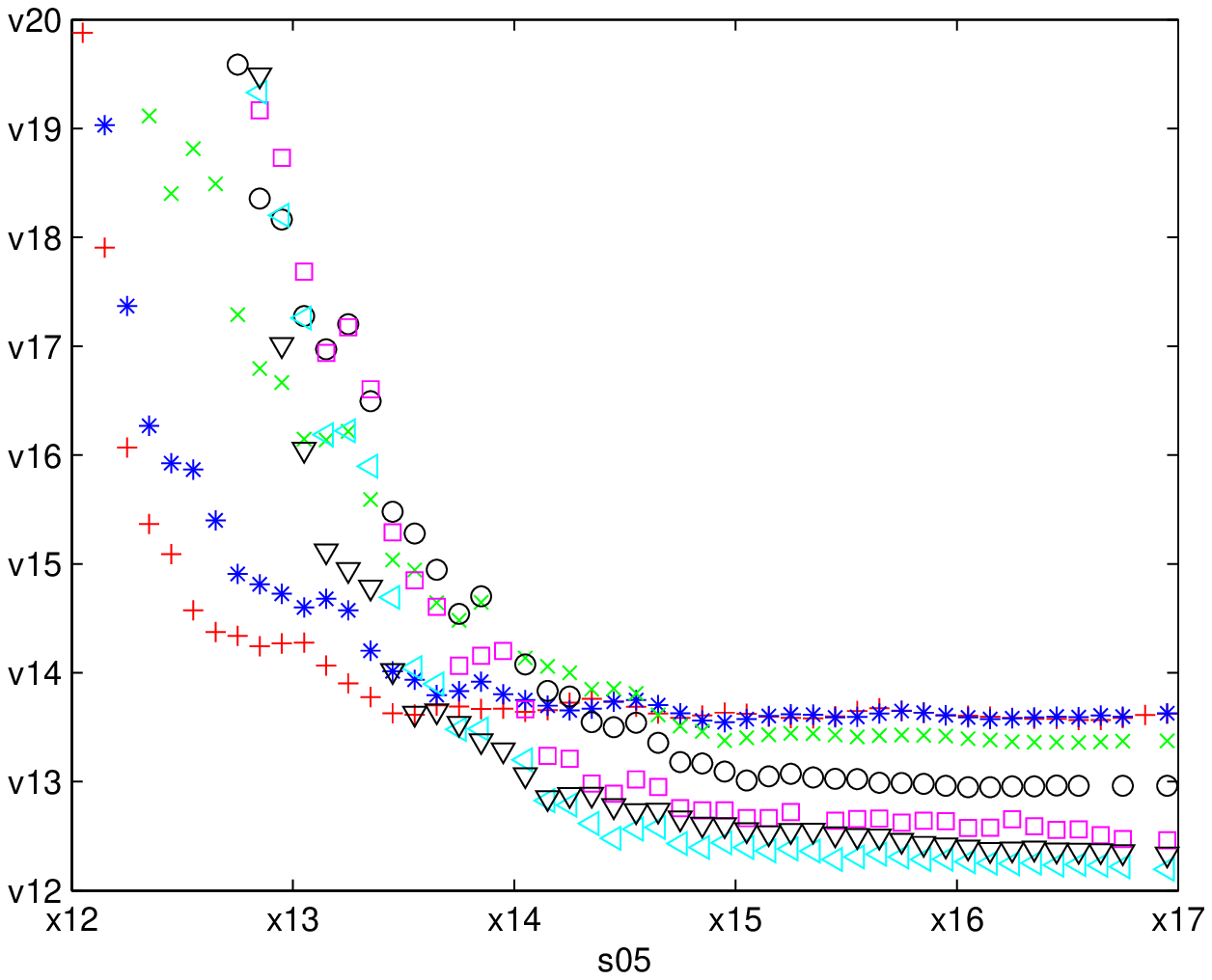}}%
\end{psfrags}%
%

}
\end{minipage}

 \caption{Distance between instantaneous reduced density matrix and thermal state 
for $N=3$, and different strengths of the Hamiltonian parameter $g$. The initial states vary from $|Y+\rangle$ to $|X+\rangle$, as depicted on the Bloch sphere.
For constant parallel field $h=0.5$, we compare the case $g=-1.05$ (upper right pane) to
$g=-0.5$ (lower left) and $g=-1.5$ (lower right).   }
\label{fig:differentG_XY}
\end{figure}

\begin{figure}[floatfix]
  \hspace{-.05\columnwidth}
  \begin{minipage}[c]{.45\columnwidth}
    \subfigure{
      \Large
%
%
\begin{psfrags}%
\psfragscanon%
%
\psfrag{s05}[t][t]{\color[rgb]{0,0,0}\setlength{\tabcolsep}{0pt}\begin{tabular}{c}\end{tabular}}%
%
\psfrag{x01}[t][t]{0}%
\psfrag{x02}[t][t]{0.1}%
\psfrag{x03}[t][t]{0.2}%
\psfrag{x04}[t][t]{0.3}%
\psfrag{x05}[t][t]{0.4}%
\psfrag{x06}[t][t]{0.5}%
\psfrag{x07}[t][t]{0.6}%
\psfrag{x08}[t][t]{0.7}%
\psfrag{x09}[t][t]{0.8}%
\psfrag{x10}[t][t]{0.9}%
\psfrag{x11}[t][t]{1}%
\psfrag{x12}[t][t]{0}%
\psfrag{x13}[t][t]{1}%
\psfrag{x14}[t][t]{2}%
\psfrag{x15}[t][t]{3}%
\psfrag{x16}[t][t]{4}%
\psfrag{x17}[t][t]{5}%
\psfrag{x18}[t][t]{6}%
\psfrag{x19}[t][t]{7}%
\psfrag{x20}[t][t]{8}%
\psfrag{x21}[t][t]{9}%
\psfrag{x22}[t][t]{10}%
%
\psfrag{v01}[r][r]{0}%
\psfrag{v02}[r][r]{0.1}%
\psfrag{v03}[r][r]{0.2}%
\psfrag{v04}[r][r]{0.3}%
\psfrag{v05}[r][r]{0.4}%
\psfrag{v06}[r][r]{0.5}%
\psfrag{v07}[r][r]{0.6}%
\psfrag{v08}[r][r]{0.7}%
\psfrag{v09}[r][r]{0.8}%
\psfrag{v10}[r][r]{0.9}%
\psfrag{v11}[r][r]{1}%
\psfrag{v12}[r][r]{-0.4}%
\psfrag{v13}[r][r]{-0.3}%
\psfrag{v14}[r][r]{-0.2}%
\psfrag{v15}[r][r]{-0.1}%
\psfrag{v16}[r][r]{0}%
\psfrag{v17}[r][r]{0.1}%
\psfrag{v18}[r][r]{0.2}%
\psfrag{v19}[r][r]{0.3}%
\psfrag{v20}[r][r]{0.4}%
%
\resizebox{\columnwidth}{!}{\includegraphics{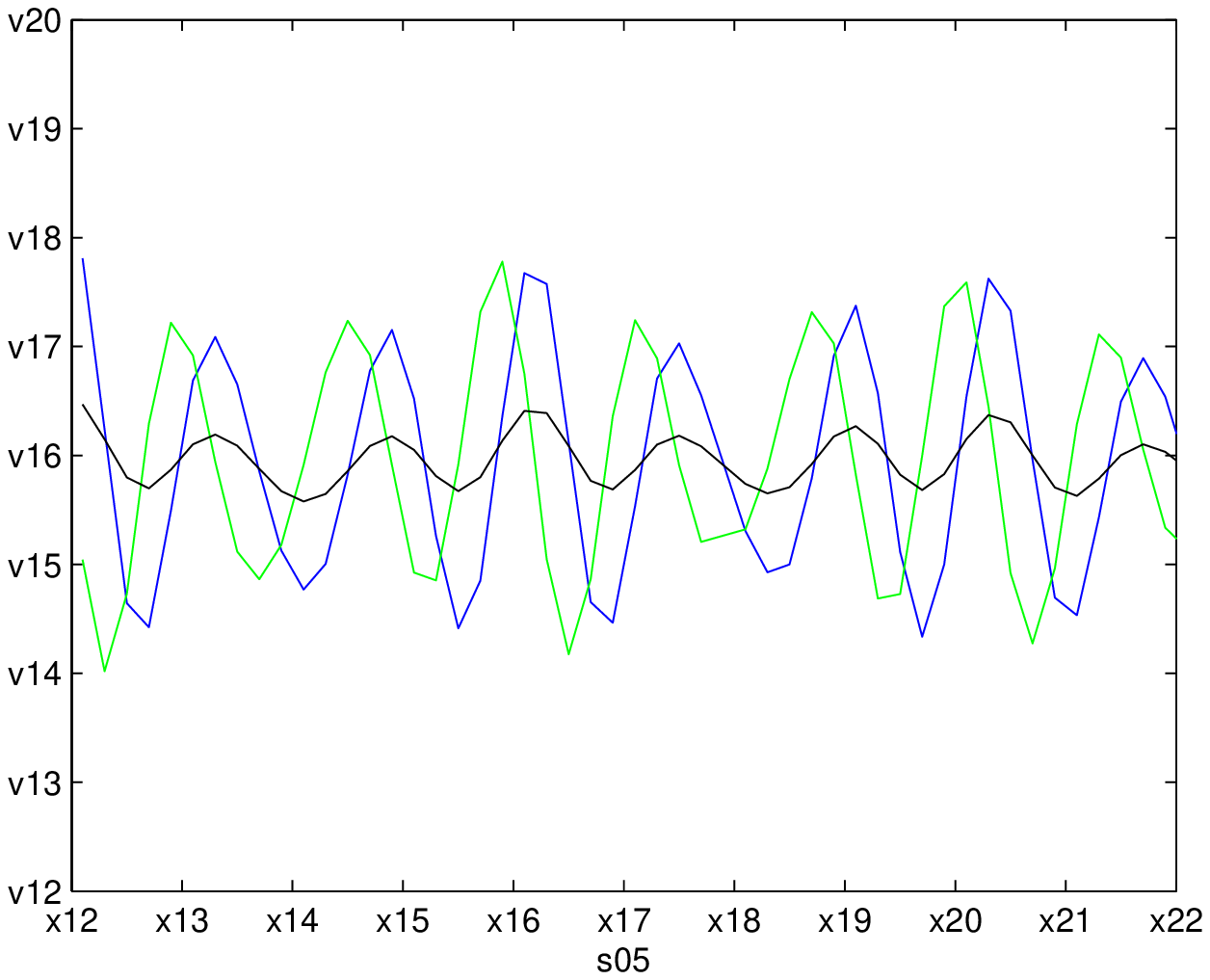}}%
\end{psfrags}%
%

    }
  \end{minipage}
  \hspace{.05\columnwidth}
  \begin{minipage}[c]{.45\columnwidth}
    \subfigure{
      \Large
%
%
\begin{psfrags}%
\psfragscanon%
%
\psfrag{s05}[t][t]{\color[rgb]{0,0,0}\setlength{\tabcolsep}{0pt}\begin{tabular}{c}\end{tabular}}%
%
\psfrag{x01}[t][t]{0}%
\psfrag{x02}[t][t]{0.1}%
\psfrag{x03}[t][t]{0.2}%
\psfrag{x04}[t][t]{0.3}%
\psfrag{x05}[t][t]{0.4}%
\psfrag{x06}[t][t]{0.5}%
\psfrag{x07}[t][t]{0.6}%
\psfrag{x08}[t][t]{0.7}%
\psfrag{x09}[t][t]{0.8}%
\psfrag{x10}[t][t]{0.9}%
\psfrag{x11}[t][t]{1}%
\psfrag{x12}[t][t]{0}%
\psfrag{x13}[t][t]{1}%
\psfrag{x14}[t][t]{2}%
\psfrag{x15}[t][t]{3}%
\psfrag{x16}[t][t]{4}%
\psfrag{x17}[t][t]{5}%
\psfrag{x18}[t][t]{6}%
\psfrag{x19}[t][t]{7}%
\psfrag{x20}[t][t]{8}%
\psfrag{x21}[t][t]{9}%
\psfrag{x22}[t][t]{10}%
%
\psfrag{v01}[r][r]{0}%
\psfrag{v02}[r][r]{0.1}%
\psfrag{v03}[r][r]{0.2}%
\psfrag{v04}[r][r]{0.3}%
\psfrag{v05}[r][r]{0.4}%
\psfrag{v06}[r][r]{0.5}%
\psfrag{v07}[r][r]{0.6}%
\psfrag{v08}[r][r]{0.7}%
\psfrag{v09}[r][r]{0.8}%
\psfrag{v10}[r][r]{0.9}%
\psfrag{v11}[r][r]{1}%
\psfrag{v12}[r][r]{-0.4}%
\psfrag{v13}[r][r]{-0.3}%
\psfrag{v14}[r][r]{-0.2}%
\psfrag{v15}[r][r]{-0.1}%
\psfrag{v16}[r][r]{0}%
\psfrag{v17}[r][r]{0.1}%
\psfrag{v18}[r][r]{0.2}%
\psfrag{v19}[r][r]{0.3}%
\psfrag{v20}[r][r]{0.4}%
%
\resizebox{\columnwidth}{!}{\includegraphics{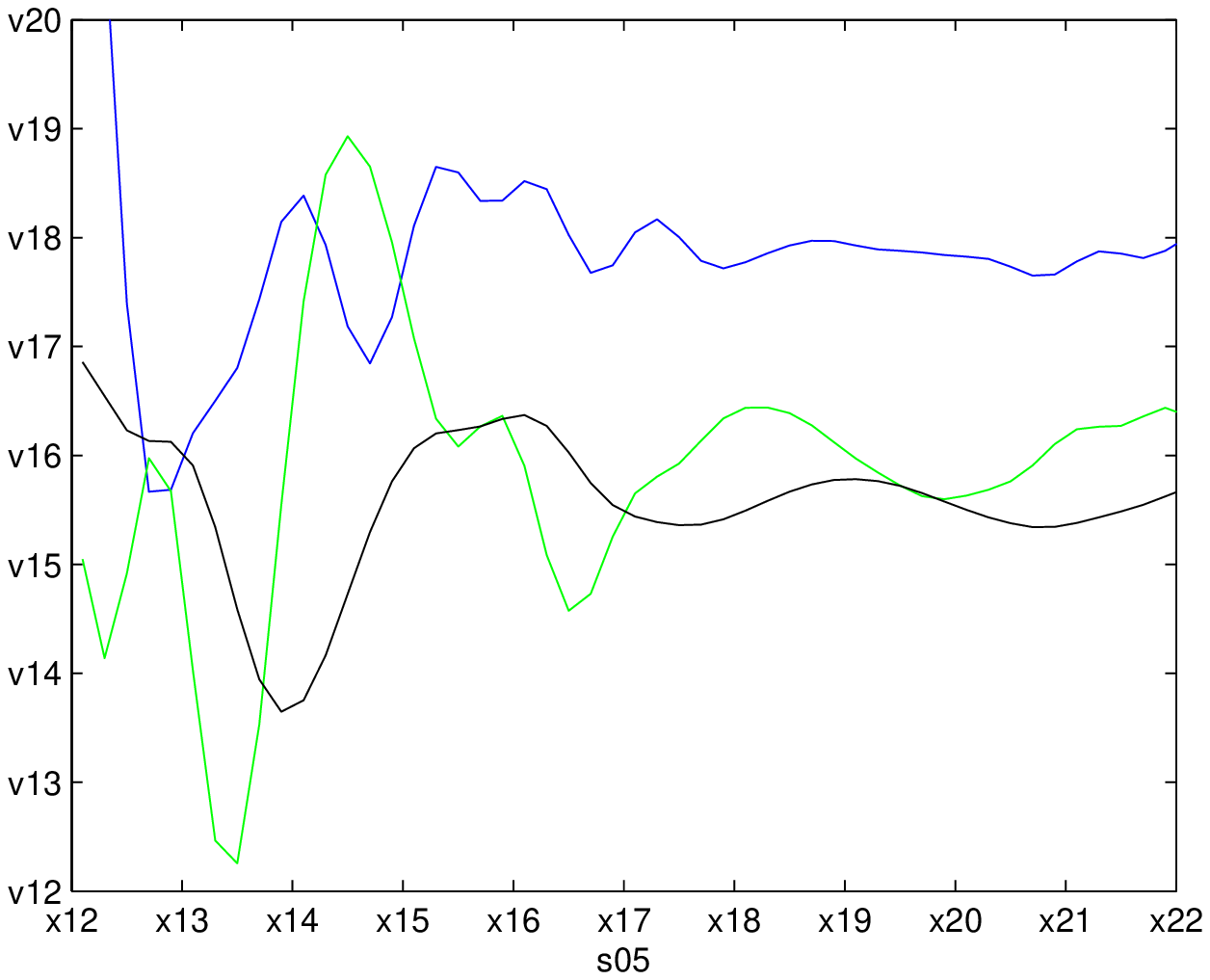}}%
\end{psfrags}%
%

    }
  \end{minipage}
  \\
  \hspace{-.05\columnwidth}
  \begin{minipage}[c]{.45\columnwidth}
    \subfigure{
      \Large
%
%
\begin{psfrags}%
\psfragscanon%
%
\psfrag{s05}[t][t]{\color[rgb]{0,0,0}\setlength{\tabcolsep}{0pt}\begin{tabular}{c}\end{tabular}}%
%
\psfrag{x01}[t][t]{0}%
\psfrag{x02}[t][t]{0.1}%
\psfrag{x03}[t][t]{0.2}%
\psfrag{x04}[t][t]{0.3}%
\psfrag{x05}[t][t]{0.4}%
\psfrag{x06}[t][t]{0.5}%
\psfrag{x07}[t][t]{0.6}%
\psfrag{x08}[t][t]{0.7}%
\psfrag{x09}[t][t]{0.8}%
\psfrag{x10}[t][t]{0.9}%
\psfrag{x11}[t][t]{1}%
\psfrag{x12}[t][t]{0}%
\psfrag{x13}[t][t]{1}%
\psfrag{x14}[t][t]{2}%
\psfrag{x15}[t][t]{3}%
\psfrag{x16}[t][t]{4}%
\psfrag{x17}[t][t]{5}%
\psfrag{x18}[t][t]{6}%
\psfrag{x19}[t][t]{7}%
\psfrag{x20}[t][t]{8}%
\psfrag{x21}[t][t]{9}%
\psfrag{x22}[t][t]{10}%
%
\psfrag{v01}[r][r]{0}%
\psfrag{v02}[r][r]{0.1}%
\psfrag{v03}[r][r]{0.2}%
\psfrag{v04}[r][r]{0.3}%
\psfrag{v05}[r][r]{0.4}%
\psfrag{v06}[r][r]{0.5}%
\psfrag{v07}[r][r]{0.6}%
\psfrag{v08}[r][r]{0.7}%
\psfrag{v09}[r][r]{0.8}%
\psfrag{v10}[r][r]{0.9}%
\psfrag{v11}[r][r]{1}%
\psfrag{v12}[r][r]{-0.4}%
\psfrag{v13}[r][r]{-0.3}%
\psfrag{v14}[r][r]{-0.2}%
\psfrag{v15}[r][r]{-0.1}%
\psfrag{v16}[r][r]{0}%
\psfrag{v17}[r][r]{0.1}%
\psfrag{v18}[r][r]{0.2}%
\psfrag{v19}[r][r]{0.3}%
\psfrag{v20}[r][r]{0.4}%
%
\resizebox{\columnwidth}{!}{\includegraphics{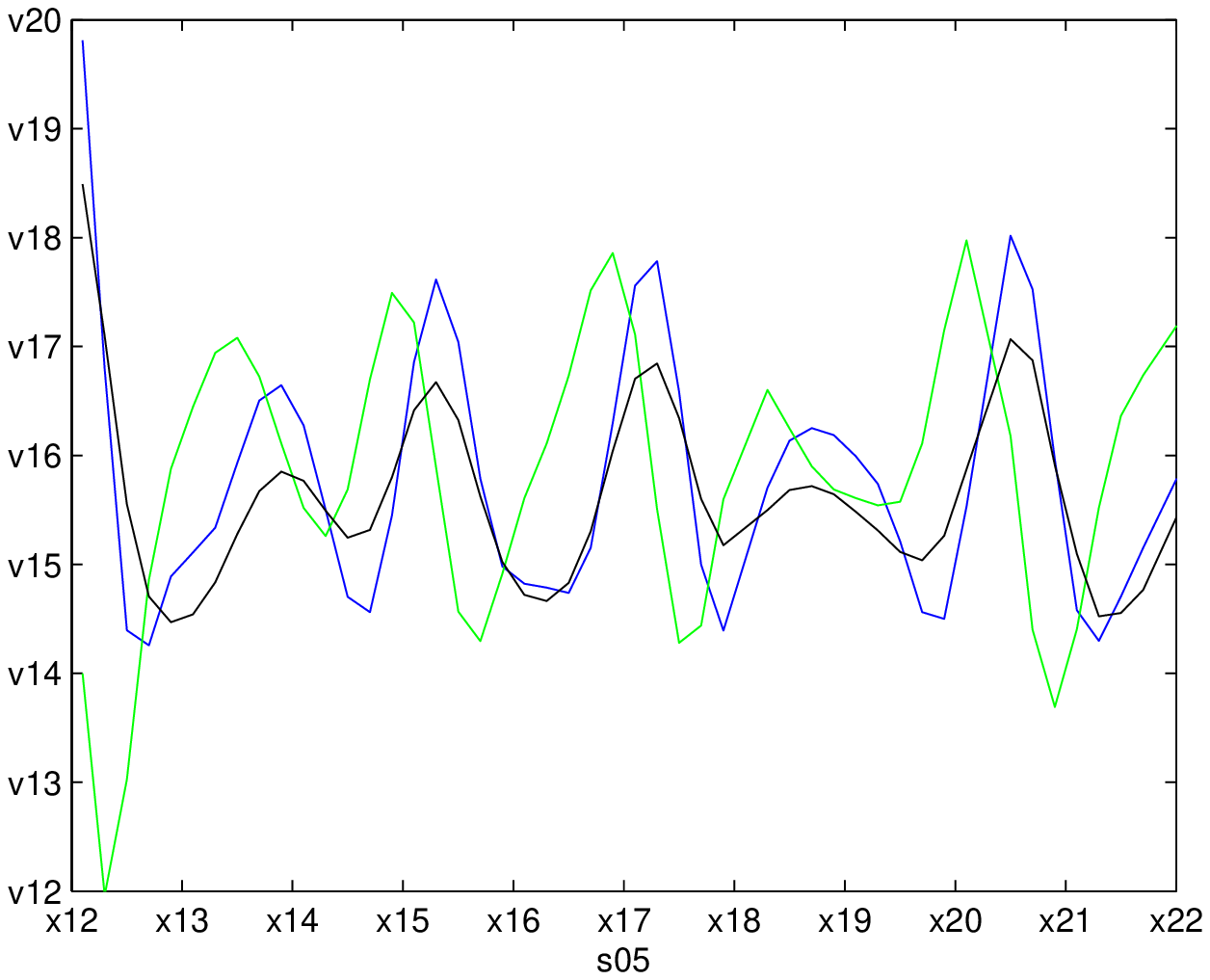}}%
\end{psfrags}%
%

    }
  \end{minipage}
  \hspace{.05\columnwidth}
  \begin{minipage}[c]{.45\columnwidth}
    \subfigure{
      \Large
%
%
\begin{psfrags}%
\psfragscanon%
%
\psfrag{s05}[t][t]{\color[rgb]{0,0,0}\setlength{\tabcolsep}{0pt}\begin{tabular}{c}\end{tabular}}%
%
\psfrag{x01}[t][t]{0}%
\psfrag{x02}[t][t]{0.1}%
\psfrag{x03}[t][t]{0.2}%
\psfrag{x04}[t][t]{0.3}%
\psfrag{x05}[t][t]{0.4}%
\psfrag{x06}[t][t]{0.5}%
\psfrag{x07}[t][t]{0.6}%
\psfrag{x08}[t][t]{0.7}%
\psfrag{x09}[t][t]{0.8}%
\psfrag{x10}[t][t]{0.9}%
\psfrag{x11}[t][t]{1}%
\psfrag{x12}[t][t]{0}%
\psfrag{x13}[t][t]{1}%
\psfrag{x14}[t][t]{2}%
\psfrag{x15}[t][t]{3}%
\psfrag{x16}[t][t]{4}%
\psfrag{x17}[t][t]{5}%
\psfrag{x18}[t][t]{6}%
\psfrag{x19}[t][t]{7}%
\psfrag{x20}[t][t]{8}%
\psfrag{x21}[t][t]{9}%
\psfrag{x22}[t][t]{10}%
%
\psfrag{v01}[r][r]{0}%
\psfrag{v02}[r][r]{0.1}%
\psfrag{v03}[r][r]{0.2}%
\psfrag{v04}[r][r]{0.3}%
\psfrag{v05}[r][r]{0.4}%
\psfrag{v06}[r][r]{0.5}%
\psfrag{v07}[r][r]{0.6}%
\psfrag{v08}[r][r]{0.7}%
\psfrag{v09}[r][r]{0.8}%
\psfrag{v10}[r][r]{0.9}%
\psfrag{v11}[r][r]{1}%
\psfrag{v12}[r][r]{-0.2}%
\psfrag{v13}[r][r]{-0.15}%
\psfrag{v14}[r][r]{-0.1}%
\psfrag{v15}[r][r]{-0.05}%
\psfrag{v16}[r][r]{0}%
\psfrag{v17}[r][r]{0.05}%
\psfrag{v18}[r][r]{0.1}%
\psfrag{v19}[r][r]{0.15}%
\psfrag{v20}[r][r]{0.2}%
%
\resizebox{\columnwidth}{!}{\includegraphics{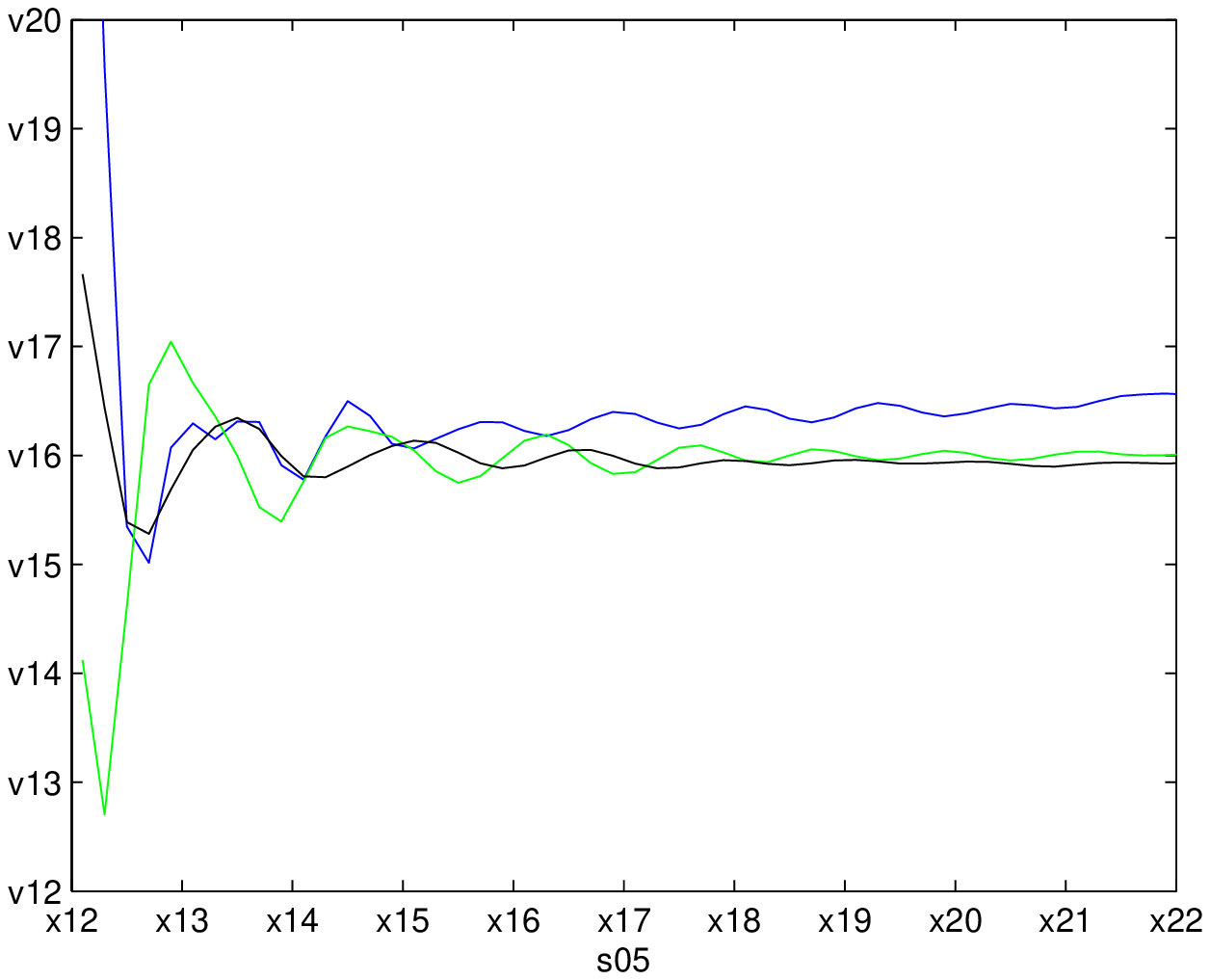}}%
\end{psfrags}%
%

    }
  \end{minipage}
\\
  \hspace{-.05\columnwidth}
  \begin{minipage}[c]{.45\columnwidth}
    \subfigure{
      \Large
      \input{figures/correlsZ+g-1.5.tex}
    }
  \end{minipage}
  \hspace{.05\columnwidth}
  \begin{minipage}[c]{.45\columnwidth}
    \subfigure{
      \Large
      \input{figures/correlsX+g-1.5.tex}
    }
  \end{minipage}
\caption{Distance between the time dependent expectation values of the one-body 
  observables $\langle\sigma_x\rangle$ (blue),
  $\langle\sigma_y\rangle$ (green), $\langle\sigma_z\rangle$ (black) and their thermal values
  for the initial state $|Z+\rangle$ (left column) and $|X+\rangle$ (right column).
  Each row corresponds to a value of the transverse magnetic field $g=-0.5$
  (uppermost), $g=-1.05$ (center) and $g=-1.5$ (bottommost).
}
\label{fig:correlsOther}
\end{figure}

Finally, we have also studied how the Hamiltonian parameters affect
the appearance of the non-thermalizing behaviour, to ensure that this
behaviour is not singular to our particular choice.

As a reference, we may compare the behavior of the same 
initial states under the chosen Hamiltonian 
and the integrable one, corresponding to $h=0$. 
In Fig.~\ref{fig:integrable} we compare the
dynamics under both models, $h=0$ and $h=0.5$, for 
and the three most representative cases.
We observe that in the integrable case, the $N=3$ reduced density matrix
appears to relax fast to a state which is not the thermal one, since 
the distance converges to a value different from zero (This could be compatible
with the generalized thermal ensemble~\cite{rigol07gge}).

We test also other values of the parameters $g$ and $h$ in the
non-integrable regime, to study how the strong and weak thermalizing
regimes appear.  Keeping $h=0.5$ constant, we observe
(Fig.~\ref{fig:differentG}) that both regimes are present for a large
range of values of $g$, but as we decrease $g$, weak thermalization
becomes dominant, while for higher values of $g$, the behaviour
approaches strong thermalization in most cases.

If we do the same now for the transition to non-thermalizing as
observed between $|Y+\rangle$ and $|X+\rangle$, we also observe that
both types of behaviour survive over a wide range of values of $g$
(see Fig.~\ref{fig:differentG_XY}).

To get a more detailed idea of the differences among the various types
of thermalization behavior, we study the individual expectation
values for different observables.  For clarity, we show the plots only
for the $N=1$ operators in Fig.~\ref{fig:correlsOther}, for constant
parallel field $h=0.5$ and varying $g=-0.5,$ $-1.05$, $-1.5$, in two
of the extreme cases, $|Z+\rangle$ and $|X+\rangle$.  We observe that
the oscillating behaviour of the initial state $|Z+\rangle$ appears
clearly correlated with the value of $g$, and for big values the
oscillations of the expectation values are clearly damped.  The
behaviour of the initial state $|X+\rangle$ is quite different, and in
all cases we observe thermalization of some observables while others
deviate from the thermal expectation value.

\end{document}